\documentclass[sigconf]{acmart}
\usepackage{graphicx}
\usepackage{makecell}
\usepackage{amsmath} 

\usepackage{amssymb}
\usepackage{subfig}
\usepackage{algpseudocode} 
\usepackage{gensymb} 
\usepackage{algorithm}

\usepackage[dvipsnames]{xcolor}

\usepackage[dvipsnames]{xcolor}

\AtBeginDocument{%
  }

\setcopyright{none}
\renewcommand\footnotetextcopyrightpermission[1]{}
\acmConference{ArXiv preprint}{April}{2026}

\begin{document}

\title{Real-Time Control of a Virtual Orchestra by Recognition of
Conducting Gestures}




\author{Mert Mermerci}
\affiliation{%
  \institution{KTH Royal Institute of Technology, Sweden}
  \country{\phantom{}\vspace{-\baselineskip}}} 
  \email{mermerci@kth.se}
\author{Emile Pascoe}
\affiliation{%
  \institution{SMASH Studios, Sweden}
  \country{\phantom{}\vspace{-\baselineskip}}}
  \email{emile@smash.studio} 
  \orcid{0009-0005-6425-2076}
\author{Fredrik Edström}
\affiliation{%
  \institution{IVAR Studios, Sweden}
  \country{\phantom{}\vspace{-\baselineskip}}}
    \email{fredrik@ivar.studio}
\author{Hedvig Kjellström}
\affiliation{%
  \institution{KTH Royal Institute of Technology, Sweden}
  \institution{Swedish e-Science Research Centre, Sweden}
  \country{\phantom{}\vspace{-\baselineskip}}}
  \email{hedvig@kth.se}
  \orcid{0000-0002-5750-9655}


\begin{abstract}

We present a museum installation in a 180° dome theater, which gives the museum visitor the experience of conducting a symphony orchestra. We have pre-recorded a short music piece performed by a professional orchestra. This recording is played back in the dome with the visitor standing in the conductor's position. The visitor's gestures are captured with a vision-based skeleton tracker, steering the recording playback pace via a gesture recognition module that translates the gestures into a time control signal. This is sent to a playback module that plays the recording in the dome at the corresponding speed. The gesture recognition module is based on a hierarchical LSTM network, trained with recorded sequences of multiple conductors with different level of expertise conducting the same recording. The system is evaluated with a quantitative study of the estimated timing accuracy, a user study evaluating the musical realism and usability of the real-time control, and a field study to evaluate the performance of the entire system with real museum visitors.

\end{abstract}




\received{20 February 2007}
\received[revised]{12 March 2009}
\received[accepted]{5 June 2009}

\maketitle

\section{Introduction}
\label{sec:intro}  

Larger classical music ensembles are led by a conductor. Conducting is essentially a complex musical sign language to communicate the conductor’s musical intentions and help coordinate the music production of an orchestra, in terms of speed, but also volume, balance, and other nuances. 

Only very few people get the opportunity to try conducting in real life, since it requires access to a live orchestra. Together with the National Museum of Science and Technology in Sweden,
we have developed an immersive interactive installation that allow users to get some of this quite extraordinary experience in a virtual environment. 
In the installation, the user stands in the middle of the museum's dome theater Wisdome Stockholm, surrounded by the $180^{\circ}$ projection of the orchestra recording. 

This paper brings forward three contributions:
\begin{itemize}
    \item A dataset of 12 different conductors conducting the same recording of a music performance (all in all 130 recordings), complete with ground truth annotations providing temporal synchronization with the audio track and music score (Sec.~\ref{sec:dataset}). 
    \item An immersive interaction system that plays back the recorded music performance in the dome theater, in a variable speed controlled by the user's gestures (Sec.~\ref{sec:installation}).
    \item A learning-based gesture recognition module that records user motion using a visual skeleton tracker, and in real-time estimates the phase (position) within a bar of the music. This estimate is transformed by a speed control module into a speed signal that steers the
    speed-controllable recording playback system (Sec.~\ref{sec:mlmodel}).
 \end{itemize}

We present related work in Sec.~\ref{sec:related_work}. The data recording is described in Sec.~\ref{sec:dataset}. We then outline the installation in Sec.~\ref{sec:installation}. The real-time gesture recognition module is described in \ref{sec:mlmodel} and both quantitative and qualitative experimental results 
are presented in Sec.~\ref{sec:evaluation}. The paper is concluded in Sec.~\ref{sec:conclusion}.

\section{Related Work}
\label{sec:related_work}

Our review of related work is divided into three parts: Earlier virtual conducting setups, methods for temporal data modeling, and systems with human motion steering.

\subsection{Virtual Conducting Installations}
The \textit{Personal Orchestra} by Borchers et al.~\cite{borchers04}, still today demonstrated at the House of Music in Vienna, provides an interactive platform for users, allowing them to conduct recordings of the Vienna Philharmonic Orchestra. The system is designed for non-professionals, and is self-instructive. The user conducts recordings, visualized on a screen in front of them.  The baton emits infrared light, received and filtered by the system to extract features influencing the expressiveness of the pre-recorded music. 

Another development in the same direction is the \textit{You're the Conductor} system \cite{youretheconductor}, designed by Lee in collaboration with Borchers et al. This interactive conduction system was presented at the Children’s Museum of Boston, and gamified conducting through hand movements. It catered to diverse user groups, from professional conductors to children. 

Both the above mentioned conducting recognition methods are designed with a rule-based methodology, not fully capturing the motion variability in a performance and the variability between different conductors; conducting is learned very much by practice, and each conductor develops a personal style. Moreover, a difficult aspect to capture with a rule-based system is that timing precision is more crucial during certain moments than others in a performance. 

In contrast to the methods above, the \textit{Home Conducting} system, introduced by Friberg et al.~\cite{homeconducting}, shifts the emphasis to fluidity. Users can interact with the system without the need for baton or gloves via optical flow features from a webcam. Musical expression and speed is derived from changes in the overall quantity of user motion.

We instead propose to use a data-driven learning-based method for capturing the user's gestures and transforming them into commands for music production timing and speed (Sec.~\ref{sec:mlmodel}). In this way, our approach constitutes a middle ground, allowing for strict timing control when appropriate but also giving room for individual variations in motion. Moreover, the presented installation will give a more immersive experience than its predecessors, thanks to the 180$^{\circ}$ projection in the dome (Sec.~\ref{sec:installation}).

\subsection{Real-Time Control in Immersive HCI}

Real-time motion control is well established in robotics, supporting tele-operation and shared autonomy in safety-critical domains such as search-and-rescue \cite{li2022} and surgical assistance \cite{rahman2020}, where latency, stability, and safety are hard requirements \cite{lasota2017survey}. In musical and artistic contexts, motion has similarly served as an expressive control modality: systems like \textit{The Wekinator} \cite{fiebrink2010wekinator} enable real-time gesture-to-sound mapping for live performance, while more recent work extends this into immersive VR 
\cite{flowwiththebeat, nordin2024}.

Immersive visualization environments combine wide field-of-view projection, spatialized audio, and interactive navigation to support exploration of complex data. Platforms such as OpenSpace \cite{openspace} demonstrate how dome theaters can serve both scientific research and public engagement through real-time phenomena exploration, yet how users maintain orientation and construct meaning within such environments remains an open challenge in HCI. 
However, users typically interact through embodied and spatial cues rather than explicit interfaces, making accurate mental model formation difficult. The \textit{Guidelines for Human–AI Interaction} \cite{guidelinesforhumanai} address this through design principles for expectation-setting, communicating system limitations, and providing interpretable feedback. In immersive contexts, opaque system responses risk disrupting presence and eroding trust, making the balance between automation and user agency particularly consequential. Evaluation must further extend beyond laboratory settings, as field studies expose systems to environmental, social, and contextual factors that controlled conditions cannot replicate \cite{IntroToHCIFieldEvaluations}.

Our system is the first interactive immersive installation in Wisdome Stockholm, enabling visitors to engage through full-body movement rather than conventional interface devices. This embodied interaction modality aligns user action with 
perception, strengthening the sense of presence within the dome environment.

\subsection{Modeling Temporal Data}
To realize the learning-based recognition of timing signals from conducting gestures, we need a method for regression of a time signal from a human motion time-series. A recent survey by Foumani et al.~\cite{foumani2024} gives an overview of different supervised approaches including convolutional neural network (CNN), recurrent neural network (RNN) and attention-based approaches. While CNN- and RNN-based approaches mainly capture local patterns, attention-based approaches also model more long-range dependencies. 

The state-of-the art method for modeling of complex time series (e.g.~word sequences in language) is the attention-based Transformer model \cite{ahmed2023,transformer}. Transformers allow for modeling of highly complex long-range dependencies -- a key to their success in language modeling -- but this comes at the cost of requiring large training data, which is problematic in the present application. 

However, the correlations in the conducting  
time series are only local, within one or two bars of music. Therefore, the full potential of a Transformer model would not be needed. We therefore use a considerably less data-hungry RNN alternative and employ long-short-term memory (LSTM) \cite{LSTM} networks, 
with a flexible memory function that allows the network to selectively remember or discard information in the history of time steps in the sequence.
LSTMs have become a de facto standard in domains such as human motion prediction \cite{OnHumanMotionPrediction}, and action recognition \cite{SkeletonBasedActionRecognition}. In particular, our architecture, described in Section \ref{sec:ConductingGestureRecognition}, is inspired from networks for gait phase estimation \cite{lu2022,sarshar2021,su2020}, a highly related task.

\section{Dataset}
\label{sec:dataset}

Two data collections were made, 
described below. The motion capture data and labels will be made available at publication.
\subsection{Orchestra and Original Conductor}
\label{sec:original}
For the purpose of our museum installation, we recorded the Swedish Radio Symphony Orchestra, led by its chief conductor Daniel Harding,
performing the opening of Beethoven Symphony no 5. 

The recording was made with a 270$^{\circ}$ wide angle camera covering the entire concert hall. The orchestra's sound technicians made a professional sound recording of the performance for the dome video soundtrack.

The body and hand 3D pose of the conductor were recorded simultaneously, using a Rokoko Smartsuit Pro V2 and Rokoko Smart Gloves. It should be noted that the conductor is the only one not captured in the video recording, since they will be replaced by the user of the museum installation. However, the conductor of the original recording is extremely important in order to shape the recorded orchestra performance. 
The conductor pose was recorded in 60 Hz, and at each timestep, a human skeleton parameterized by joint angles was obtained from the suit sensor readings using Rokoko Studio. The joint angle representation was transformed into a representation with 3D joint locations using Blender. 

Let $\mathbf{f}^0$ be the pose sequence of the original conductor. The frame at timestep $t$ is denoted $\mathbf{f}^0_t \in \mathbb{R}^{3N}$, i.e., the 3D positions of each of the $N=61$ joints in the model skeleton, expressed in a fixed world coordinate system. Each timestep $t$ corresponds to a time position $t/60$ seconds into the audio and video files, which are temporally synchronized with $\mathbf{f}^0$. 

\begin{figure}[t]
    \centering
    \includegraphics[width=0.9\linewidth]{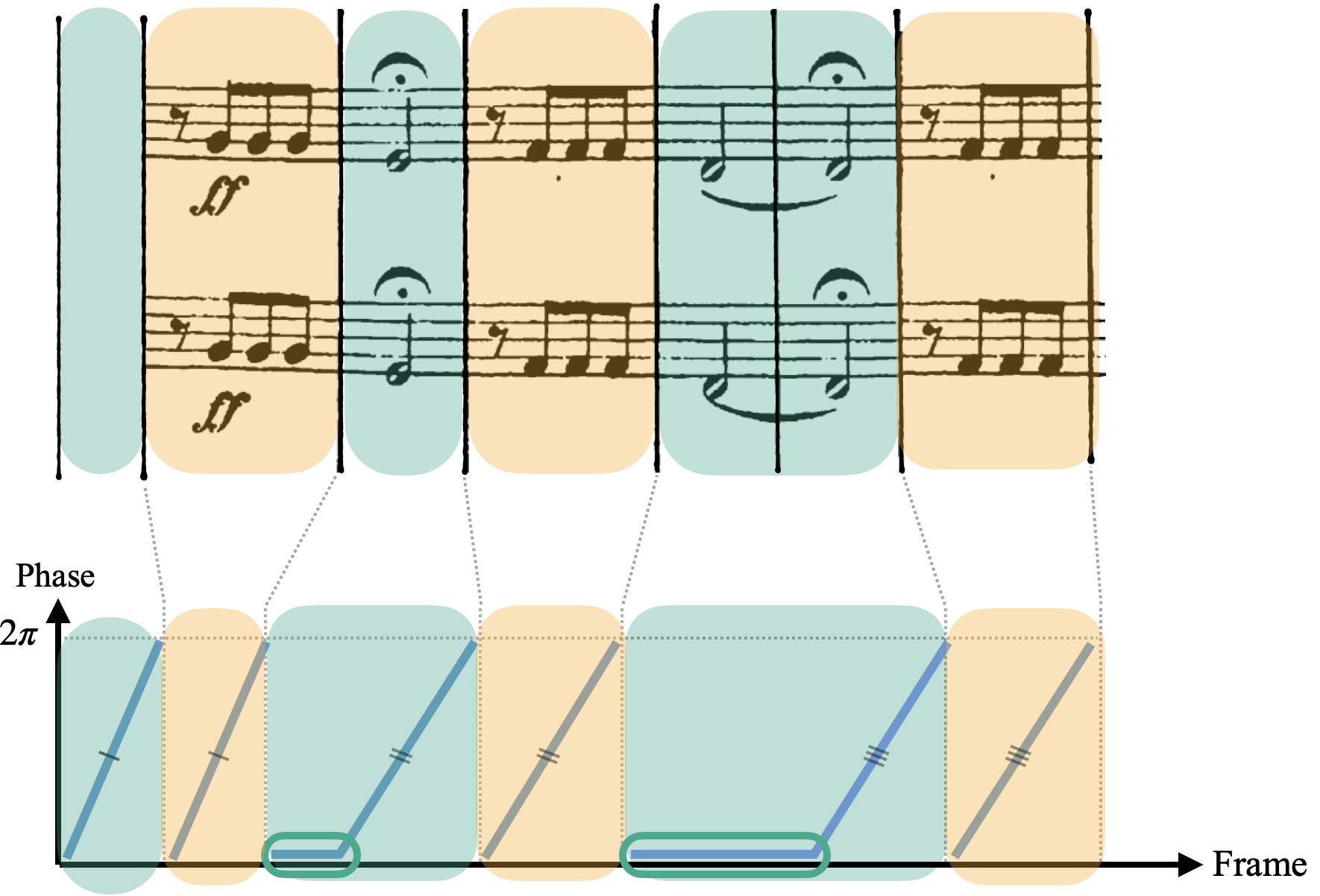}
    \caption{Time labeling. Regular bars are yellow and bars (including bar 0) containing a waiting time followed by a silent upbeat are green. Waiting times during fermatas are circled with green.}
    \label{fig:fermataill}
\end{figure}

The video, audio and pose recordings were manually synchronized with the music score (see Fig.~\ref{fig:fermataill}), defining the notes to be played for each instrument and divided into 122 numbered bars.\footnote{Originally 124 bars, but in our labeling we have merged bars 4-5 and 23-24 since they in practice are treated as one bar by both the conductor and orchestra.} This piece is conducted in one-beat which means that the conductor gives one (down)beat per bar, marking the start of the bar. 

The progression through each bar can be represented as a phase variable, progressing from $0$ to $2\pi$ as shown in Fig.~\ref{fig:fermataill}, lower graph. This means that for each time frame $t$, there is a corresponding ground truth label $(b^\mathrm{gt}_t,\varphi^\mathrm{gt}_t)$ where $b^\mathrm{gt}_t$ is the bar number at time $t$, and $\varphi^\mathrm{gt}_t$ is the phase in this bar. 

In addition, when the orchestra is starting -- at the very start or after a fermata\footnote{A fermata is a waiting point in the music, denoted \includegraphics[width=3mm,height=1.6mm]{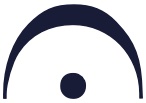} where the orchestra will wait until given an upbeat by the conductor.} -- the conductor will give a silent upbeat to indicate to the orchestra when it is time to play and at what pace to play the next bar. Bars $B=\{2, 4, 20, 22\}$ have a fermata in them, all other bars are meant to be played in tempo. A professional conductor used the software Sonic Visualizer to mark silent upbeats and the following regular bar start (5 instances in total) with audible clicks on the audio file to enable other conductors to predict them when listening to the audio. 

\subsection{Additional Conductors}
\label{sec:additional}
At this point we had one conductor recording $\mathbf{f}^0$. In order to train the gesture recognition system presented in Sec.~\ref{sec:mlmodel}, more examples of conducting motion are needed. 

We therefore used a mock-up procedure and made 130 recordings from 12 additional individuals while they conducted the recording, instructed to move as they would have done in order to produce the recorded performance. We used the same motion capture suit as for the original conductor. Let $\mathbf{f}^i, i\in[1,130]$ denote the mocap sequences of the additional conductors. 

The subjects (six men and six women) included six professional orchestra conductors, one professional choir conductor, two conducting students, and three amateur musicians. The subjects had the score in front of them and prepared by listening to the recording multiple times. They did not get to see the original conductor motion. Each conductor was recorded 7 to 17 times. The recordings of the four left-handed subjects were mirrored left-to-right. The more experienced subjects made some recordings at slightly slower and faster pace 
to increase the variability of the training set. 

\paragraph{Normalization, velocity and acceleration.}
The sequences $\mathbf{f}^i$ were normalized by moving the world origin to the hip joint position in each frame $t$, and normalizing the subject height to 1.
Moreover, the velocity and acceleration of frame $t$ in sequence $i$ were:
\begin{equation}
\begin{aligned}
\mathbf{v}^i_t &= \mathbf{f}^i_t - \mathbf{f}^i_{t-1} ,\\
\mathbf{a}^i_t &= \mathbf{f}^i_t - 2\mathbf{f}^i_{t-1} + \mathbf{f}^i_{t-2} .
\end{aligned}
\end{equation}
Frame $t$ of sequence $i$ is denoted $\mathbf{x}_t^i = (\mathbf{f}_t^i,\mathbf{v}_t^i,\mathbf{a}_t^i) \in \mathbb{R}^{9N}$.

\paragraph{2D upper body pose.}
\label{sec:upperbody}
 In the museum installation (Sec.~\ref{sec:installation}), the user's upper body pose in 2D is captured with MediaPipe Pose Landmarker,\footnote{The MediaPipe framework, including the Pose Landmarker, is available as an open-source platform on Google's GitHub repository.} which is very robust and thus suitable in a public museum setting. 9 keypoints are retained that show the most stable performance (r/l shoulder, r/l elbow, r/l wrist, r/l hand, hip center computed as mean of r/l hip). 
To emulate this representation, a low-dimensional version of each Rokoko pose frame $\mathbf{f}^i_t$ was created by maintaining the $x$ and $y$ coordinates of $n=9$ joints (r/l shoulder, r/l elbow, r/l wrist, r/l hand,  hip center). This produces a 2D pose frame $\mathbf{f}^{\mathrm{2D},i}_t \in \mathbb{R}^{2n}$ and a sequence frame $\mathbf{x}^{\mathrm{2D},i}_t \in \mathbb{R}^{6n}$.
Please see Appendix \ref{sec:appendixRecording} for more information on the dataset.
\section{Dome Installation}
\label{sec:installation}
\label{sec:Dome Installation}

The installation we have developed allows users to experience conducting a symphony orchestra, standing in the center of a dome theater, surrounded by the $180^{\circ}$ projection of the orchestra recording. The application was built in Unreal Engine 5.5 \footnote{Unreal Engine 5.5 (\url{https://www.unrealengine.com}), with the Off World Live Streaming Toolkit (\url{https://offworld.live}) and MetaHuman (\url{https://www.unrealengine.com/metahuman}) plugins.}, with a dome version using the Off World Live plugin to combine rendered images into a single dome projection frame, streamed to the venue's rendering software via NDI. In addition to the recorded orchestra video, users can optionally display a conductor avatar for guidance (the movements of the conductor from the original recording ) and a gender-neutral MetaHuman avatar representing themselves, animated from upper-body pose tracking. A musical stave scrolling left-to-right provides visual feedback on progress and tempo, with fermata bars highlighted; an end screen summarizes the user's conducted tempo relative to the original recording. An example dome screen is presented in Figure \ref{fig:domeescreen} in Appendix \ref{sec:appendixInstallation}.

The interactive system operates at 20 Hz (governed by the available hardware and the computational requirements of pose capture, gesture recognition, and dome visualization). In each timestep of $k$ of the interaction session,\footnote{We use $t$ to denote a frame number in the recorded conducting sequences (i.e., the timing of the original performance), and $k$ to denote a time-step in the real-time user interaction with the system (i.e., the timing of the recreated virtual performance).}the user's pose and motion $\mathbf{x}^\mathrm{2D}_k$ is recorded and computed as described in Sec.~\ref{sec:upperbody}, and fed into the gesture recognition module, see Sec.~\ref{sec:ConductingGestureRecognition}, resulting in an estimated phase $\hat{\varphi}_k$. This phase is fed into the control module, see Sec.~\ref{sec:RealTimeController}, which estimates a recording playback speed $s_k$, the signal by which the user steers the pace of the recording. In addition, as described in Sec.~\ref{sec:dataset}, this piece contains a number of fermata bars, which can take an arbitrarily long time. We also implemented this notion in our audio playback system. In these bars, the playback continues in the present speed; if it reaches the end of the bar, the playback stops. When a new speed command $s_k$ is sent from the speed control module, the playback jumps to the beginning of next bar and continues in the new speed. Audio time-stretching uses the WOLA algorithm via an Unreal Engine plugin, which minimised the artefacts present in simpler pitch-shift approaches. For more information on the installation please check Appendix \ref{sec:appendixInstallation}. 

\begin{figure*}[t]
    \centerline{\includegraphics[width=0.96\linewidth]{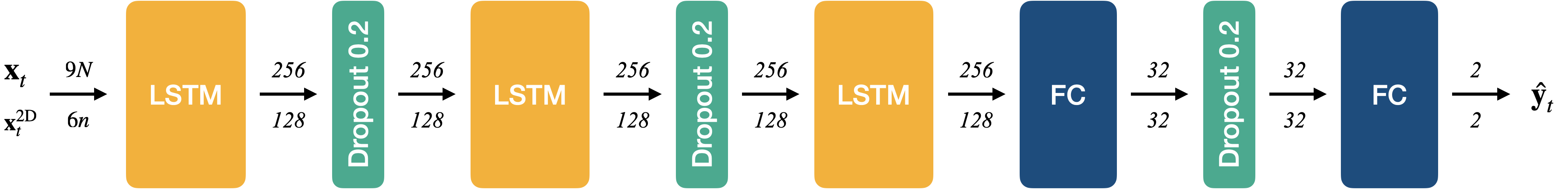}}
    \caption{Architecture of the gesture recognition network. }
    \label{fig:network_illustration}
\end{figure*}





\subsection{Rudimentary Setup for the User Study}


We also design a rudimentary audio playback application using the Python libraries \texttt{librosa} and \texttt{sounddevice}. Like the dome video application, this application takes in a recording playback speed $s_k$ and plays the (compressed) audio file at this variable speed. While not as immersive as the actual setup, this lightweight implementation is adequate for testing the control feedback loop and offers an additional advantage: it does not mask or compensate for errors in the control signal, making deviations more directly observable during user evaluation. However, this uncompensated response led to some irritation due to the sound artefacts. 

\section{Real-Time Conducting Gesture Recognition}
\label{sec:mlmodel}

We now describe the 
method for estimating the desired recording playback speed from the user's pose and motion. The method consists of two steps, described below.

\subsection{Conducting Gesture Recognition}
\label{sec:ConductingGestureRecognition}

We employ a network architecture with LSTM layers to capture temporal patterns in the conducting motion (see Fig.~\ref{fig:network_illustration}).

The input is the pose and motion of the user at time $t$, either the 3D full body and hand pose and motion $\mathbf{x}_t$ in Rokoko format (Sec.~\ref{sec:additional}), or the 2D upper body pose and motion $\mathbf{x}^\mathrm{2D}_t$ in MediaPipe format (Sec.~\ref{sec:upperbody}). We also use two temporal resolutions. One variant of the network operates in 60 Hz, the training sequence frequency. For the purpose of the real-time controller, we also develop a variant that operates in 20 Hz. These alternatives are evaluated in Sec.~\ref{sec:evaluation_grm}.

The network includes three stacked LSTM layers, followed by two fully connected linear layers. To mitigate overfitting due to the limited dataset size, dropout layers are applied between certain layers. As described in Sec.~\ref{sec:related_work}, this choice of architecture is inspired by gait analysis applications \cite{lu2022,sarshar2021,su2020} with similar temporal pattern recognition tasks. A comparative analysis with a Kalman filter \cite{welch1995introduction} is provided in Appendix \ref{sec:kalman2}, showing that the LSTM network is able to capture fast phase velocity changes better.

Although the data are labeled with both bar number $b$ and phase within the bar $\varphi$, the network only estimates phase. The reasons are that 1) the correlation between beating pattern and bar number is very weak which makes the estimation problem much more difficult, 2) such a model would require much more training data, 3) such a model would be specific to this particular piece while the present model can be used for any piece in one-beat.

As shown in Fig.~\ref{fig:fermataill}, the phase signal $\varphi$ contains sharp changes from $2\pi$ to $0$. From design, a neural network will model a smooth signal better than a signal with sharp changes. We therefore let the network estimate two smooth signals that contain the same information $\varphi$: 
$\hat{\mathbf{y}}_t = (\hat{\sigma}_t, \hat{\kappa}_t)$
where $\sigma_t \simeq \sin(\varphi_t)$ and $\kappa_t \simeq \cos(\varphi_t)$. 
The estimated phase is
$\hat{\varphi}_t = 
    \mathrm{arctan}(\frac{\hat{\sigma}_{t}}{\hat{\kappa}_{t}}) + \pi$. 

We employ a loss function minimizing MSE over time: 
\begin{equation}
L_\mathrm{MSE} = \frac{1}{2T} \sum_{t=1}^{T}  \left( \hat{\mathbf{y}}_t - \mathbf{y}^\mathrm{gt}_t \right) \left( \hat{\mathbf{y}}_t - \mathbf{y}^\mathrm{gt}_t \right)^\mathrm{T}
\label{eq:mseloss}
\end{equation}
where $\mathbf{y}^\mathrm{gt}_t = (\sin(\mathbf{\varphi}^\mathrm{gt}_t), \cos(\mathbf{\varphi}^\mathrm{gt}_t))$, $T$ the sequence length.

We also encourage monotonicity by introducing a loss on negative phase change over time:
\begin{equation}
 L_\mathrm{mono}=
\frac{1}{T} \sum_{t=1}^{T}  \Delta t^\mathrm{neg}_t
\label{eq:monotonicityloss}
\end{equation}
where $\Delta t^\mathrm{neg}_t$ is defined (handling $0 \equiv 2\pi$) as:
\begin{equation}
\begin{aligned}
\Delta t_t &= \mathrm{mod}(\hat{\varphi}_t - \hat{\varphi}_{t-1} + \pi, 2\pi) - \pi \\
\Delta t^\mathrm{neg}_t &= 
\begin{cases}
-\Delta t_t, & \Delta t_t < \epsilon,\\ 0, & \Delta t_t \geq \epsilon 
\end{cases} ,\hspace{8mm}  \epsilon = -10^{-7} .
\label{eq:delta}
\end{aligned}
\end{equation}

The total loss combines these two terms using a factor $\beta$:
\begin{equation}
L = L_{\mathrm{MSE}} + \beta \times L_{\mathrm{mono}}.
\label{eq:total_loss}
\end{equation}

\subsubsection{Training procedure for gesture recognition evaluation}
In order to enable evaluation of the generalized performance of the network on motion sequences from unseen individuals, we conduct a leave-one-subject-out cross-validation with twelve subjects, where data from ten subjects are used for training, one subject is reserved for validation, and the remaining subjects used for testing. All combinations of full 3D pose $\mathbf{f}_t$ or emulated 2D pose $\mathbf{f}^\mathrm{2D}_t$, 60 or 20 Hz are used in training. The results are discussed in Sec.~\ref{sec:evaluation_grm} and further details are in Appendix C.2.

\subsubsection{Training procedure for real-time deployment}
For the purpose of the real-time controller (Sec.~\ref{sec:RealTimeController}), 
the (relatively limited) dataset is used maximally, only one subject (number 8) is withheld for validation, all others used for training. 
In the installation, pose is captured with MediaPipe Pose Landmarker\footnotemark[3] in 20 Hz as described in Sec.~\ref{sec:installation}. We therefore use the 2D pose and 20 Hz settings in the real-time LSTM training.


\subsection{Real-Time Controller}
\label{sec:RealTimeController}

At each lap of the installation loop at timestep $k$,\footnotemark[4] the input to the controller is the estimated phase $\hat{\varphi}_k$ and the resulting output is the desired playback speed $s_k$, a factor indicating how much to speed up or slow down the original recording playback; $s_k=1.0$ corresponds to original tempo. 

To compress the signal from the user and make the system robust to noise in the phase estimate, 
upbeats and downbeats (Sec.~\ref{sec:original}) are detected from the phase signal according to the following logic:
\begin{align}
    \text{Upbeat at } k &\iff \hat{\varphi}_k - \hat{\varphi}_{k-1} > \Delta \varphi_{\text{up}}, \label{eq:upbeat}\\
    \text{Downbeat at } k &\iff \hat{\varphi}_{k-1} > \varphi_{\text{high}} \;\; \wedge \;\; \hat{\varphi}_k < \varphi_{\text{low}}, \label{eq:downbeat}
\end{align}
where $\Delta \varphi_{\text{up}} = 0.5$, $\varphi_{\text{high}} = 3.8$, and $\varphi_{\text{low}} = 2.5$, selected empirically to achieve a responsive controller but still with some robustness to accidental motions. The times of detected beats are incrementally saved in a beat history array $\mathbf{k}$. 

\begin{figure}[t]
    \centerline{\includegraphics[width=0.9\linewidth]{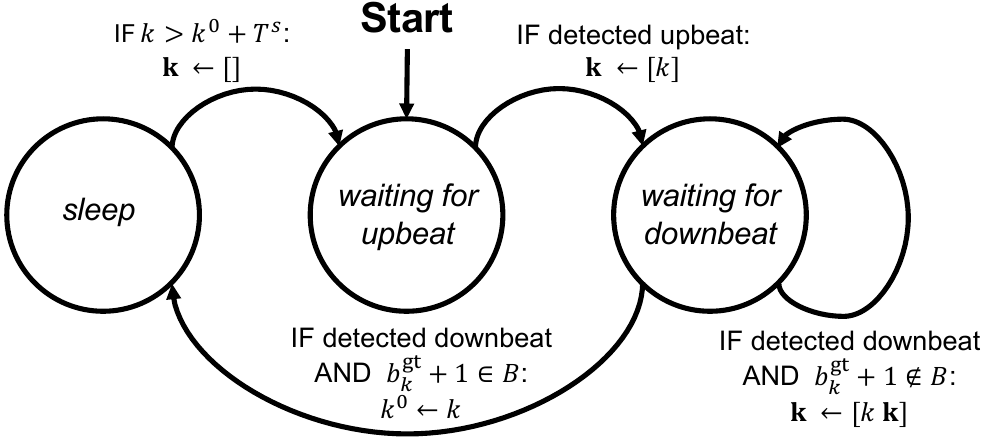}}
    \caption{The behavior of the real-time controller is governed by a finite state machine. $k$ is the current timestep, $\mathbf{k}$ the history of detected beat timings,  $b^\mathrm{gt}_k$ the current bar number, $B$ the set of fermata bars in this piece (Sec.~\ref{sec:original}), and $T^\mathrm{s}$ minimum waiting time at fermatas.}
    \label{fig:controlfsm}
\end{figure}

The behavior of the controller is governed by a finite state machine (see Fig.~\ref{fig:controlfsm}) with three states, taking one step at each lap of the installation loop. The algorithmic details are provided in 
Appendix~\ref{sec:appendixController}.
The controller starts in the $\textit{waiting for upbeat}$ state with a speed estimate $s_k = 0$. In the states \textit{waiting for upbeat} and \textit{sleep}, $s_k$ continues as $0$, while in \textit{waiting for downbeat}, $s_k$ is updated from the beat history $\mathbf{k}$ as follows:

Let $\Delta\mathbf{k} \gets \mathbf{k}_{1:end-1}-\mathbf{k}_{2:end}$ be the history of time intervals between beats. 
The interval history is combined in different ways to obtain an average interval $\Delta k$. In addition, let $\Delta t_b$ be the duration of the next bar in the original recording. The speed factor is
\begin{equation}
s_k = \text{speed}(\mathbf{k}) = \frac{\Delta k}{\Delta t_b} 
\label{eq:speed}
\end{equation}
where $\Delta k$ is estimated using one of three strategies, which are compared and evaluated in Sec.~\ref{sec:handsonperformance}:

{\em Raw:} $\Delta k = \Delta\mathbf{k}_1$, interval between two latest beats.

{\em Median:} If length of beat history $|\mathbf{k}|<4$, use \textit{Raw}, else $\Delta k = \text{median} (\Delta\mathbf{k}_{1:3})$, i.e., median of the three latest intervals. 

{\em Average:} If length of beat history $|\mathbf{k}|<4$, use \textit{Raw}, else $\Delta k = \sum_{i=1}^3 \mathbf{w}_i\Delta\mathbf{k}_i$ where $\mathbf{w} = [\frac{1}{2}~\frac{1}{3}~\frac{1}{6}]$, i.e., weighted mean of the three latest intervals.

\section{Evaluation}
\label{sec:evaluation}

\begin{table}[b]
\centering
\small
\caption{Phase error (MSPE, Eq.~(\ref{eq:MSAE})) for different pose input settings, average performance over 12 different subjects.}
\begin{tabular}{lcc}
\toprule
\makecell{} & \makecell{60 Hz} & \makecell{20 Hz} \\
\midrule
    Full 3D pose: $\mathbf{x}_t$  & 1.42 $\pm$ 0.34 & 1.64 $\pm$ 0.39   \\
    Upper-body 2D pose: $\mathbf{x}_t^\mathrm{2D}$ & 1.60 $\pm$ 0.42 & 1.65 $\pm$ 0.35 \\
\bottomrule
\end{tabular}
\label{tab:cv_metrics_by_test}
\end{table}


This section presents a quantitative assessment of the gesture recognition module, a  user study of the real-time controller, and a field study conducted in the dome theater.

\subsection{Conducting Gesture Recognition Evaluation}
\label{sec:evaluation_grm}

\begin{figure*}[t]
    \centering
    \subfloat[\label{fig:fermataperformancea}\centering Fermata bars, 60 Hz using $\mathbf{x}_t$]{%
        \includegraphics[width=5.8cm]{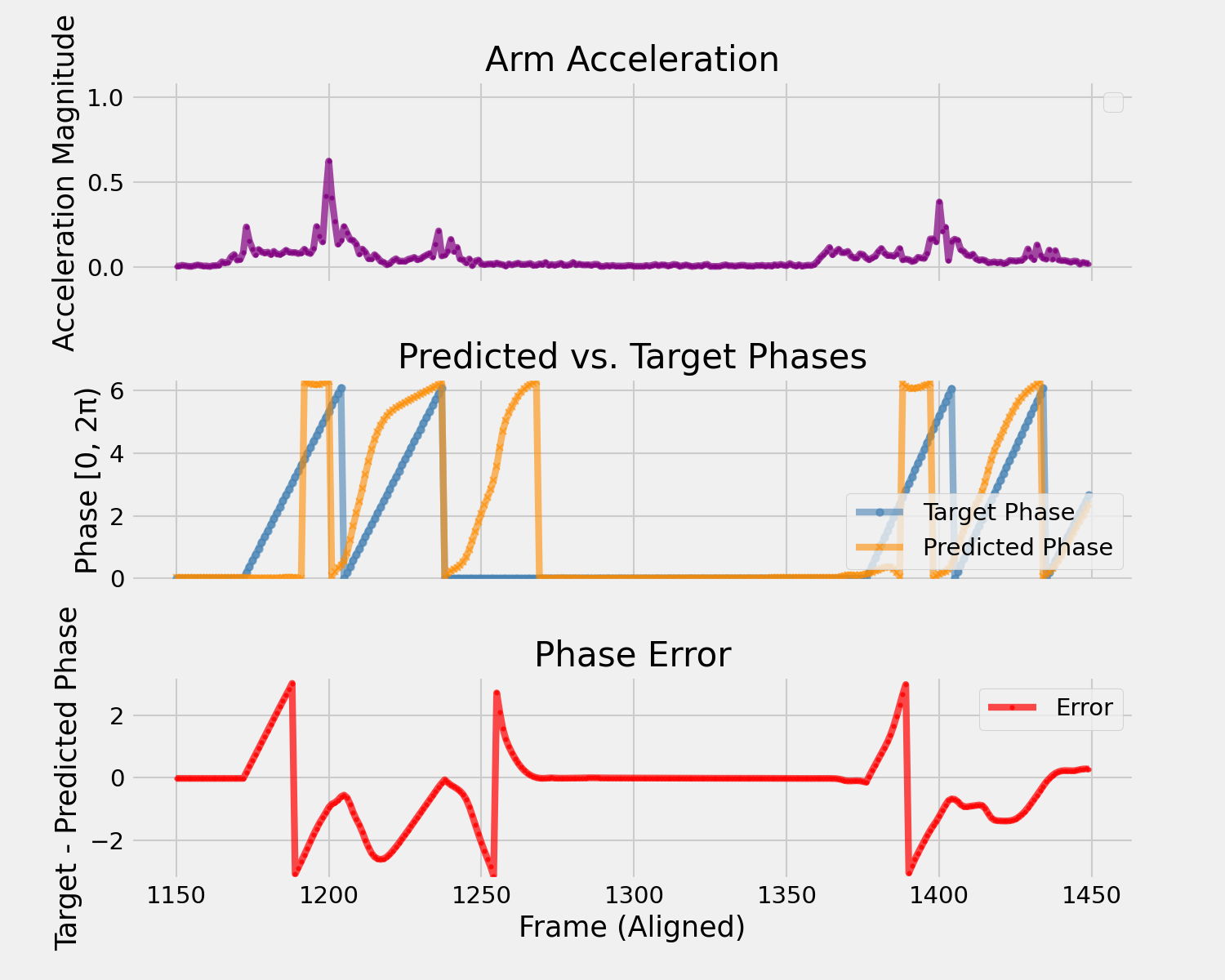}
        } 
    \subfloat[\label{fig:fermataperformanceb}\centering Fermata bars, 20 Hz using $\mathbf{x}_t^\mathrm{2D}$]{%
         \includegraphics[width=5.8cm]{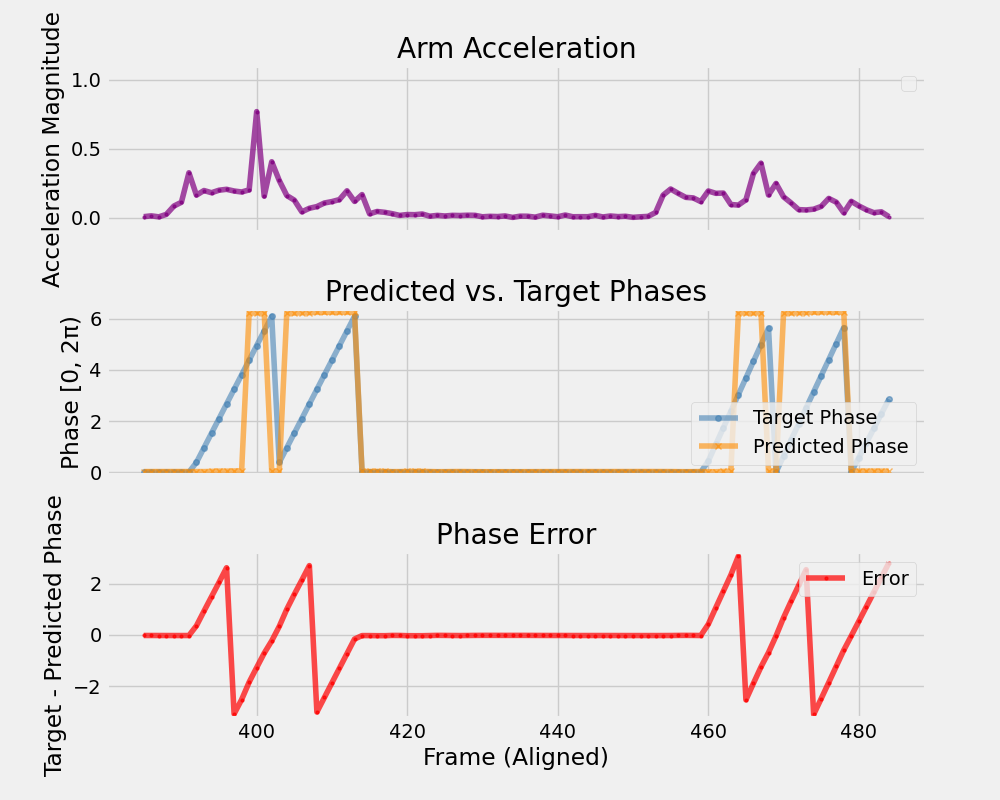}
         }
    \subfloat[\label{fig:fermataperformancec}\centering Fermata bars, 20 Hz using $\mathbf{x}_t^\mathrm{2D}$, $\beta = 1$]{%
        \includegraphics[width=5.8cm]{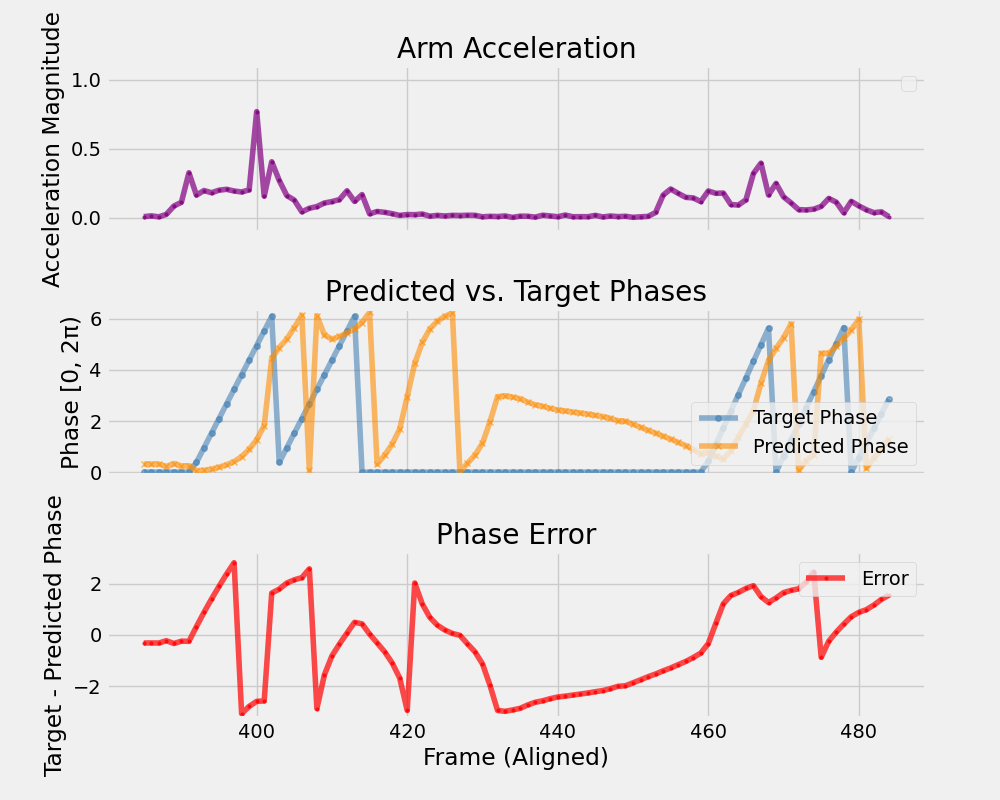}}

    \subfloat[\label{fig:fermataperformanced}\centering Regular bars, 60 Hz using $\mathbf{x}_t$]{%
        \includegraphics[width=5.8cm]{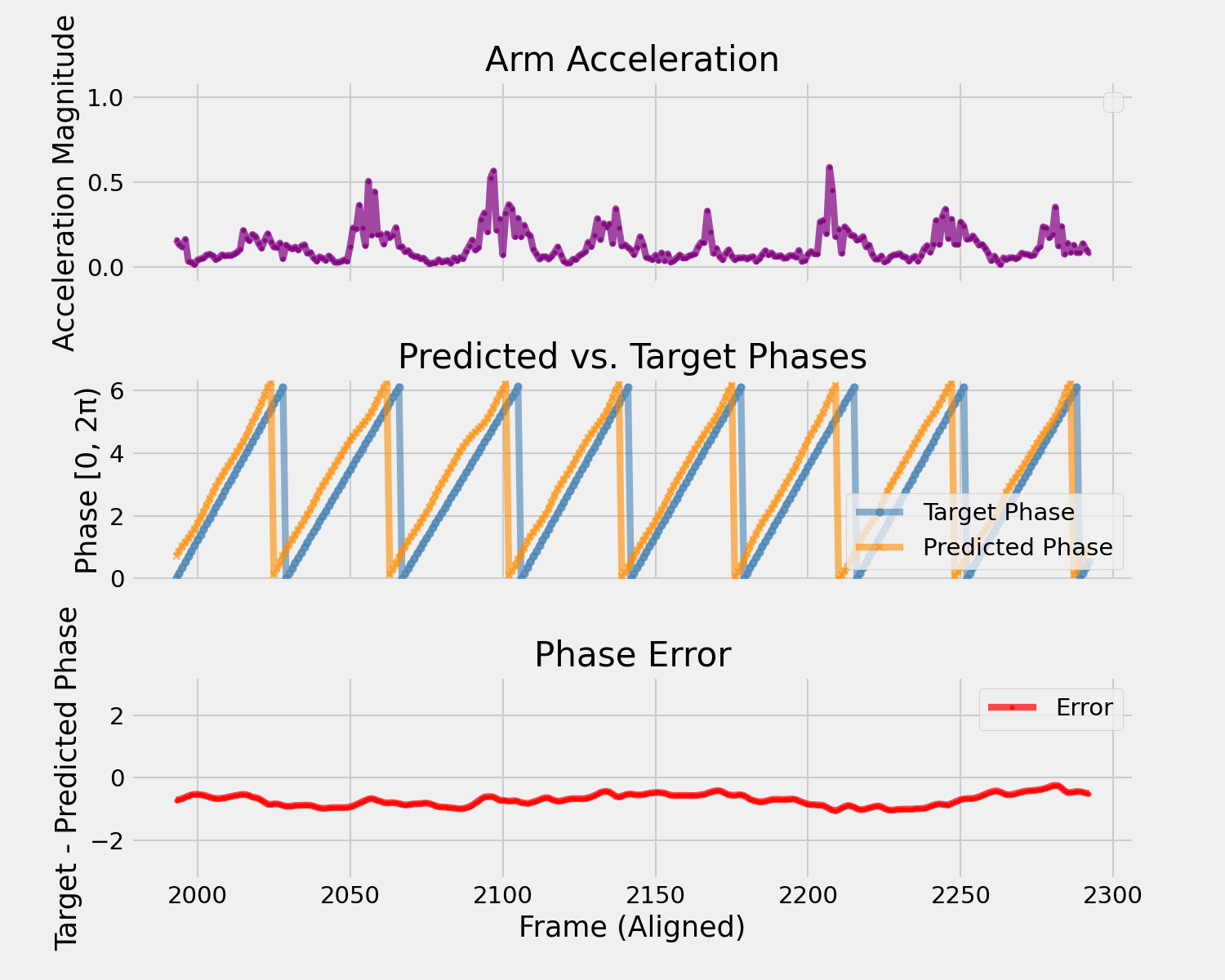}
        } 
    \subfloat[\label{fig:fermataperformancee}\centering Regular bars, 20 Hz using $\mathbf{x}_t^\mathrm{2D}$]{%
        \includegraphics[width=5.8cm]{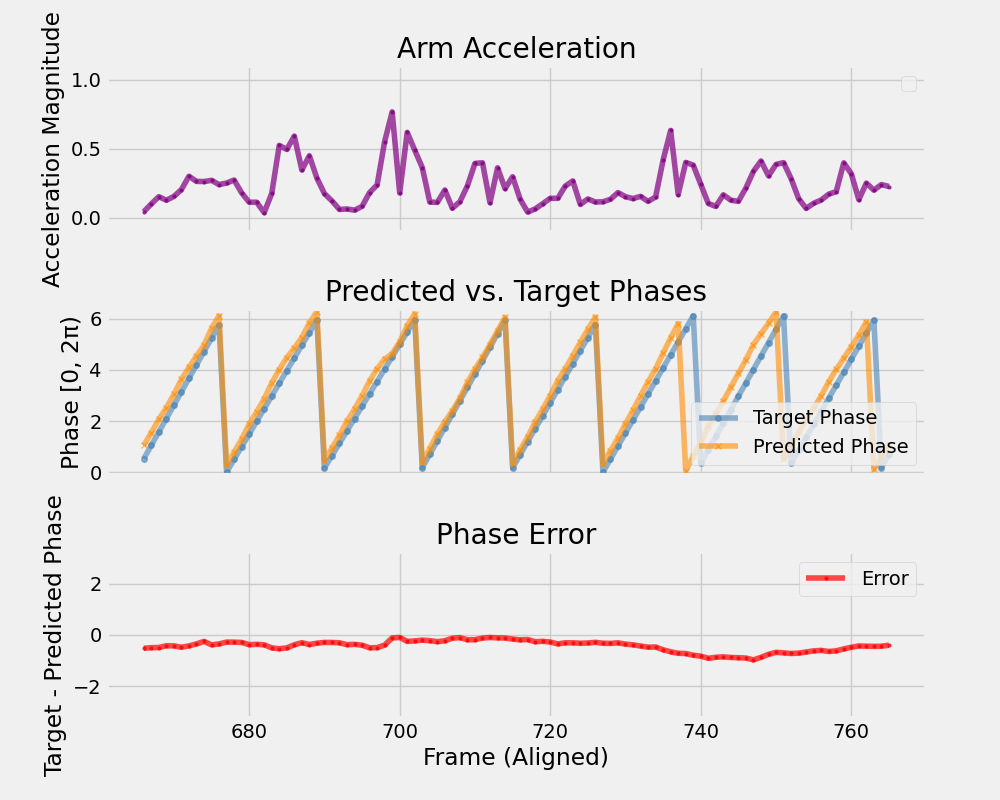}
        }
    \subfloat[\label{fig:fermataperformancef}\centering Regular bars, 20 Hz using $\mathbf{x}_t^\mathrm{2D}$, $\beta = 1$]{%
        \includegraphics[width=5.8cm]{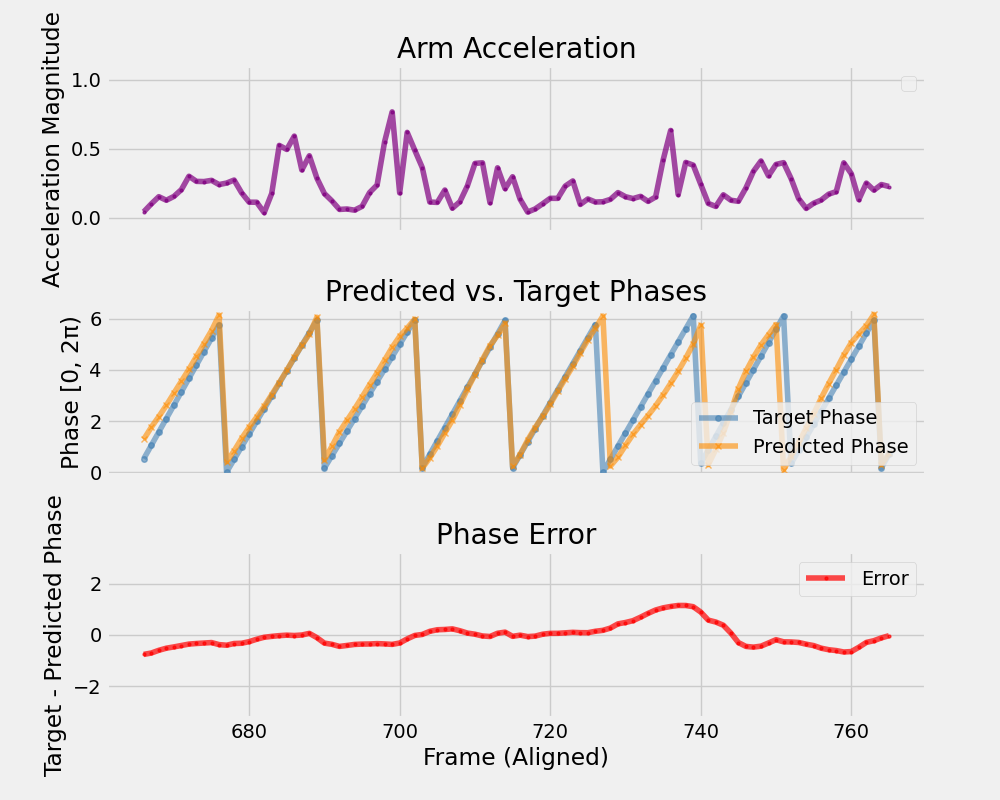}}

        \caption{Example of phase estimation for Subject 1. (a) and (d) correspond to column 1 row 1 in Table \ref{tab:cv_metrics_by_test}, (b) and (e) correspond to column 2 row 2 in Table \ref{tab:cv_metrics_by_test}. (c) and (f) are a variant of (b) and (e) but with $\beta$ increased from 0.3 to 1. Top: Arm acceleration $\rho_t$. Middle: Ground truth phase $\varphi^\mathrm{gt}_t$ and estimated phase $\hat{\varphi}_t$. Bottom: Phase error $\varphi^\mathrm{gt}_t - \hat{\varphi}_t$. The three graphs are temporally aligned. Further examples are found in Appendix \ref{sec:appendixGesture}.}
    \label{fig:fermataperformance}
\end{figure*}

The purpose is to investigate the performance of the gesture recognition network on subjects not seen during training.

The evaluation metric for a sequence $i$ is the mean squared phase error (MSPE), defined as:
\begin{equation} \delta_\mathrm{MSPE}^i = \frac{1}{T} \sum_{t=1}^{T} (\varphi^\mathrm{gt}_t - \hat{\varphi}_t)^2 \label{eq:MSAE} \end{equation}
where $T$ is the total number of frames in each recording.

The MSPE is computed for each of the 130 sequences. The MSPE mean and standard deviation are presented in Table \ref{tab:cv_metrics_by_test}, for four different input spaces, as explained in Sec.~\ref{sec:additional}. The MSPE per test subject is presented in Appendix D.1.

We observe that the accuracy is not severely affected by lowering the input resolution from 60 to 20 Hz and from $\mathbf{x}_t$ to $\mathbf{x}_t^{2D}$. However, the table shows a certain error increase for lower resolution input. 

What are then the causes of error in the phase estimation? 
In regular in-tempo bars, the error is low. As illustrated in Fig.~\ref{fig:fermataperformanced}, which depicts consecutive bars in a regular tempo, the estimated and ground truth phase are well aligned, demonstrating 1) that the recorded Subject 1 was able to make their beats in the correct pace and timing, 2) that the network performs reliably when the tempo is constant. 
This performance remains largely similar when the input space is changed from $\mathbf{x}_t$ in 60 Hz to  $\mathbf{x}_t^{2D}$ in 20 Hz, as shown in Fig.~\ref{fig:fermataperformancee}. 
As can be seen in Fig.~\ref{fig:fermataperformancea} and \ref{fig:fermataperformanceb}, the network successfully identifies the fermata before frame 1380 (460), indicating that it is not only predicting constant speed from the original recording, but is also able to recognize a signal to stop. However, the same figures show that the upbeats around frames 1170 (390) and 1380 (460) are detected with some delay compared to the ground truth (conveyed to the subject by click sounds).
The error is not in the subject recording: The arm acceleration (derived from the input signal, and only provided for visualization, not used in the estimation) shows that the subject indeed provided upbeats on time. 

There are likely two reasons for this. Firstly, we saw a large variability in the styles of upbeats among our subjects. A dataset of only 12 subjects is probably insufficient to capture the full variability of upbeat gestures. Furthermore, the data collection setup was somewhat artificial: participants were asked to pretend to conduct the playback of a recording, rather than leading the orchestra producing the audio, which created delays especially in upbeats.

For the usability of the installation, it is of course important that upbeats are detected in a robust manner. In order to make the system more sensitive to gestures connected to moving forward, we trained a version of the LSTM with increased monotonicity loss weight $\beta$ (Eq.~(\ref{eq:total_loss})). 
This biased the model towards 'wanting to proceed' whenever movement was detected. As illustrated in Fig.~\ref{fig:fermataperformancec}, this  improved the detection of upbeats considerably. Although fermatas were predicted less accurately, importantly, the high precision in downbeat detection remained stable, as shown in Fig.~\ref{fig:fermataperformancef}.

\subsection{User Study with a Rudimentary Setup}
\label{sec:handsonperformance}

\begin{figure*}
\centering
    \subfloat[\label{fig:userstudy1a_median}\centering Fermata bars]{%
        \includegraphics[width=5.5cm]{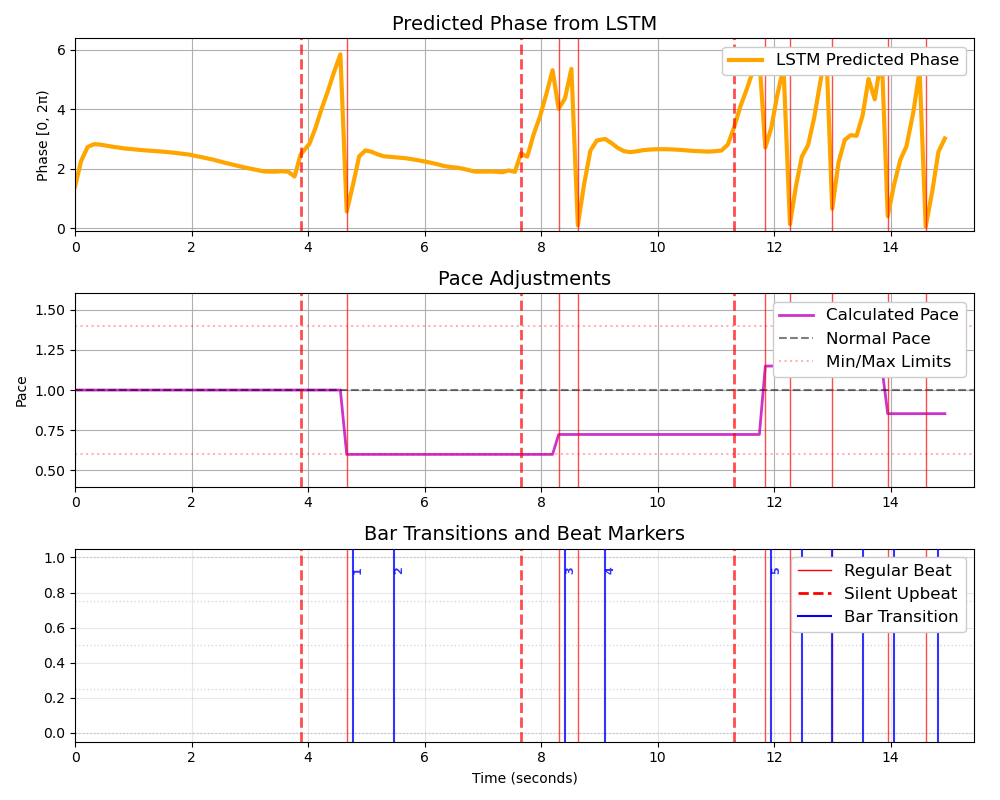}
    } 
    \subfloat[\label{fig:userstudy1b_median}\centering Regular bars, steady pace]{%
        \includegraphics[width=5.5cm]{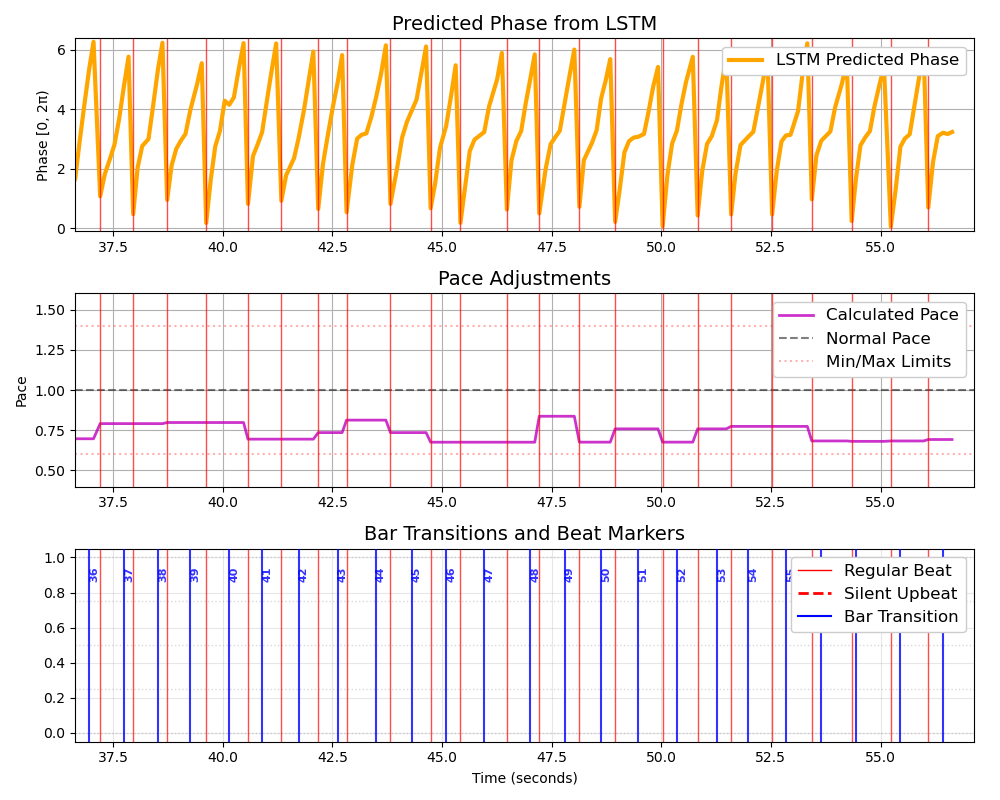}
    } 
    \subfloat[\label{fig:userstudy1c_median}\centering Regular bars, varying pace]{%
        \includegraphics[width=5.5cm]{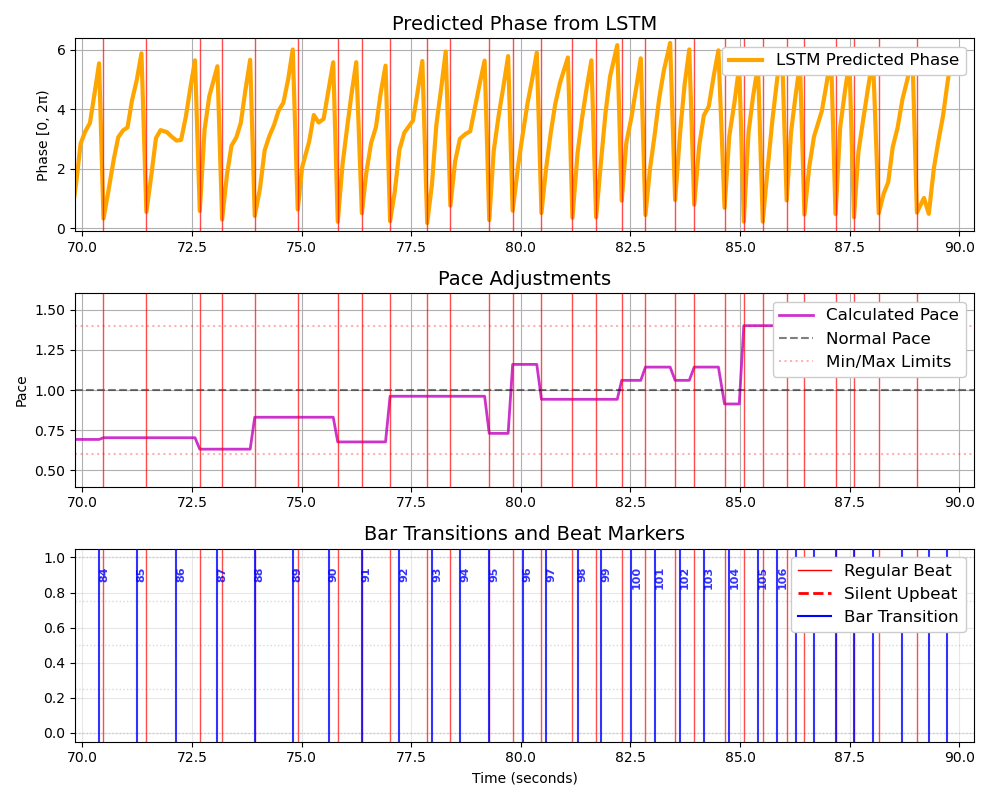}
    }
    \caption{Real-time interaction of User 1 in Median setting. Top: Predicted phase $\hat{\varphi}^i_t$. Middle: Pace adjustments according to the detected beats. Bottom: Beats and bar beginnings. Further examples are found in Appendix \ref{sec:appendixRealtime}.}
    \label{fig:userstudy1_median}
\end{figure*}

We then evaluate the entire control loop including gesture recognition, real-time speed estimation and audio feedback.

The evaluation of the system is both qualitative and quantitative, and was conducted through a user study with three participants: an orchestra musician (User 1), a professional orchestra conductor (User 2), and a non-musician (User 3). Each participant tested the three control strategies (Raw, Median, Average) explained in Sec.~\ref{sec:RealTimeController}. For each strategy we did two recordings; in the first, the participants were instructed to conduct at a steady tempo of their choice, and in the second, they were asked to induce tempo changes, alternating between slowing down and speeding up.

\begin{table}[b]
\centering
\small
\caption{Diff between beats and nearest bar starts (s) for different users and settings (S: steady pace, V: varying pace).}
\begin{tabular}{lccc}
\toprule
& \makecell{User 1} & \makecell{User 2} & \makecell{User 3}\\
\midrule
{\footnotesize Raw S} & 0.197 $\pm$ 0.132 &  0.159 $\pm$ 0.098 & 0.137 $\pm$ 0.108\\
{\footnotesize Raw V} & 0.176 $\pm$ 0.122 & 0.169 $\pm$ 0.104 & 0.183 $\pm$ 0.112\\
{\footnotesize Median S\hspace{-3mm}} & 0.239 $\pm$ 0.118 & 0.168 $\pm$ 0.109 &  0.257 $\pm$ 0.136\\
{\footnotesize Median V\hspace{-3mm}} & 0.172 $\pm$ 0.127 &  0.130 $\pm$ 0.111 & 0.147 $\pm$ 0.113\\
{\footnotesize Average S\hspace{-3mm}} & 0.178 $\pm$ 0.126 & 0.152 $\pm$ 0.110 & 0.207 $\pm$ 0.122\\
{\footnotesize Average V\hspace{-3mm}} & 0.189 $\pm$ 0.123 & 0.153 $\pm$ 0.101 & 0.172 $\pm$ 0.115\\
\midrule
\textbf{\footnotesize All settings\hspace{-3mm}} & 0.192 $\pm$ 0.125 & \textbf{0.155 $\pm$ 0.106}& 0.184 $\pm$ 0.118\\
\bottomrule
\end{tabular}
\label{tab:beat-bar-distances}
\end{table}

The first question to be answered is whether the interaction of the control loop resembles that of a real orchestra. We first define a succesful interaction as one where the orchestra start each bar in direct connection to a beat from the conductor, i.e., that the orchestra follows the conductor. This can be measured as the coherence (temporal difference) between the detected beats and the bar starts in the playback. We hypothesized that if the system behaves as a real orchestra, then the professional orchestra conductor should be most familiar with that, and achieve the best results when interacting with it. Indeed, as shown in Table~\ref{tab:beat-bar-distances}, the lowest average bar-beat distance was achieved by User~2, the professional conductor. This finding suggests that the system response resembles that of a real orchestra to some degree.

\begin{table}[b]
\centering
\small
\caption{Standard deviation of speed factor $s_k$ in non-fermata part of the music, takes with steady pace.}
\begin{tabular}{lccc}
\toprule
& \makecell{Raw Steady} & \makecell{Median Steady} & \makecell{Average Steady} \\
\midrule
User 1 & 0.157 &  0.079 &  0.103\\
User 2 & 0.256 &  0.200  & 0.167  \\
User 3 & 0.232 &  0.149  & 0.181\\
\midrule
\textbf{All users} & 0.215 & \textbf{0.143} & 0.150 \\
\bottomrule
\end{tabular}
\label{tab:pace_stability}
\end{table}

The second question to be answered is which of the three control strategies (Raw, Median, Average) should be selected for the museum installation. We hypothesize that the best strategy would be the one that yields the most stable pace when users are asked to perform at a steady tempo. To evaluate this, we calculated the standard deviation of the estimated speed factor $s_k$ (see Eq.~(\ref{eq:speed})) after bar 25 (i.e., after all fermata bars) in each take, and compared the results across control settings. As shown in Table~\ref{tab:pace_stability}, the Median setting produced the lowest average standard deviation, indicating that it produced a robust speed estimate.
This is also supported by the qualitative feedback given by the users. The Raw setting was consistently described as unstable and difficult to control, compared to an inexperienced or unresponsive orchestra, or a 'goldfish orchestra' because it did not remember anything, just reacted to the latest that was happening. In contrast, the Median setting was repeatedly characterized as steady, intuitive, and closest to the experience of conducting a real ensemble. When seeing the Median formula afterwards, both User 1 and 2 noted that this is what musicians in an ensemble would do: Ignore an occasional strange or missing beat, but react fast to a consistent change of pace. The Average setting was perceived as solid and responsive but less flexible than the Median setting.
We provide examples of the Median interaction of User 1 in Fig.~\ref{fig:userstudy1_median}.
Further user interactions are visualized in Fig.~22-29 in Appendix \ref{sec:appendixRealtime}.

However, the recordings in variable pace revealed a notable caveat. The system reacts slowly to changes in tempo, not due to error but as a direct consequence of the design: A tempo change can only be detected after two (in the Raw case) to four (in the Median and Average case) beats. In a real orchestra-conductor interaction, the orchestra can sense the tempo change from the conducting gesture in not more than a beat, so the delay of up to four beats is far too long to be natural. This posed less of a problem to User 1 (see Fig.~\ref{fig:userstudy1_median}c) who did not have conducting training and simply persisted in a higher pace to increase the speed of the recording, with the side effect that the beats and bar starts where out of synchrony for a while. However, the professional conductor, User 2 (Fig.~25c in Appendix \ref{sec:appendixRealtime}) tried to maintain synchrony with the orchestra while showing higher speed (as you would with a real orchestra). The result was that there was no speed change, since the system did not react fast enough to maintain synchrony through the speed change. User 2 commented that 'they are not listening'. 
This points to the need for a future control strategy that takes into account the entire phase progression. 

Another observation that can be made in Fig.~\ref{fig:userstudy1_median} is that the phase estimation for User 1 is more unstable than for the subjects in the conducting dataset (Sec.~\ref{sec:additional}). The reason can be that the motion of User 1 -- being a layperson -- deviated from the range of motions captured in the training set. 
However, beats are reliably detected despite this, which indicates that the gesture recognition will be adequate for the museum installation.

\subsection{Field Study in the Dome Theater}
\label{sec:field_study}


\begin{figure*}
\centering
    \subfloat[\label{fig:userstudy1a_median}\centering Case 1]{%
        \includegraphics[width=5.5cm]{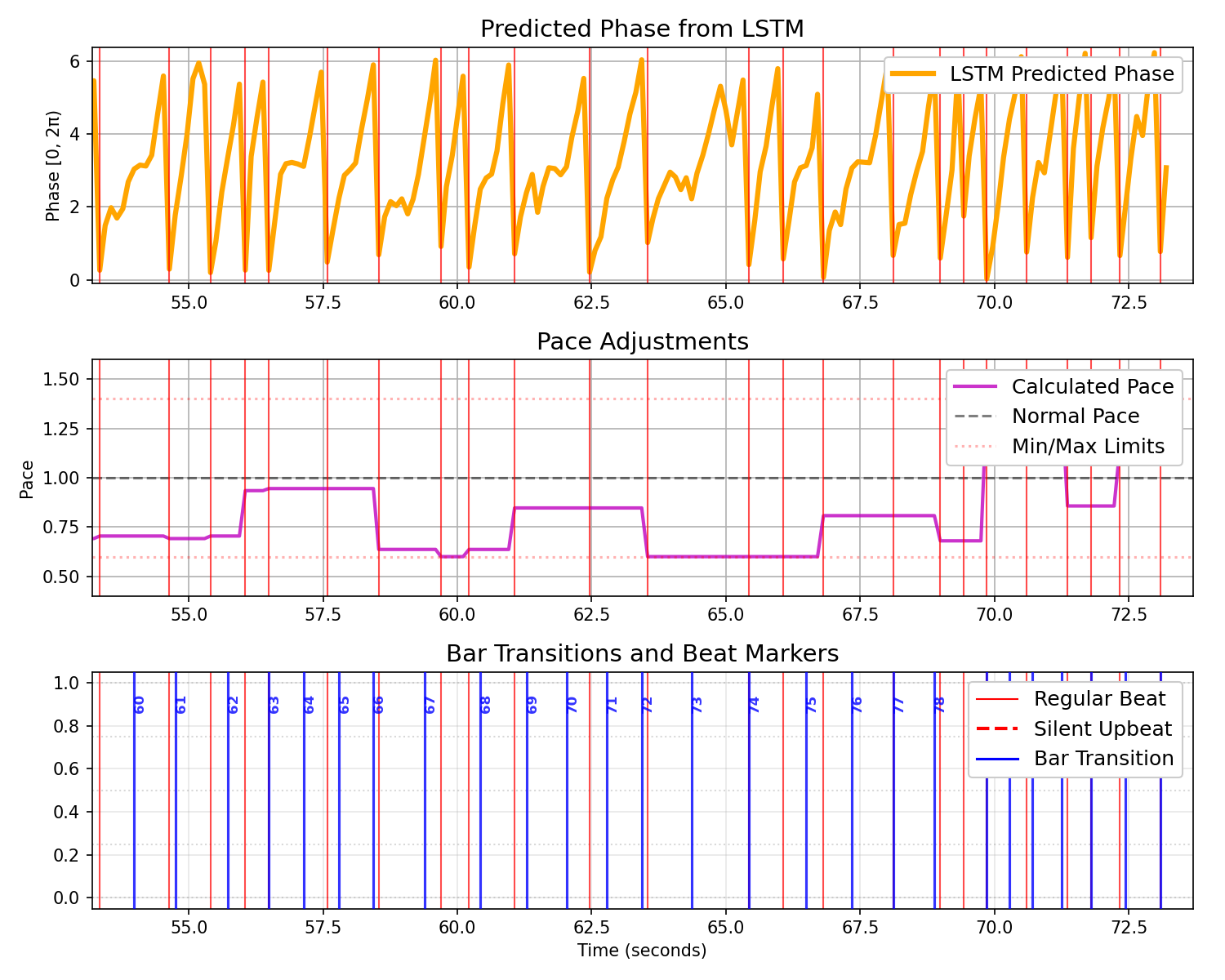}
        \label{fig:fieldstudycases_case1}
    } 
    \subfloat[\label{fig:userstudy1b_median}\centering Case 2]{%
        \includegraphics[width=5.5cm]{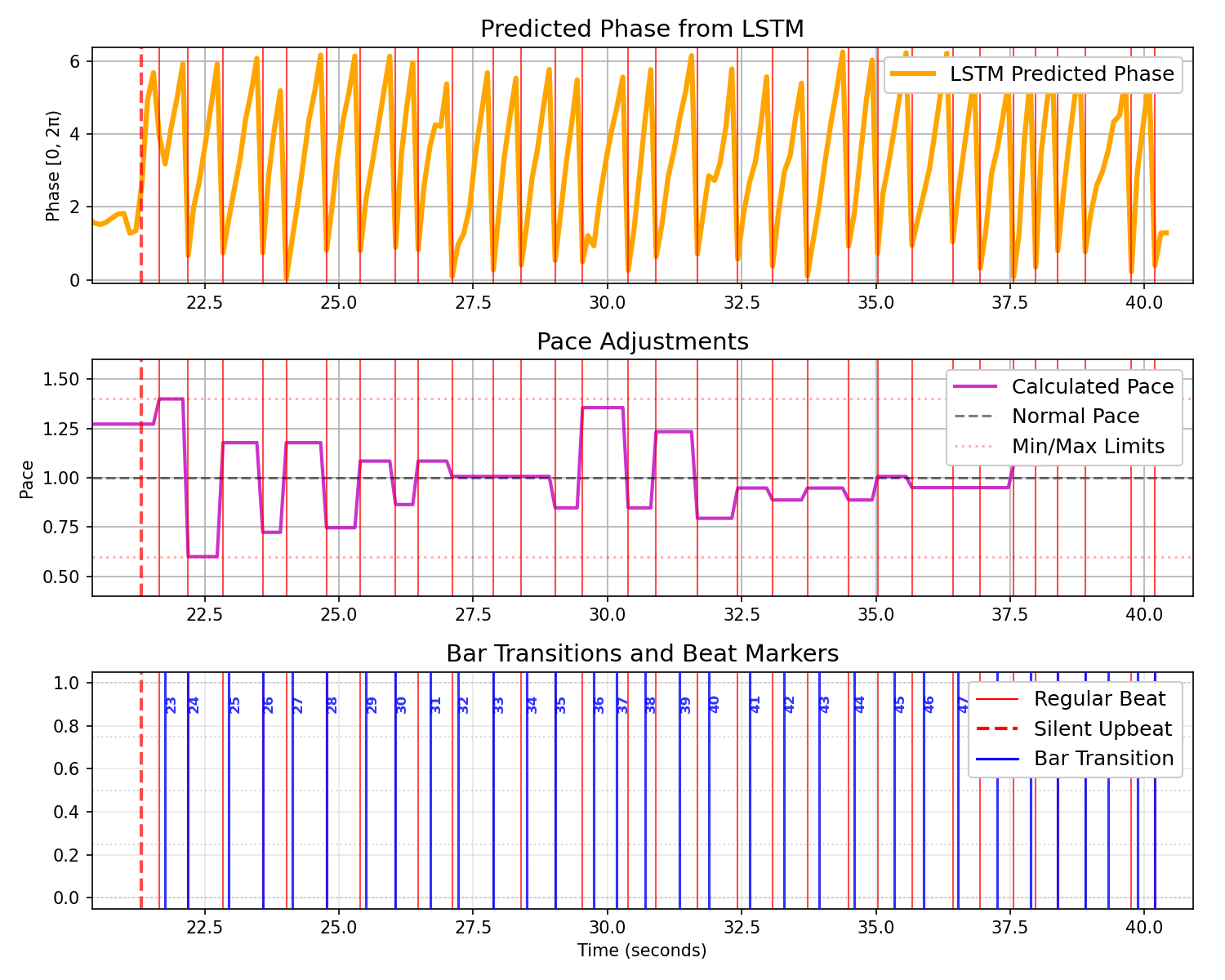}
        \label{fig:fieldstudycases_case2}
    } 
    \subfloat[\label{fig:userstudy1c_median}\centering Case 3]{%
        \includegraphics[width=5.5cm]{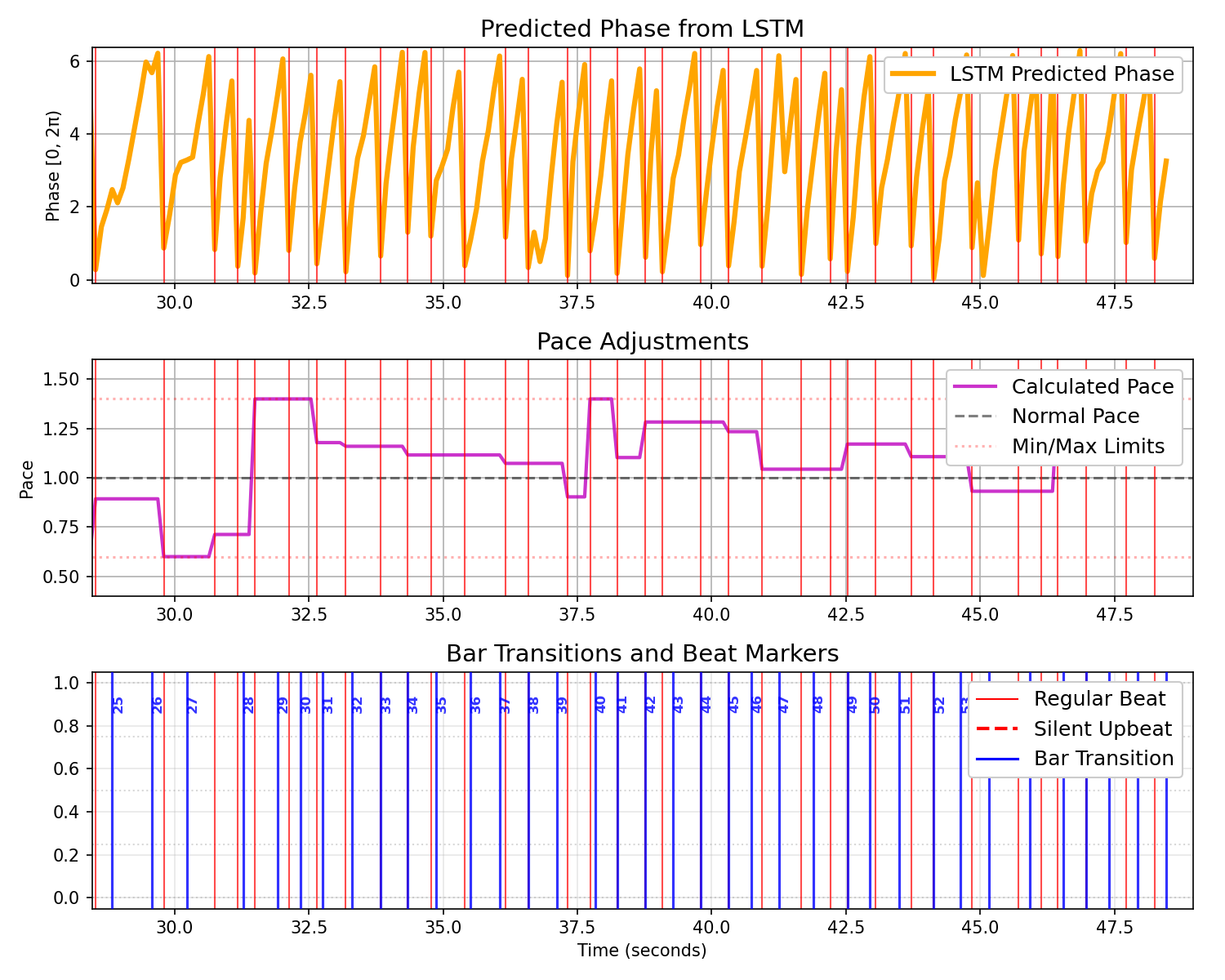}
        \label{fig:fieldstudycases_case3}
    }
    \caption{3 cases from 3 different subjects from the field study. Top: Predicted phase $\hat{\varphi}^i_t$. Middle: Pace adjustments according to the detected beats. Bottom: Beats and bar beginnings.}
    \label{fig:fieldstudycases}
\end{figure*}

As a final evaluation, we conducted a field study during a public showing in the dome theater Wisdome Stockholm with 23 participants ranging from children to senior adults, spanning musical expertise from laypeople to professional conductors. The goal was to assess system performance under real-world conditions and identify practical bottlenecks not visible in lab settings.

Three representative cases were examined. In Case 1 (Fig.~6a), a participant with very limited musical background produced smooth and indecisive movements with low acceleration. The gesture recognition module struggled to reliably estimate the phase, resulting in unstable predictions with frequent instances of reverse phase progression. A possible reason for this difficulty is that the participant's movements fell outside the distribution of the training data; since the gesture recognition module is trained to map motion to in-bar phase using sequences from a specific dataset, movements that do not resemble those in the training set, being out-of-domain, may yield poor estimates.

In Case 2 (Fig.~6b), a participant with limited musical exposure -- but with prior experience with the system -- produced movements more consistent with the training distribution. This resulted in noticeably more stable phase predictions compared to Case 1, illustrating that even brief familiarity with the interaction paradigm can meaningfully improve performance.

In Case 3 (Fig.~6c), a child participant achieved good performance on their first attempt. This case is particularly noteworthy as it demonstrates that effective system use requires neither adulthood nor formal musical training, highlighting the accessibility and ease of use of the system for a broad range of users.

A key observation across these cases is that the critical factor governing performance is not age or musical background, but how closely a user's movements align with the training data, meaning the system's performance ceiling is largely shaped by the coverage and diversity of that data. Additionally, lower frame rates introduced noise in velocity and acceleration features. This limitation can not be addressed at a network level, as these signals are inherently sensitive to temporal resolution. Finally, the results suggest that users can get better at controlling the system, emphasizing the importance of designing an onboarding session. 

Further quantitative evaluation and results from the field study are presented in Appendix \ref{sec:AppendixFieldStudy}.

\section{Conclusions}
\label{sec:conclusion}


We have presented the development and evaluation of an immersive, interactive museum installation, allowing users to get the experience of conducting a symphony orchestra. Our three contributions are the creation of a unique multi-conductor motion capture dataset (Sec.~\ref{sec:dataset}), an immersive interaction system that plays back the recorded music performance in a speed controlled by the user (Sec.~\ref{sec:installation}), and an LSTM-based gesture recognition network connected to a real-time control system, which enables users to control the playback speed of the music recording (Sec.~\ref{sec:mlmodel}). 

The  system was evaluated through a quantitative timing accuracy study, a user study assessing musical realism and usability, and a field study with real museum visitors (Sec.~\ref{sec:evaluation}).

\subsection{Future Work}
\label{sec:future}
Several directions are envisioned to improve the adaptivity, robustness, and responsiveness of 
the system. 

First, the dataset should be expanded with recordings collected in natural conditions, 
incorporating tempo changes, accents, and different beating patterns, enabling the network to 
generalize better to expressive real-world conducting gestures; the main hurdle being the 
labeling of beat timings for each unique take. 

Second, the frame rate of the real-time pose 
tracker should be increased, reducing temporal aliasing and improving derivative estimates, 
thereby yielding smoother LSTM input trajectories and more precise phase predictions -- though 
this is not feasible in the current setup, where the $180^\circ$ visualization is highly 
computationally demanding. 

Finally, the control module should move beyond discrete beat detection 
and instead leverage the continuous phase trajectory to adjust $s_k$ based on the phase 
derivative, yielding smoother and more responsive control dynamics. This would require a higher phase prediction precision, made possible by the improvements above.

\subsection*{Safe and Responsible Innovation Statement} 
The societal impact of this work lies in its creation of an immersive virtual conducting experience, offering novel opportunities for public engagement, education, and accessibility in cultural spaces such as museums. All recordings were obtained with informed consent and used solely for research. Responsible deployment includes transparency in data usage, protection of user privacy, and attention to accessibility. As the system evolves, care will be taken to ensure the system fosters inclusive experiences and actively supports equitable access to digital cultural participation.

\bibliographystyle{ACM-Reference-Format}
\bibliography{main}


\clearpage

\appendix

\twocolumn[
\begin{center}
    {\Huge\bfseries Supplementary Material}
    \vspace{1em}
\end{center}
]


\section{Additional Dataset Details}
\label{sec:appendixRecording}
This section provides further details on the collected dataset described in Sec.~\ref{sec:dataset}.

The number of recordings in the dataset per subject is presented in  Table \ref{tab:dist_of_record}. Subjects with longer experience in conducting or musical performance were recorded across more tempo variations and repetitions. For less experienced subjects, the recordings were limited primarily to the original recording tempo.

\begin{table}[h]
\centering
\small
\caption{Distribution of recordings by subject and tempo}
\begin{tabular}{lccccc}
\toprule
Subject & Total & 1.0 & 0.8 & 1.2 & Other \\
\midrule
Subject 1  & 17 & 8 & 2 & 2 & \makecell{0.6 : 1 \\ 1:3 : 1 \\ only first 25 bars: 3}  \\
Subject 2  & 11 & 7 & 2 & 2 & 0 \\
Subject 3  & 13 & 9 & 2 & 2 & 0 \\
Subject 4  & 10 & 6 & 2 & 2 & 0 \\
Subject 5  & 7  & 7 & 0 & 0 & 0 \\
Subject 6  & 10 & 6 & 2 & 2 & 0 \\
Subject 7  & 12 & 8 & 2 & 2 & 0 \\
Subject 8  & 10 & 6 & 2 & 2 & 0 \\
Subject 9  & 12 & 7 & 2 & 3 & 0 \\
Subject 10 & 7  & 7 & 0 & 0 & 0 \\
Subject 11 & 10 & 6 & 2 & 2 & 0 \\
Subject 12 & 11 & 7 & 2 & 2 & 0 \\
\midrule
\textbf{Total} & 130 & 84 & 20 & 21 & 5 \\
\bottomrule
\end{tabular}
\label{tab:dist_of_record}
\end{table}

\begin{figure}[t]
    \centerline{\includegraphics[width=\linewidth]{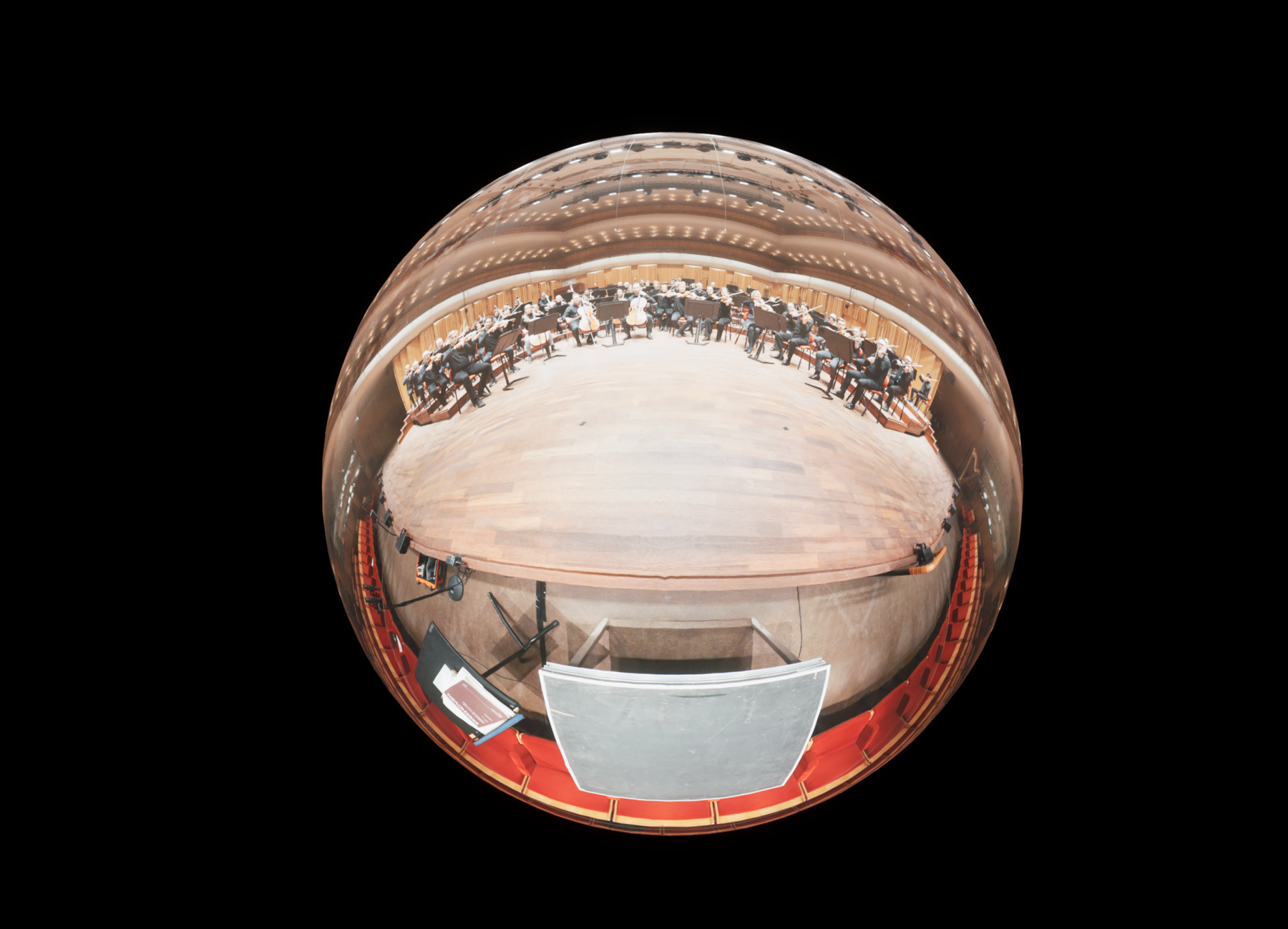}}
    \caption{Recording view.}
    \label{fig:recordingview}
\end{figure}

Figure \ref{fig:recordingview} shows how the video is captured during the recordings.

\section{Additional Dome Installation Details}
\label{sec:appendixInstallation}

This section provides further details on the dome installation described in Sec.~\ref{sec:installation}.

\begin{figure}[t]
    \centerline{\includegraphics[width=\linewidth]{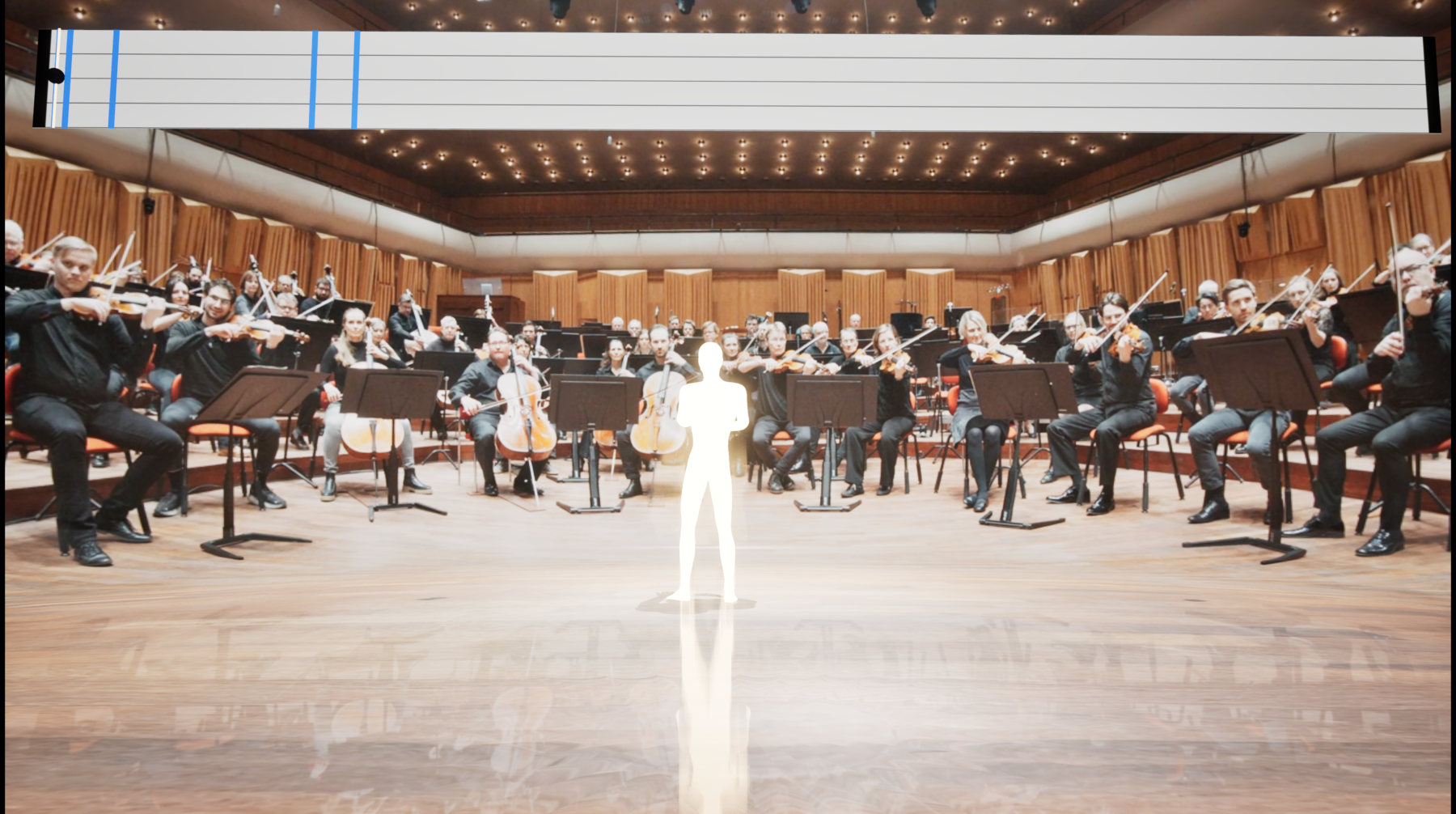}}
    \caption{Flat screen with conductor guide overlay. Musical stave on top of the screen.}
    \label{fig:flatscreen}
\end{figure}

\begin{figure}[t]
    \centerline{\includegraphics[width=\linewidth]{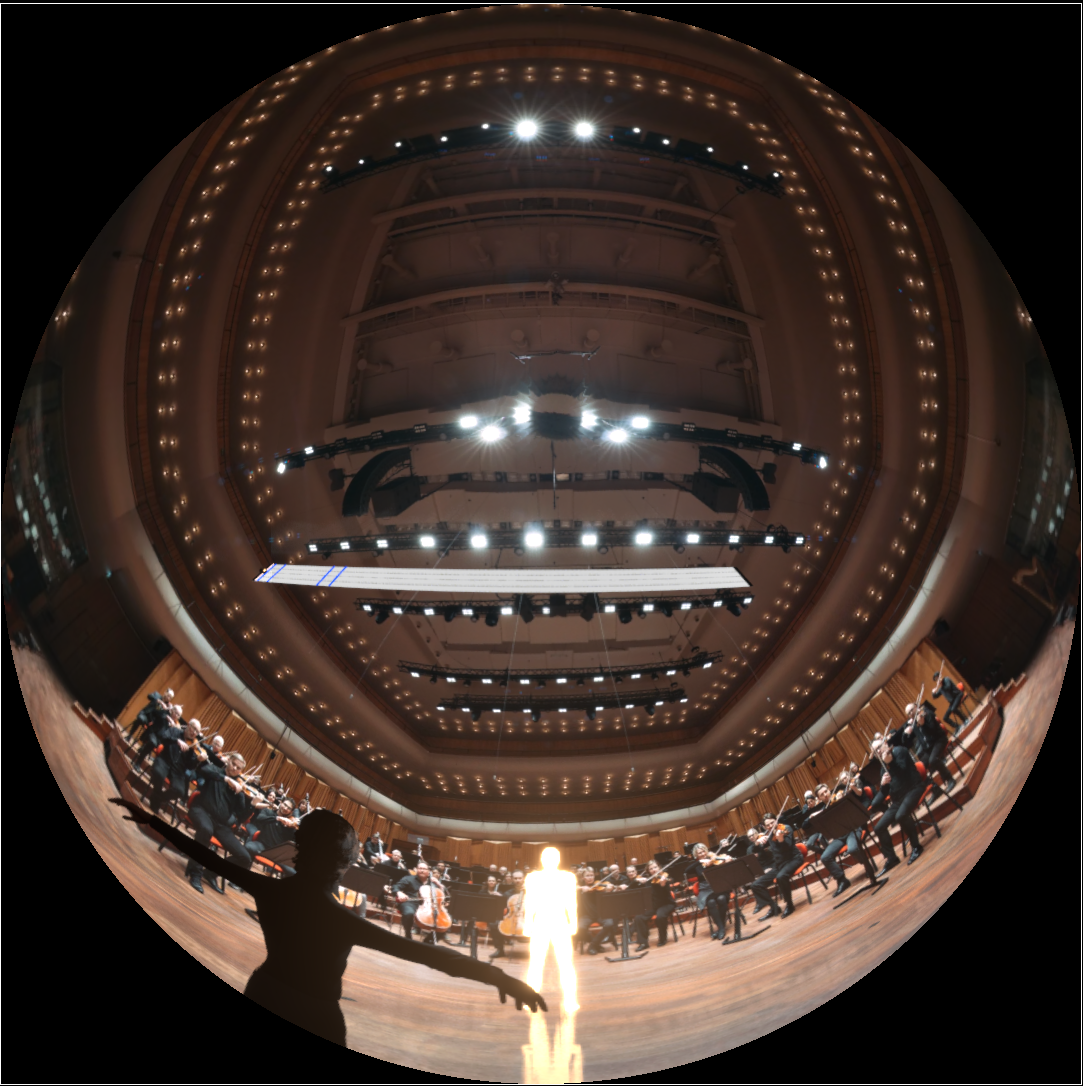}}
    \caption{Dome screen with avatar and conductor guide overlay. Musical stave on top of the screen.}
    \label{fig:domeescreen}
\end{figure}

Two separate versions of the application were created: a standard screen version using a 
standalone Unreal Engine executable rendering directly to the display (Fig.~\ref{fig:flatscreen}), and the dome version 
described in Sec.~\ref{sec:installation} (Figure \ref{fig:domeescreen}). Communication between the AI model and the Unreal 
Engine application was handled via UDP. Commands to start the neural network, play, stop, and 
restart the experience, as well as activate recording, were exchanged bidirectionally over the 
same protocol.

\subsection{Audio Time-Stretching}

Three approaches to real-time audio time-stretching were explored. The first pre-rendered the 
music at different speeds and switched between versions upon a tempo change, allowing the use of 
a higher-quality offline algorithm; however, this introduced noticeable synchronization issues 
and audible transitions between files. The second approach applied a simple real-time pitch-shift 
algorithm: while acceptable at higher speeds, this produced a metallic quality at lower speeds. 
The final approach, adopted in the installation, uses the WOLA Time Stretch algorithm via an 
Unreal Engine audio plugin, which produced the least perceptible artefacts of the three. During 
fermata bars, both audio and video pause, with the audio looping until a new speed command is 
received.

\subsection{Visual Feedback and User Interface}

A tutorial screen introduces the piece - its name, performers, and recording location - and 
explains how to conduct in both regular and fermata sections, with two animated demonstrations 
of the required gestures. During the experience, a musical stave scrolling from right to left 
provides tempo feedback, with fermata bars highlighted in blue and a numerical tempo indicator; 
the word 'fermata' appears on screen when the user enters a fermata bar. An end screen 
displays the original recording duration, the user's conducted duration, and the percentage 
difference. 

\section{Additional Real-Time Conducting Gesture Recognition Details}
\label{sec:appendixCondGestRecog}

This section complements Sec.~\ref{sec:mlmodel}, providing additional details on the phase state representation, LSTM training, and the Kalman Filter as an alternative architecture to the LSTM, including a comparison between the two, and more details on the playback speed controller.

\subsection{Phase State Representation}

We provide more motivation on the internal state representation in the gesture recognition module.

\paragraph{Unit circle representation.}
The key insight is to represent phase not as a scalar angle $\varphi_t$, but as a 2D vector on the unit circle:
\begin{equation}
\mathbf{x}_t = \begin{bmatrix} \sin(\varphi_t) \\ \cos(\varphi_t) \end{bmatrix} \in \mathbb{R}^2
\end{equation}

This representation has several advantages:
\begin{itemize}
    \item \textbf{Linear state transition:} Phase progression becomes a rotation matrix
    \item \textbf{Linear observation model:} Pose features can be linearly related to $[\sin(\varphi), \cos(\varphi)]$
    \item \textbf{Natural wrapping:} The unit circle naturally handles phase wrapping from $2\pi$ to $0$
\end{itemize}

\paragraph{Phase extraction.}
Given the state $\mathbf{x}_t = [\sin(\varphi_t), \cos(\varphi_t)]^T$, the phase angle can be recovered using:
\begin{equation}
\varphi_t = \mathrm{arctan2}(\sin(\varphi_t), \cos(\varphi_t)) = \mathrm{arctan2}(x_{t,1}, x_{t,2})
\end{equation}

\subsection{LSTM Training Details}
\label{sec:appendixLSTM}
To avoid early overfitting in the training of the LSTM (Sec.~\ref{sec:ConductingGestureRecognition}), we used a gradual learning-rate increase strategy with a cyclic learning-rate scheduler paired with the AdamW optimizer \cite{adamw}. Additionally, we introduced a ramp-up mechanism for the monotonicity loss weight. Initially, we allow the MSE to guide the network's early tuning stages without any other constraints. After a certain period, we progressively increase the monotonicity weight $\beta$ to discourage unnatural behavior, thus improving overall accuracy. During the ramp-up stage, a temporary weight $\beta^{\text{temp}}$ is recalculated at each epoch $e$ according to the ramp-up duration $r$, and then assigned to $\beta$:
\begin{equation}
\beta \rightarrow\beta^{\text{temp}}=
\begin{cases}
0, & e \le \frac{r}{5},\\
\beta \min\left(\dfrac{e}{r},\,1\right), & e > \frac{r}{5}.
\end{cases}
\label{eq:mono_weight}
\end{equation}

The gesture recognition network is implemented in Pytorch Lightning \footnote{The Pytorch LIghtning framework is available as an open-source platform on GitHub.}, trained and tested on Nvidia RTX 3080. We use \texttt{Tuner} class of Pytorch Lightning to set the maximum learning rate (the base learning rate is $10^{-7}$ in all trainings) of cyclical learning rate policy. The sliding window size of the LSTMs is set to 500. For other hyper-parameters ($\beta= 0.3$, $r=40$) we apply a grid search approach. Each model is trained for 200 epochs with an early stop if the model performance did not increase during consecutive 50 epochs. 

For each test subject in the dataset, two different validation subjects are used. This procedure results in 24 different models with slightly different training and validation data, as presented by the two leftmost columns of Table \ref{tab:cv_metrics_by_test}. Through this 24-fold strategy, each subject served as a test subject of two different model variants, enabling a comprehensive evaluation of the model's ability to generalize to unseen individuals.


\begin{table*}[t]
\centering
\small
\caption{Phase error (MSPE, Eq.~(\ref{eq:MSAE})) for different test subjects and pose input settings.}
\begin{tabular}{lcccc}
\toprule
\makecell{} & \makecell{60 Hz using $\mathbf{x}_t$} & \makecell{60 Hz using $\mathbf{x}_t^\mathrm{2D}$} & \makecell{20 Hz using $\mathbf{x}_t$} & \makecell{20 Hz using $\mathbf{x}_t^\mathrm{2D}$} \\
\toprule
    Subject 1  & 1.48 $\pm$ 0.43 & 1.95 $\pm$ 0.50 & 1.49 $\pm$ 0.43 & 1.87 $\pm$ 0.38 \\
    Subject 2  & 1.16 $\pm$ 0.34 & 1.29 $\pm$ 0.33 & 1.36 $\pm$ 0.43 & 1.47 $\pm$ 0.47 \\
    Subject 3  & 1.21 $\pm$ 0.17 & 1.27 $\pm$ 0.19 & 1.56 $\pm$ 0.24 & 1.60 $\pm$ 0.26 \\
    Subject 4  & 1.32 $\pm$ 0.29 & 1.52 $\pm$ 0.39 & 1.82 $\pm$ 0.38 & 1.33 $\pm$ 0.27 \\
    Subject 5  & 0.91 $\pm$ 0.16 & 1.20 $\pm$ 0.24 & 1.14 $\pm$ 0.23 & 1.45 $\pm$ 0.24 \\
    Subject 6  & 2.64 $\pm$ 0.55 & 2.90 $\pm$ 0.83 & 2.96 $\pm$ 1.07 & 2.65 $\pm$ 0.73 \\
    Subject 7  & 1.04 $\pm$ 0.28 & 1.50 $\pm$ 0.49 & 1.44 $\pm$ 0.31 & 1.05 $\pm$ 0.24 \\
    Subject 8  & 2.07 $\pm$ 0.24 & 2.17 $\pm$ 0.60 & 2.19 $\pm$ 0.48 & 2.48 $\pm$ 0.15 \\
    Subject 9  & 0.89 $\pm$ 0.12 & 1.02 $\pm$ 0.15 & 1.13 $\pm$ 0.13 & 1.00 $\pm$ 0.16 \\
    Subject 10 & 1.18 $\pm$ 0.34 & 1.19 $\pm$ 0.39 & 1.46 $\pm$ 0.16 & 1.26 $\pm$ 0.29 \\
    Subject 11 & 1.56 $\pm$ 0.46 & 1.64 $\pm$ 0.38 & 1.54 $\pm$ 0.35 & 2.02 $\pm$ 0.46 \\
    Subject 12 & 1.60 $\pm$ 0.73 & 1.60 $\pm$ 0.52 & 1.55 $\pm$ 0.55 & 1.62 $\pm$ 0.57 \\
\midrule
\textbf{All subjects} & 1.42 $\pm$ 0.34 & 1.60 $\pm$ 0.42 & 1.64 $\pm$ 0.39 & 1.65 $\pm$ 0.35 \\
\bottomrule
\end{tabular}
\label{tab:cv_metrics_by_test_detail}
\end{table*}

 
\subsection{Comparison of Gesture Recognition with LSTM vs Kalman Filtering}
\label{sec:kalman2}

As an alternative to the LSTM-based gesture recognition module, we implement a 
lightweight baseline using a linear Kalman filter. The key idea is to represent the conducting 
phase not as a scalar angle, but as a 2D vector on the unit circle $[\sin(\varphi_k), 
\cos(\varphi_k)]^T$. This allows phase progression to be modeled as a linear rotation, making 
the filter predict the next phase state by rotating the current estimate by a fixed mean phase 
increment $\omega$ estimated from training data, and then corrects this prediction using the 
incoming pose observation $\mathbf{x}^\mathrm{2D}_k$ via a learned linear observation model. 
The model parameters - including the observation matrix, noise covariances, and mean phase 
increment - are estimated from training data. For a thorough treatment of the Kalman filter 
we refer the reader to~\cite{welch1995introduction}.

The input and output of this module are identical to those of the LSTM-based module: it receives 
the 2D pose signal $\mathbf{x}^\mathrm{2D}_k$ and produces an estimated conducting phase 
$\hat{\varphi}_k$, making it a drop-in replacement within the system pipeline.

The Kalman filter represents uncertainty about the next phase state using a Gaussian probability distribution and assumes that the phase evolves smoothly over time according to a linear dynamical model. Under regular musical bars, where phase progression is continuous and approximately linear, these assumptions are satisfied, allowing the Kalman filter to estimate the phase with high accuracy, even with limited set of joints with low sampling rate, as it is demonstrated in Figures \ref{fig:app_fermataperformanceg}, \ref{fig:app_fermataperformanceh} .

\begin{figure*}[t]
    \centering
    \subfloat[\label{fig:app_fermataperformancea}\centering LSTM, Fermata bars, 60 Hz using $\mathbf{x}_t$]{%
            \includegraphics[width=0.48\linewidth]{images/PerformanceFigures/Subject1/outputs_test_David_17_val_Sofia_new_data_accel_phase_figure_20250429_152740.png}
        } 
    \subfloat[\label{fig:app_fermataperformanceb}\centering LSTM, Fermata bars, 20 Hz using $\mathbf{x}_t^\mathrm{2D}$, $\beta = 1$]{%
        \includegraphics[width=0.48\linewidth]{images/PerformanceFigures/Subject1/test_David_17_val_Sofia_fermata_upper_2D_high_mono.png}
        }

    \subfloat[\label{fig:app_fermataperformancec}\centering Kalman Filter, Fermata bars, 60 Hz using $\mathbf{x}_t$]{%
        \includegraphics[width=0.48\linewidth]{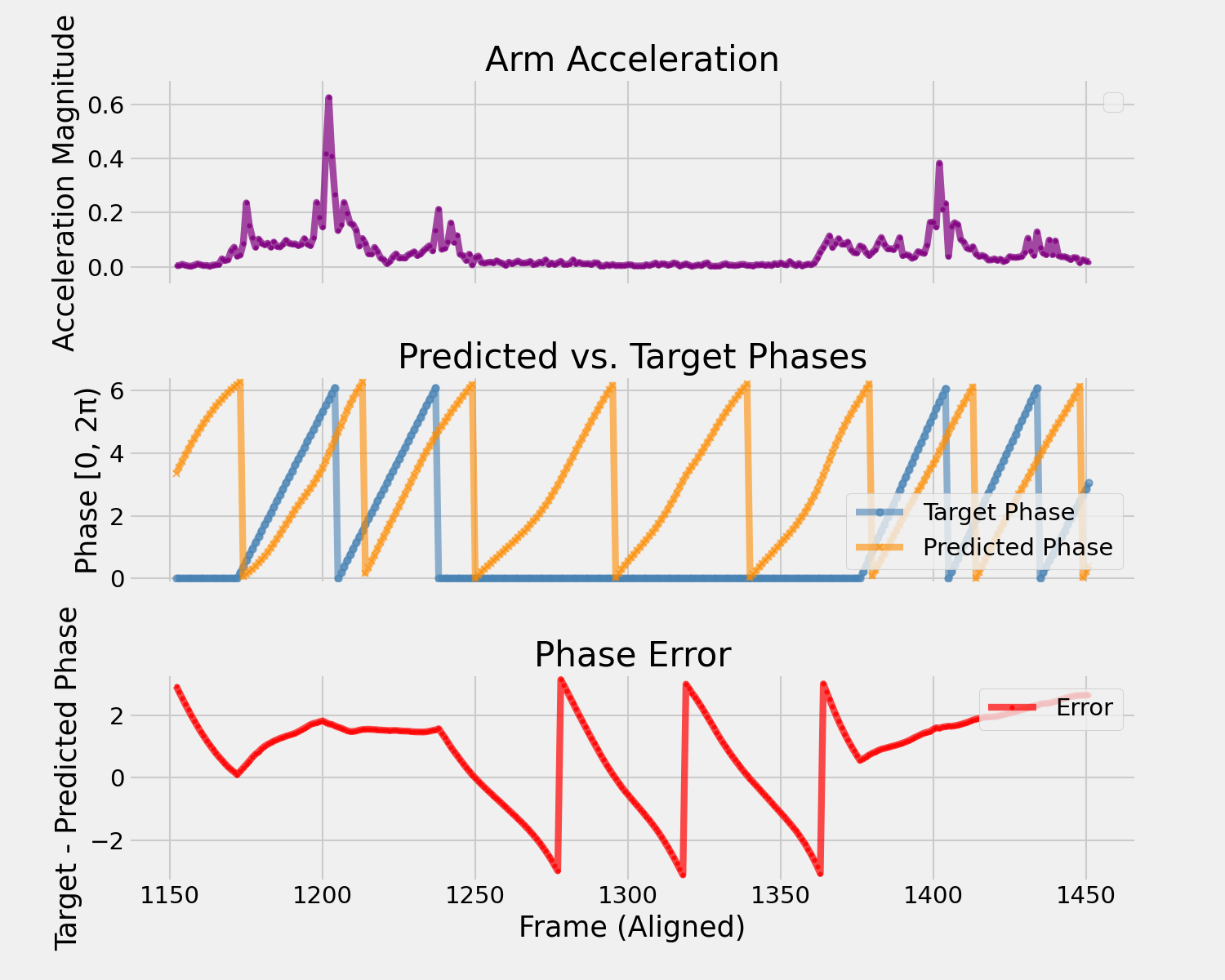}
    }
    \subfloat[\label{fig:app_fermataperformanced}\centering Kalman Filter, Fermata bars, 20 Hz using $\mathbf{x}_t^\mathrm{2D}$]{%
        \includegraphics[width=0.48\linewidth]{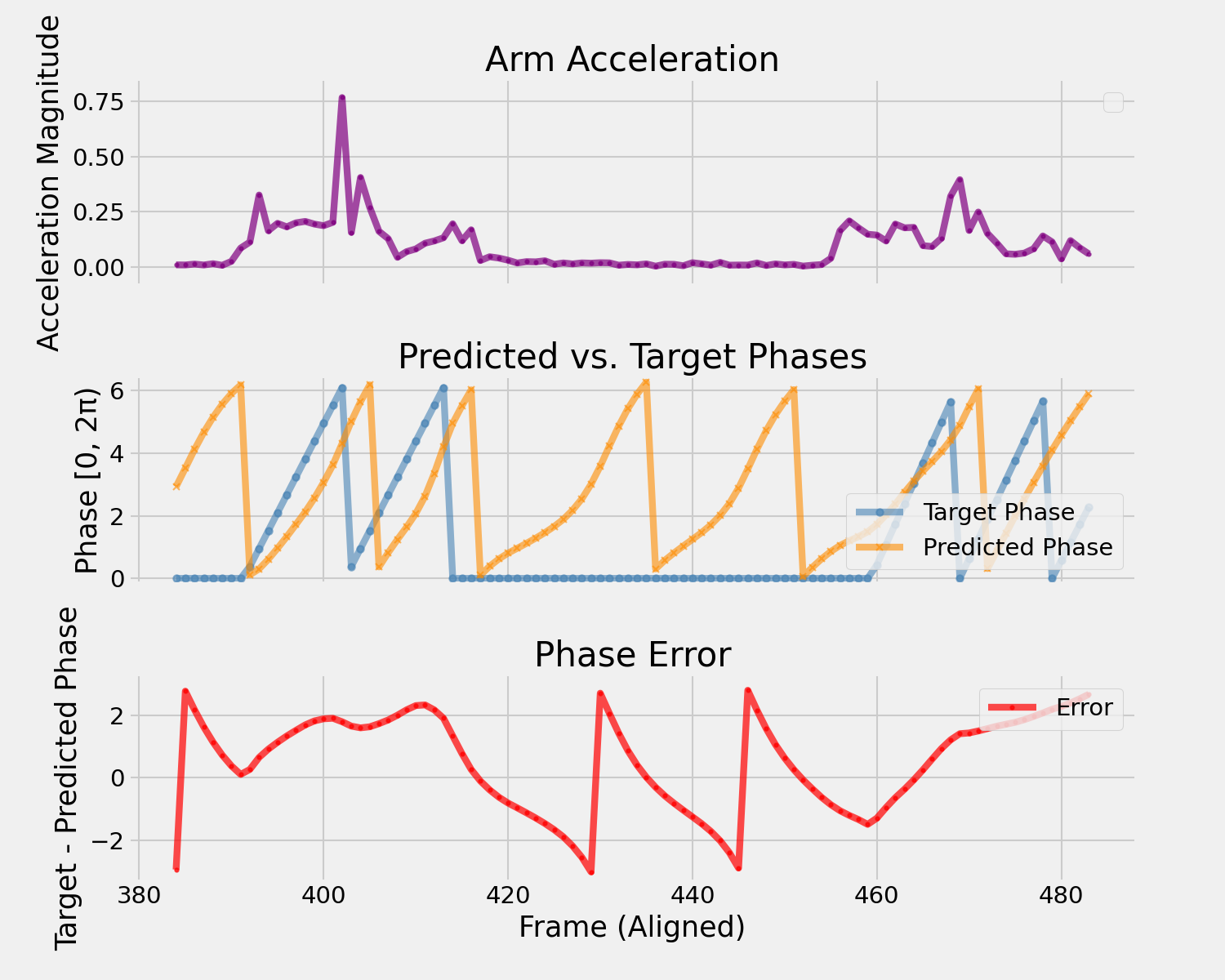}
    }

          \caption{Example of phase estimation for Subject 1. {\em Subfigures (e)-(h) on next page.} (a), (b), (e) and (f) demonstrates the performance of LSTM in regular and fermata bars. (c), (d), (g) and (h) demonstrate the performance of Kalman filter in regular and fermata bars. Top: Arm acceleration $\rho_t$. Middle: Ground truth phase $\varphi^\mathrm{gt}_t$ and estimated phase $\hat{\varphi}_t$. Bottom: Phase error $\varphi^\mathrm{gt}_t - \hat{\varphi}_t$. The three graphs are temporally aligned.}
    \label{fig:app_fermataperformance}
\end{figure*}

\begin{figure*}[t]
\setcounter{subfigure}{4}
    \centering
    \subfloat[\label{fig:app_fermataperformancee}\centering LSTM, Regular bars, 60 Hz using $\mathbf{x}_t$]{%
        \includegraphics[width=0.48\linewidth]{images/PerformanceFigures/Subject1/outputs_test_David_17_val_Sofia_new_data_accel_phase_figure_20250429_152835.png}
        } 
    \subfloat[\label{fig:app_fermataperformancef}\centering LSTM, Regular bars, 20 Hz using $\mathbf{x}_t^\mathrm{2D}$, $\beta = 1$]{%
        \includegraphics[width=0.48\linewidth]{images/PerformanceFigures/Subject1/test_David_17_val_Sofia_regular_upper_2D_high_mono.png}
    }

    \subfloat[\label{fig:app_fermataperformanceg}\centering Kalman Filter, Regular bars, 60 Hz using $\mathbf{x}_t$]{%
        \includegraphics[width=0.48\linewidth]{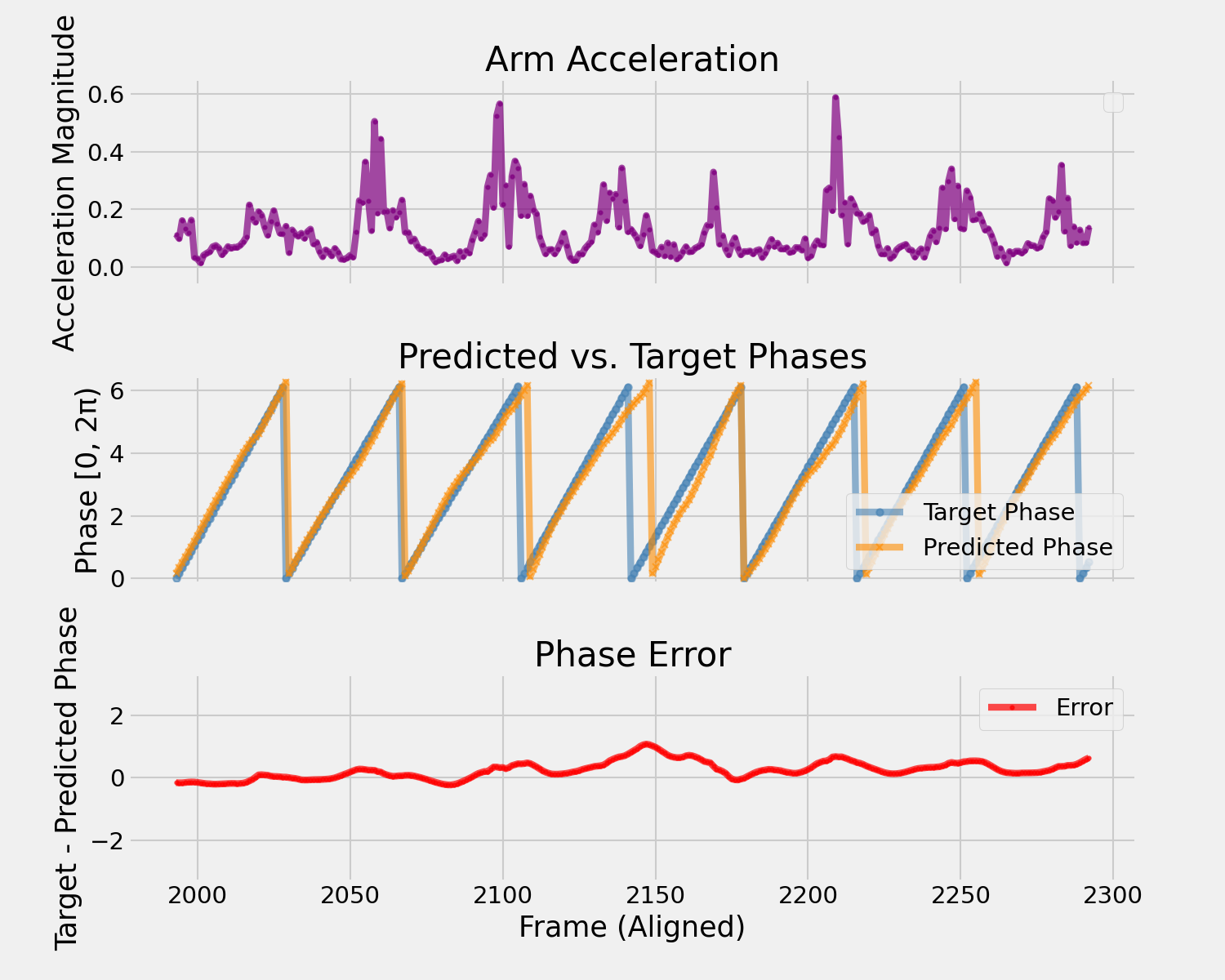}
    }
    \subfloat[\label{fig:app_fermataperformanceh}\centering Kalman Filter, Regular bars, 20 Hz using $\mathbf{x}_t^\mathrm{2D}$]{%
        \includegraphics[width=0.48\linewidth]{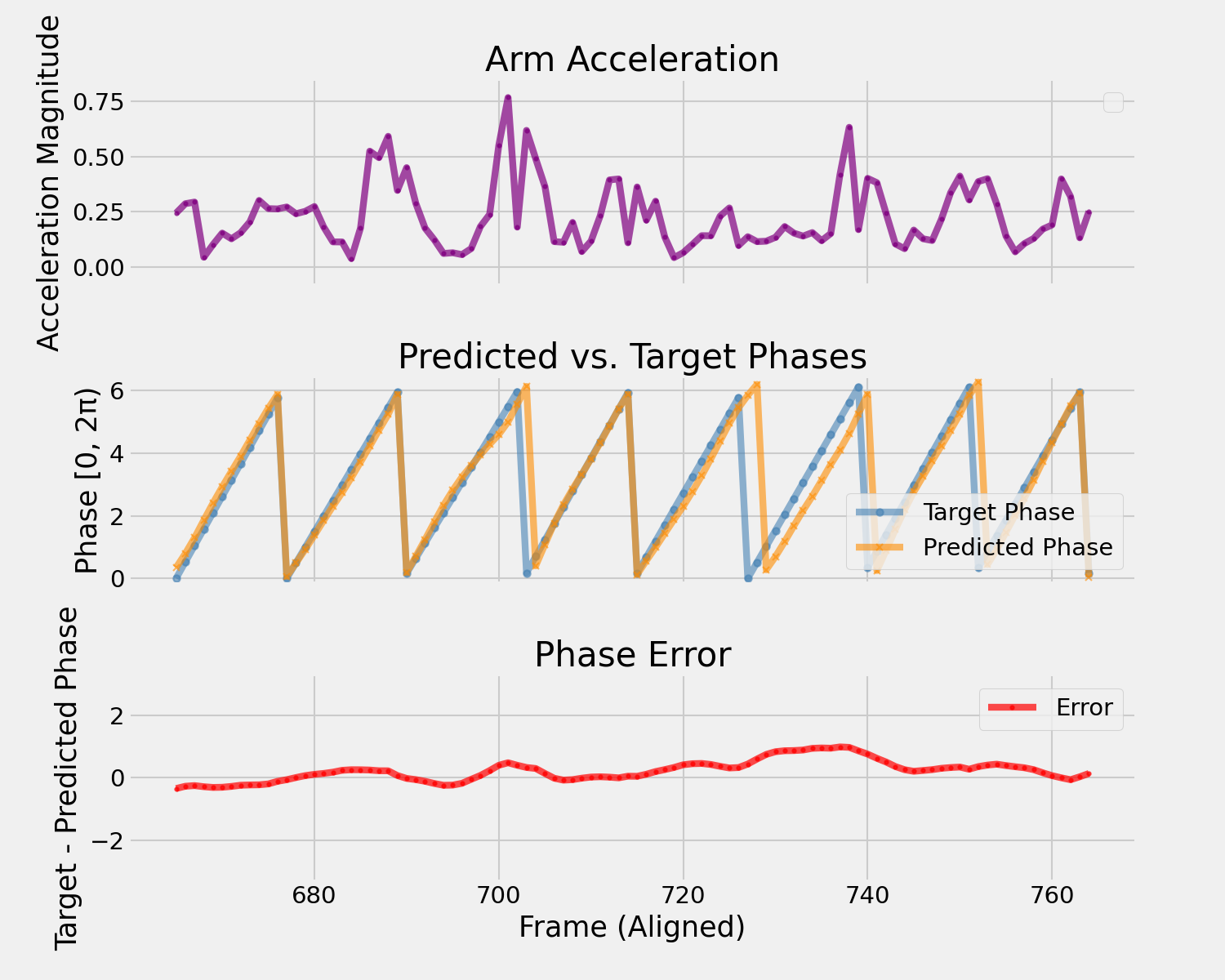}
    }
  
\end{figure*}

In fermata bars, however, the phase progression changes abruptly and may effectively stop for an unknown duration. This behavior violates the smoothness assumption of the Kalman filter’s state transition model. Consequently, the predicted Gaussian distribution places very little probability mass near zero progression, making it unlikely for the filter to correctly infer a halted or near-stationary phase from the observations alone. The performance of Kalman filter in fermata bars is demonstrated in Figures \ref{fig:app_fermataperformancec}, \ref{fig:app_fermataperformanced}. These results indicate that the Kalman filter continues to operate under the same dynamical assumptions as in regular bars, without adapting to the constraints of fermata bars.

In contrast, the LSTM model demonstrates robust performance across both regular and fermata bars. By not relying on explicit assumptions of linearity or smooth state evolution, the LSTM learns non-linear temporal relationships directly from data through internal memory states that capture both short-term dynamics and longer-term contextual information. This learned, piece-specific mapping between pose features and phase progression enables the model to accurately represent continuous phase evolution in regular bars, as in Figures \ref{fig:app_fermataperformancee}, \ref{fig:app_fermataperformancef}, while also recognizing and accommodating the characteristic of fermata bars as in Figures \ref{fig:app_fermataperformancea}, \ref{fig:app_fermataperformanceb}, resulting in more reliable phase estimates throughout the sequence.

\subsection{Additional Real-Time Controller Details}
\label{sec:appendixController}
This section provides further algorithmic details to the description of the finite state machine in Sec.~\ref{sec:RealTimeController}.

One step of the controller looks as follows:
\begin{algorithm}
\begin{algorithmic}[1]
\If{$\text{state} = \textit{waiting for upbeat}$} 
    \If{detected upbeat} \Comment{See Eq.~(\ref{eq:upbeat})}
        \State $\mathbf{k} \gets [k]$ \Comment{Start beat history}
        \State $\text{state} \gets \textit{waiting for downbeat}$ 
    \EndIf
\Else
    \If{$\text{state} = \textit{waiting for downbeat}$}
        \If{detected downbeat} \Comment{See Eq.~(\ref{eq:downbeat})}
            \If{$b^\mathrm{gt}_k + 1 \in B$} \Comment{Fermata bar\footnote{Checks the upcoming bar. If $k$ is just in the beginning of the current bar, it instead checks if the current bar has a fermata: $b^\mathrm{gt}_k \in B$.} }
                \State $\text{state} \gets \textit{sleep}$
                \State $k^0 \gets k$ \Comment{Set fermata start time}
            \Else
                \State $\mathbf{k} \gets [k~\mathbf{k}]$ \Comment{Add to beat history}
                \State $s_k \gets \text{speed}(\mathbf{k})$ \Comment{See Eq.~(\ref{eq:speed})}
            \EndIf
        \EndIf
    \Else
        \If{$\text{state} = \textit{sleep}$}
            \If{$k > k^0 + T^\mathrm{s}$} \Comment{Waited $T^\mathrm{s}$}
                \State $\text{state} \gets \textit{waiting for upbeat}$
                \State $\mathbf{k} \gets []$ \Comment{Empty beat history}
            \EndIf
        \EndIf
    \EndIf
\EndIf 
\end{algorithmic}
\smallskip
{\small where $B$ is the set of fermata bars (Sec.~\ref{sec:original}).}
\end{algorithm}


\section{Additional Evaluation Details}
This section provides further details on the gesture recognition evaluation, additional gesture recognition examples, additional real-time interaction examples from the user study, and quantitative results from the field study, complementing the evaluation presented in Sec.~\ref{sec:evaluation}.


\begin{table*}[t]
\centering
\caption{Mean distance, standard deviation (in seconds), average bar length, and mean distance as a percentage of average bar length per subject from the field study.}
\begin{tabular}{lccc}
\toprule
 & Mean Distance $\pm$ Std (s) & Avg Bar Length (s) & \% of Bar Length \\
\midrule
Take 1 (Case 1 in Sec.~6.3) & $0.179 \pm 0.122$ & 0.726 & 24.7\% \\
Take 2  & $0.157 \pm 0.108$ & 0.726 & 21.6\% \\
Take 3  & $0.125 \pm 0.115$ & 0.654 & 19.1\% \\
Take 4  & $0.181 \pm 0.111$ & 0.715 & 25.3\% \\
Take 5  & $0.186 \pm 0.122$ & 0.757 & 24.6\% \\
Take 6  & $0.177 \pm 0.130$ & 0.790 & 22.4\% \\
Take 7  & $0.163 \pm 0.108$ & 0.574 & 28.4\% \\
Take 8  & $0.160 \pm 0.115$ & 0.685 & 23.4\% \\
Take 9  & $0.134 \pm 0.109$ & 0.581 & 23.1\% \\
Take 10 & $0.214 \pm 0.142$ & 0.877 & 24.4\% \\
Take 11 & $0.169 \pm 0.133$ & 0.792 & 21.3\% \\
Take 12 & $0.180 \pm 0.111$ & 0.667 & 27.0\% \\
Take 13 & $0.182 \pm 0.131$ & 0.810 & 22.5\% \\
Take 14 & $0.147 \pm 0.107$ & 0.608 & 24.2\% \\
Take 15 & $0.145 \pm 0.112$ & 0.611 & 23.7\% \\
Take 16 & $0.162 \pm 0.105$ & 0.590 & 27.5\% \\
Take 17 & $0.183 \pm 0.108$ & 0.660 & 27.7\% \\
Take 18 (Case 2 in Sec.~6.3) & $0.144 \pm 0.104$ & 0.643 & 22.4\% \\
Take 19 (Case 3 in Sec.~6.3) & $0.181 \pm 0.118$ & 0.729 & 24.8\% \\
Take 20 & $0.169 \pm 0.122$ & 0.704 & 24.0\% \\
Take 21 & $0.151 \pm 0.116$ & 0.618 & 24.4\% \\
Take 22 & $0.183 \pm 0.128$ & 0.762 & 24.0\% \\
Take 23 & $0.151 \pm 0.091$ & 0.567 & 26.6\% \\
\midrule
Mean $\pm$ Std & $0.166 \pm 0.020$ & $0.689 \pm 0.085$ & $24.2 \pm 2.2$\% \\
\bottomrule
\end{tabular}
\label{tab:field_study}
\end{table*}

\subsection{Additional Gesture Recognition Examples}
\label{sec:appendixGesture}

Table \ref{tab:cv_metrics_by_test_detail} shows the phase error for different test subjects and pose input settings. An observation we can make is that the phase estimation accuracy variation among subjects is not very large. This indicates that the network generalizes well and is quite robust to individual variation in motion pattern.

\begin{figure*}[t]
    \centering
    \subfloat[\label{fig:fermataperformanceSubject2a}\centering Fermata bars, 60 Hz using $\mathbf{x}_t$]{%
        \includegraphics[width=0.48\linewidth]{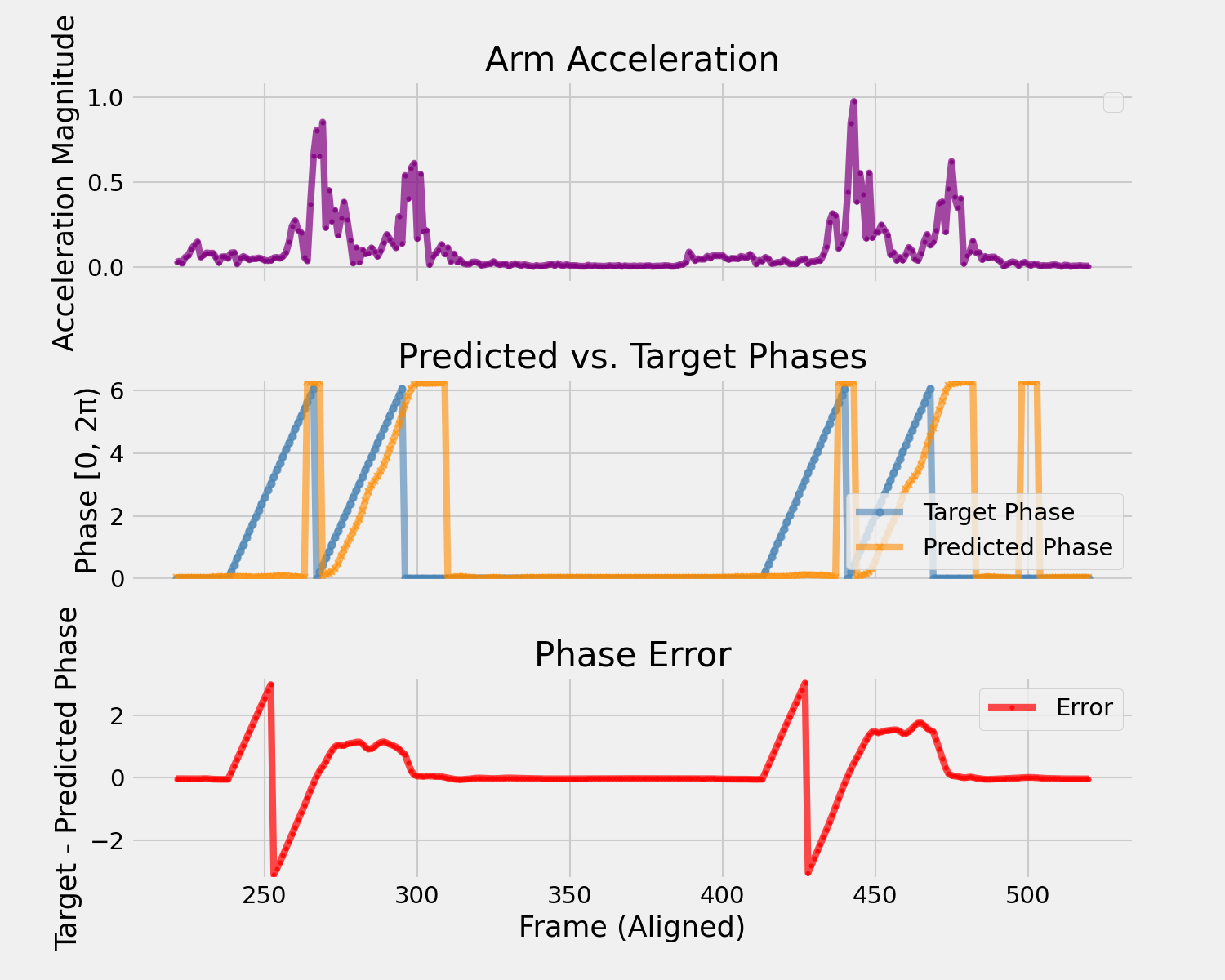}
    } 
    \subfloat[\label{fig:fermataperformanceSubject2b}\centering Regular bars, 60 Hz using $\mathbf{x}_t$]{%
        \includegraphics[width=0.48\linewidth]{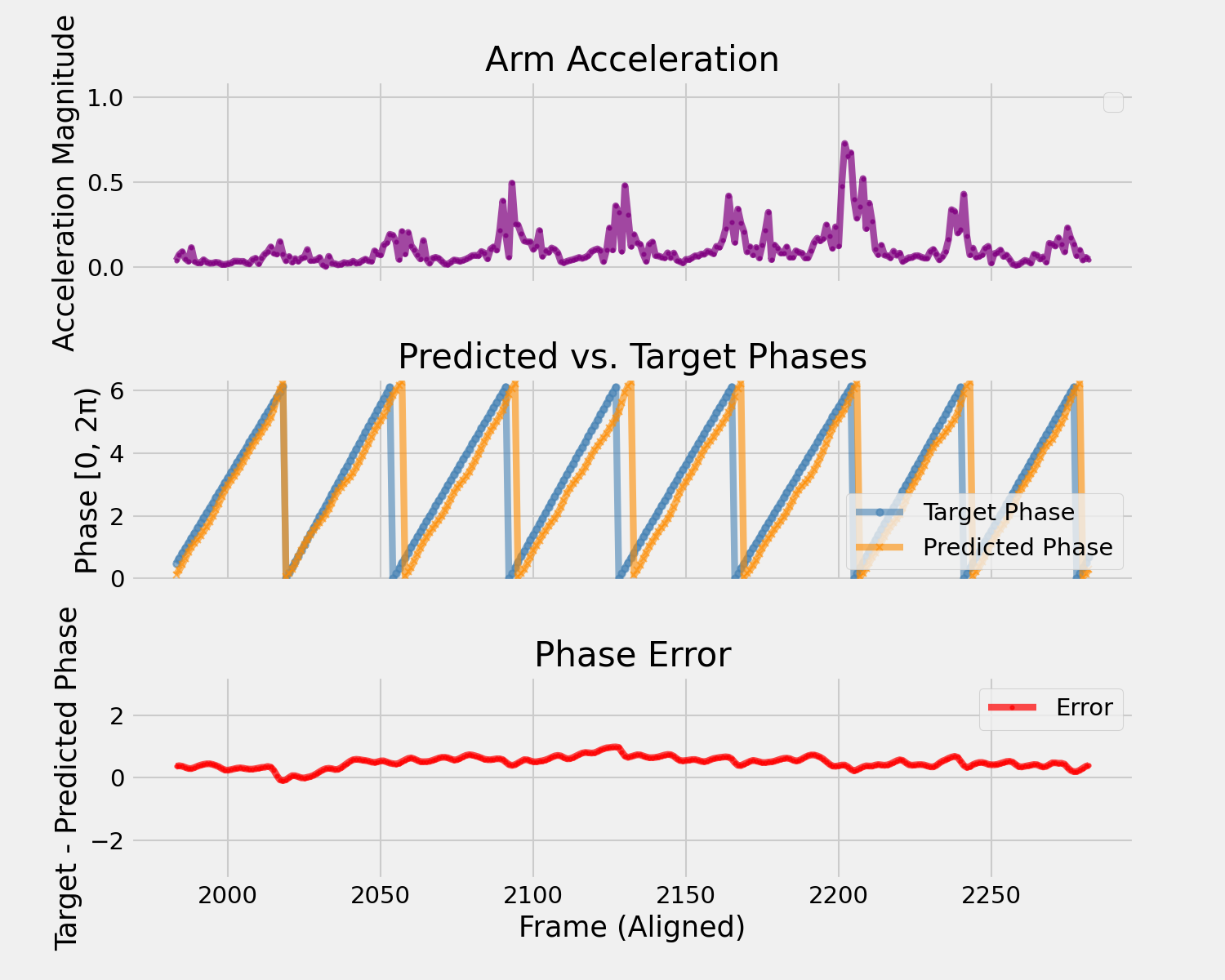}
    } \\ 

    \subfloat[\label{fig:fermataperformanceSubject2c}\centering Fermata bars, 20 Hz using $\mathbf{x}_t^\mathrm{2D}$]{%
        \includegraphics[width=0.48\linewidth]{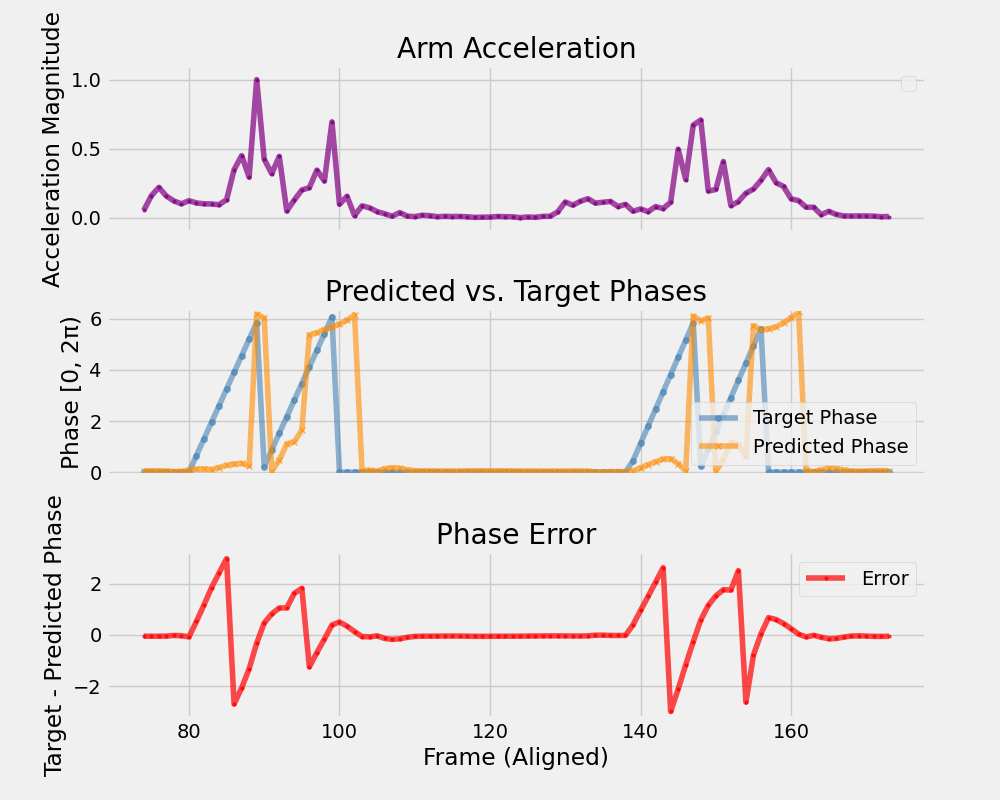}
        }
    \subfloat[\label{fig:fermataperformanceSubject2d}\centering Regular bars, 20 Hz using $\mathbf{x}_t^\mathrm{2D}$]{%
        \includegraphics[width=0.48\linewidth]{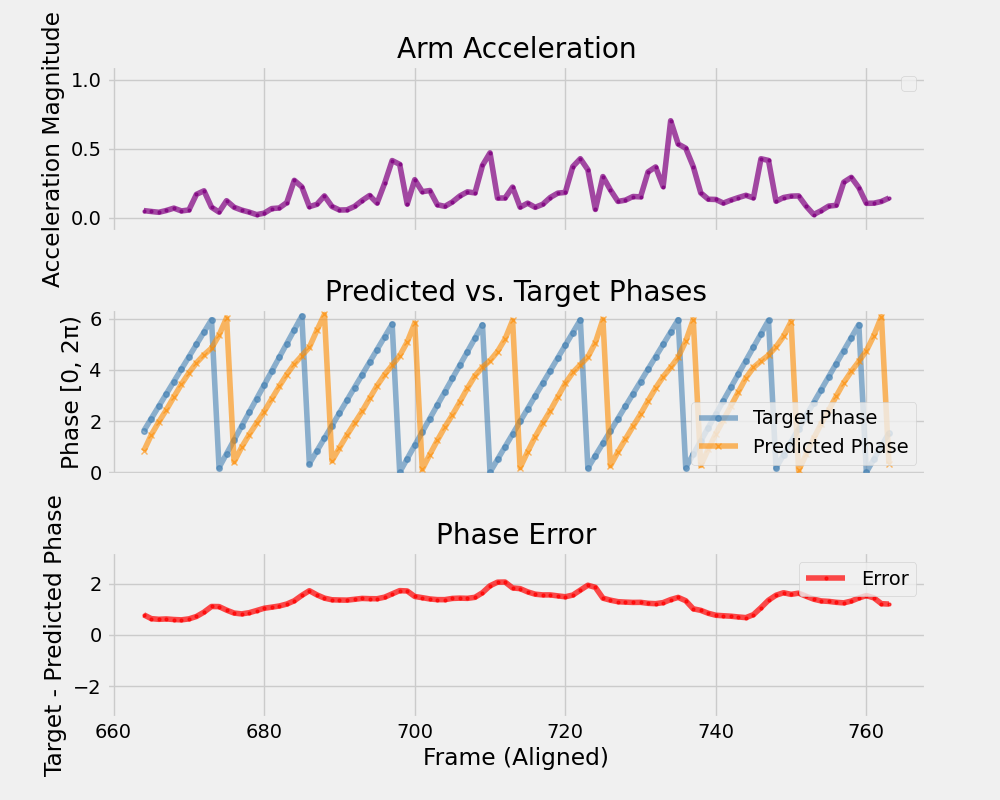}}
    
    \caption{Example of phase estimation for Subject 2. (a) and (b) correspond to column 1 in Table \ref{tab:cv_metrics_by_test_detail}, (c) and (d) correspond to column 4 in Table \ref{tab:cv_metrics_by_test_detail}. Top: Arm acceleration $\rho_t$. Middle: Ground truth phase $\varphi^\mathrm{gt}_t$ and estimated phase $\hat{\varphi}_t$. Bottom: Phase error $\varphi^\mathrm{gt}_t - \hat{\varphi}_t$. The three graphs are temporally aligned.}
    \label{fig:fermataperformanceSubject2}
\end{figure*}
\begin{figure*}[t]
    \centering
    \subfloat[\label{fig:fermataperformanceSubject3a}\centering Fermata bars, 60 Hz using $\mathbf{x}_t$]{%
        \includegraphics[width=0.48\linewidth]{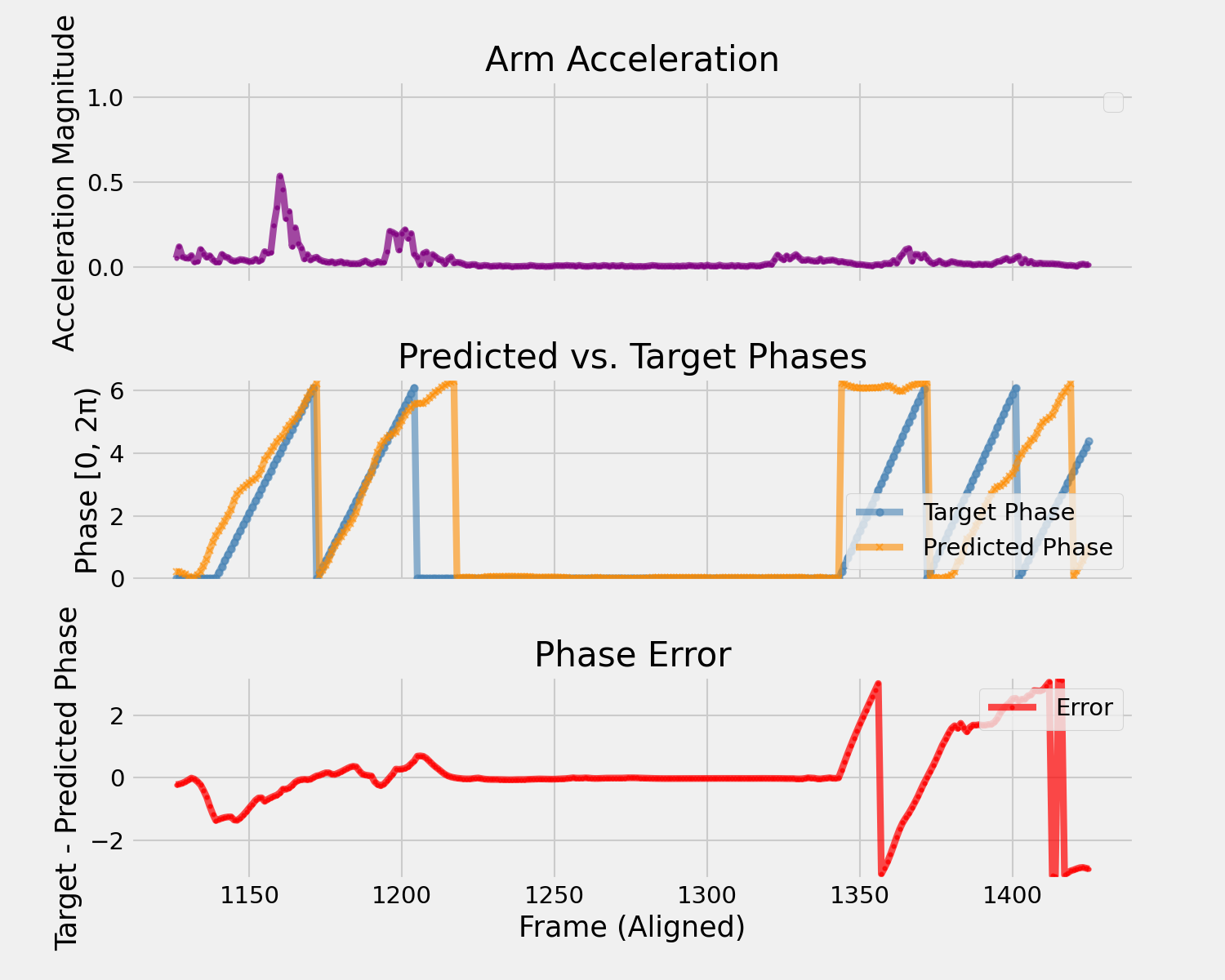}
    } 
    \subfloat[\label{fig:fermataperformanceSubject3b}\centering Regular bars, 60 Hz using $\mathbf{x}_t$]{%
        \includegraphics[width=0.48\linewidth]{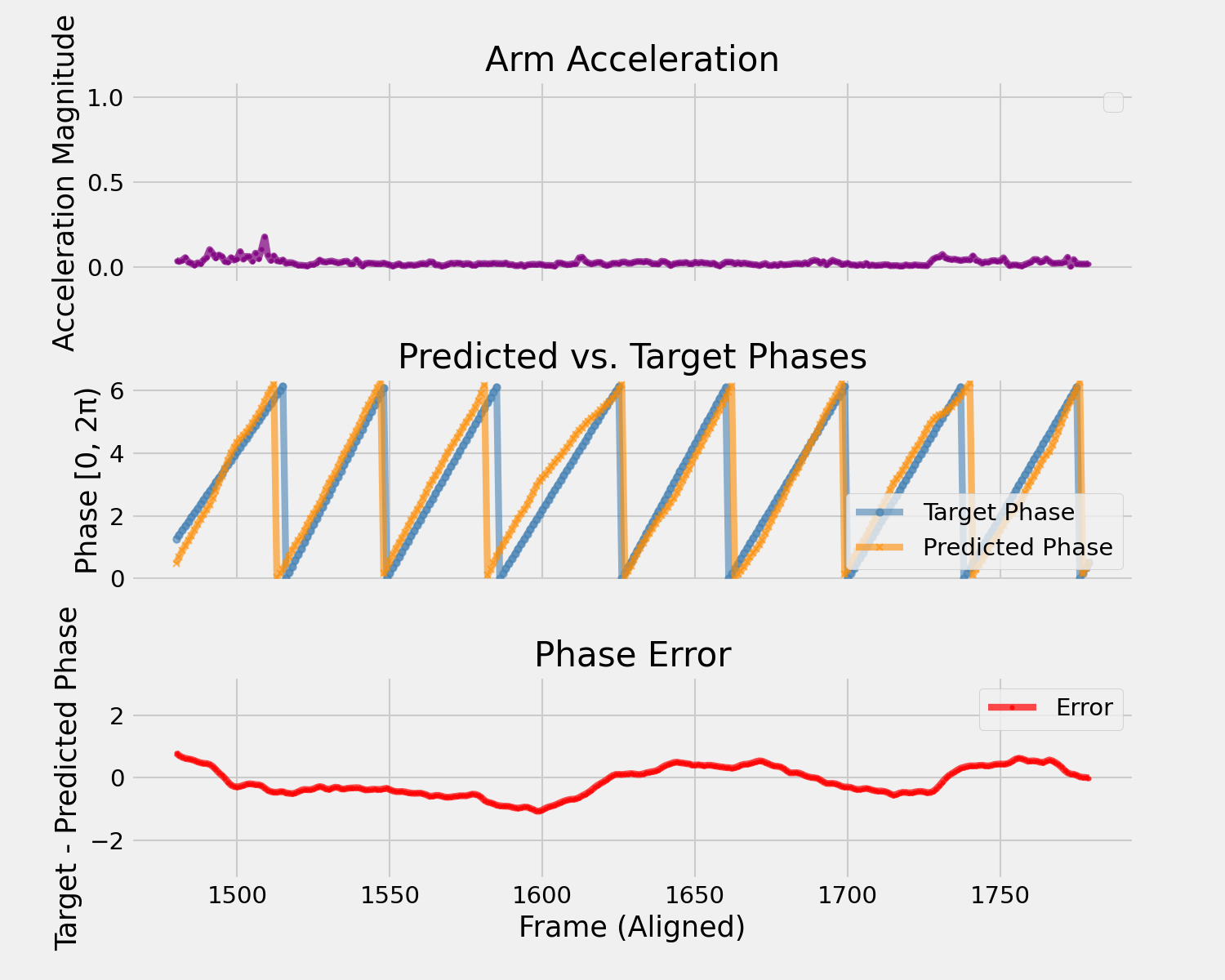}
    } \\ 

    \subfloat[\label{fig:fermataperformanceSubject3c}\centering Fermata bars, 20 Hz using $\mathbf{x}_t^\mathrm{2D}$]{%
        \includegraphics[width=0.48\linewidth]{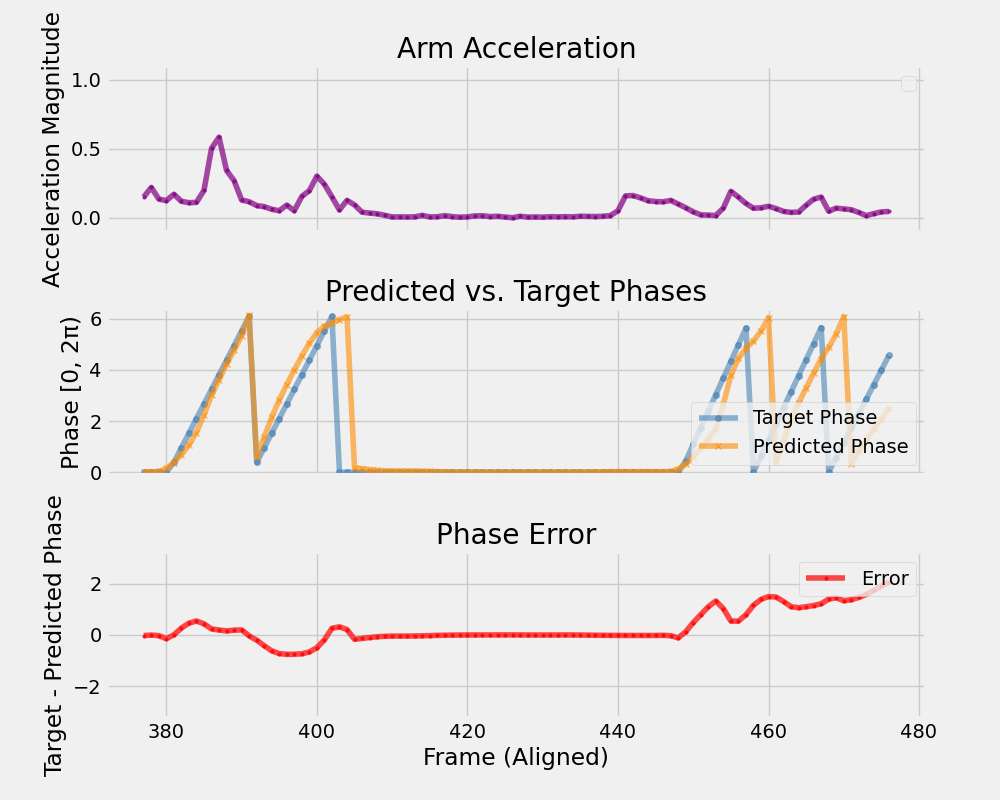}
        }
    \subfloat[\label{fig:fermataperformanceSubject3d}\centering Regular bars, 20 Hz using $\mathbf{x}_t^\mathrm{2D}$]{%
        \includegraphics[width=0.48\linewidth]{images/PerformanceFigures/Subject3/outputs_test_Gudrun_11_val_Ellinor_new_data_accel_phase_figure_20250428_143820.png}}
    
    \caption{Example of phase estimation for Subject 3. (a) and (b) correspond to column 1 in Table \ref{tab:cv_metrics_by_test_detail}, (c) and (d) correspond to column 4 in Table \ref{tab:cv_metrics_by_test_detail}. Top: Arm acceleration $\rho_t$. Middle: Ground truth phase $\varphi^\mathrm{gt}_t$ and estimated phase $\hat{\varphi}_t$. Bottom: Phase error $\varphi^\mathrm{gt}_t - \hat{\varphi}_t$. The three graphs are temporally aligned.}
    \label{fig:fermataperformanceSubject3}
\end{figure*}
\begin{figure*}[t]
    \centering
    \subfloat[\label{fig:fermataperformanceSubject4a}\centering Fermata bars, 60 Hz using $\mathbf{x}_t$]{%
        \includegraphics[width=0.48\linewidth]{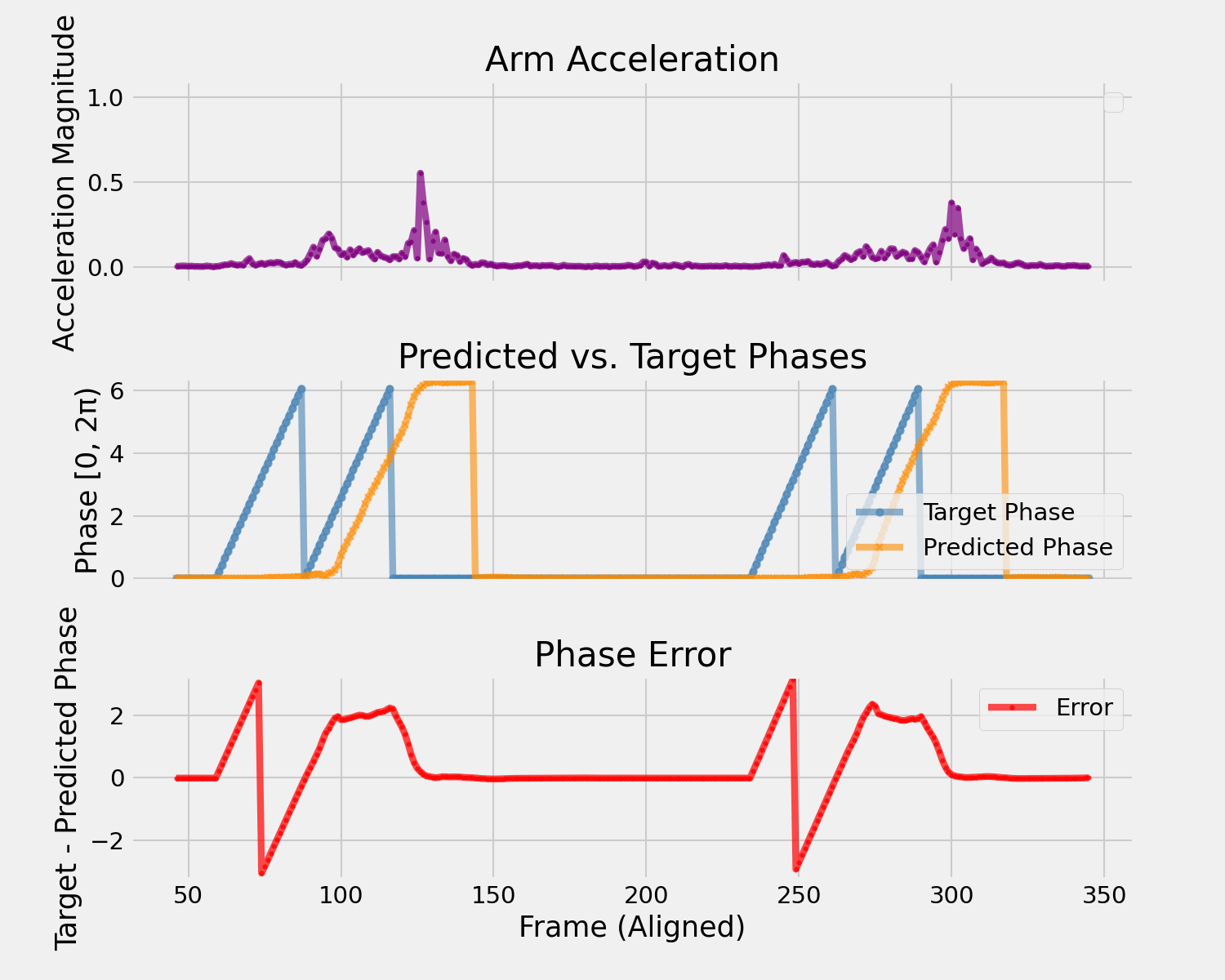}
    } 
    \subfloat[\label{fig:fermataperformanceSubject4b}\centering Regular bars, 60 Hz using $\mathbf{x}_t$]{%
        \includegraphics[width=0.48\linewidth]{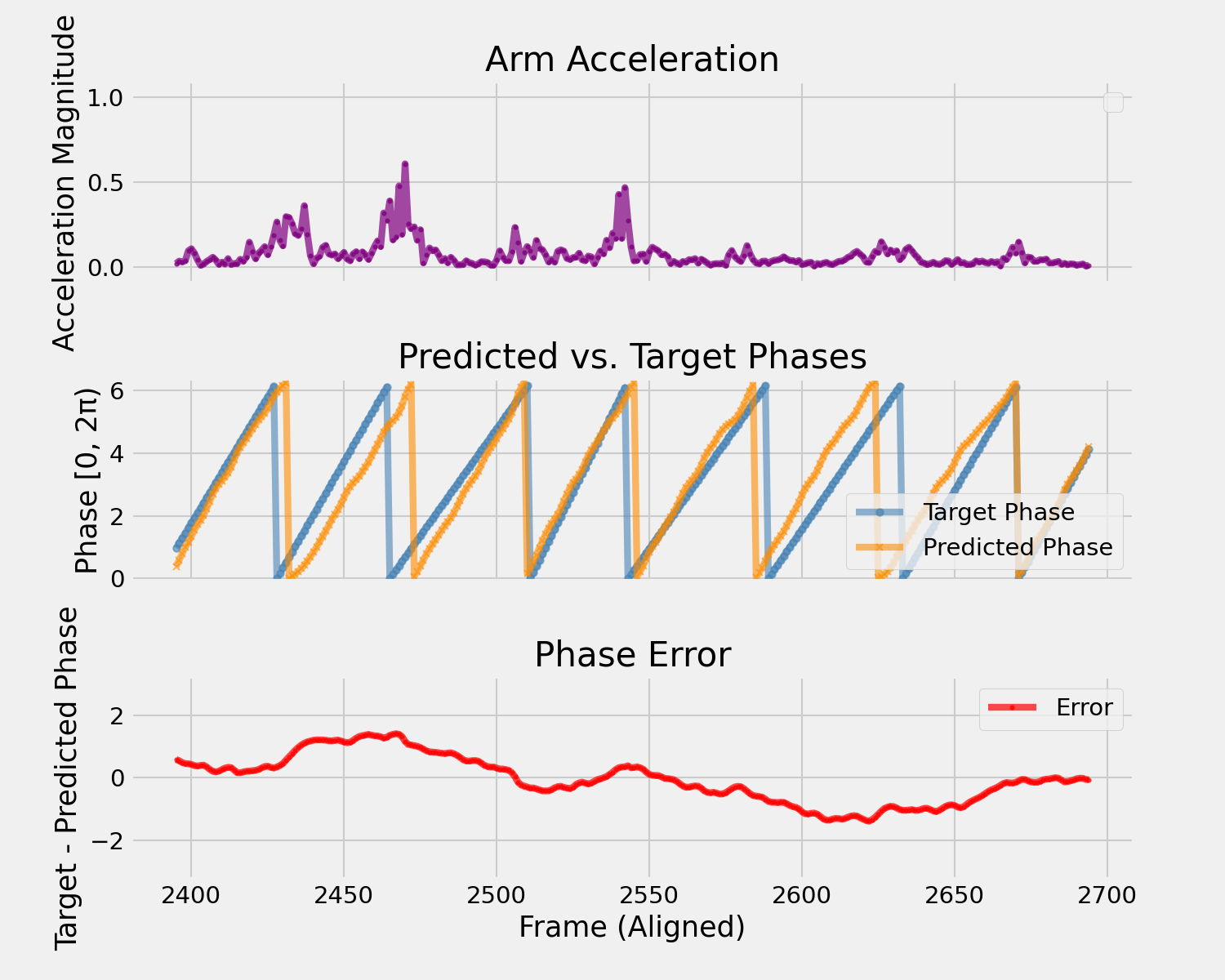}
    } \\ 

    \subfloat[\label{fig:fermataperformanceSubject4c}\centering Fermata bars, 20 Hz using $\mathbf{x}_t^\mathrm{2D}$]{%
        \includegraphics[width=0.48\linewidth]{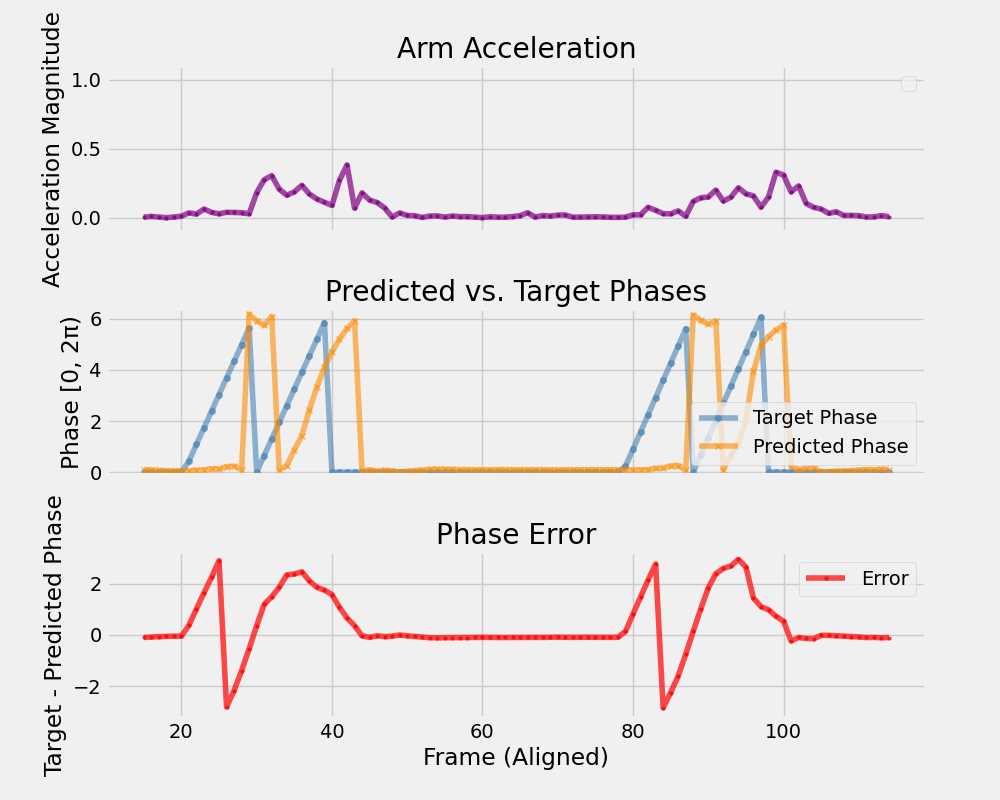}
        }
    \subfloat[\label{fig:fermataperformanceSubject4d}\centering Regular bars, 20 Hz using $\mathbf{x}_t^\mathrm{2D}$]{%
        \includegraphics[width=0.48\linewidth]{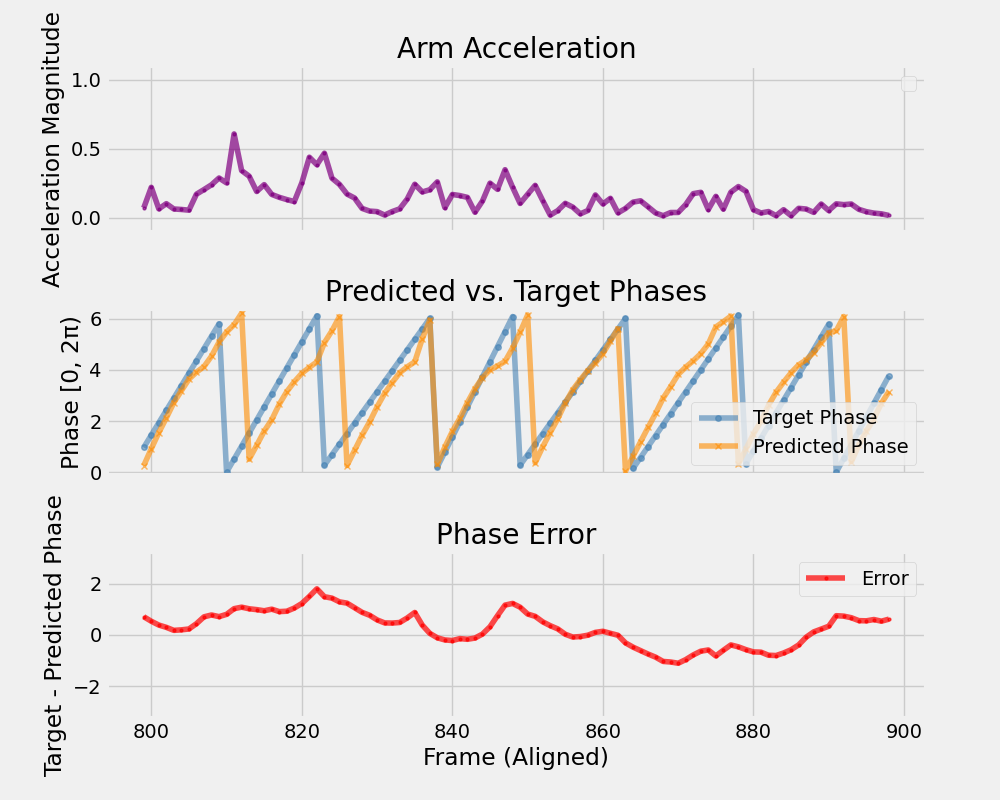}}
    
    \caption{Example of phase estimation for Subject 4. (a) and (b) correspond to column 1 in Table \ref{tab:cv_metrics_by_test_detail}, (c) and (d) correspond to column 4 in Table \ref{tab:cv_metrics_by_test_detail}. Top: Arm acceleration $\rho_t$. Middle: Ground truth phase $\varphi^\mathrm{gt}_t$ and estimated phase $\hat{\varphi}_t$. Bottom: Phase error $\varphi^\mathrm{gt}_t - \hat{\varphi}_t$. The three graphs are temporally aligned.}
    \label{fig:fermataperformanceSubject4}
\end{figure*}
\begin{figure*}[t]
    \centering
    \subfloat[\label{fig:fermataperformanceSubject5a}\centering Fermata bars, 60 Hz using $\mathbf{x}_t$]{%
        \includegraphics[width=0.48\linewidth]{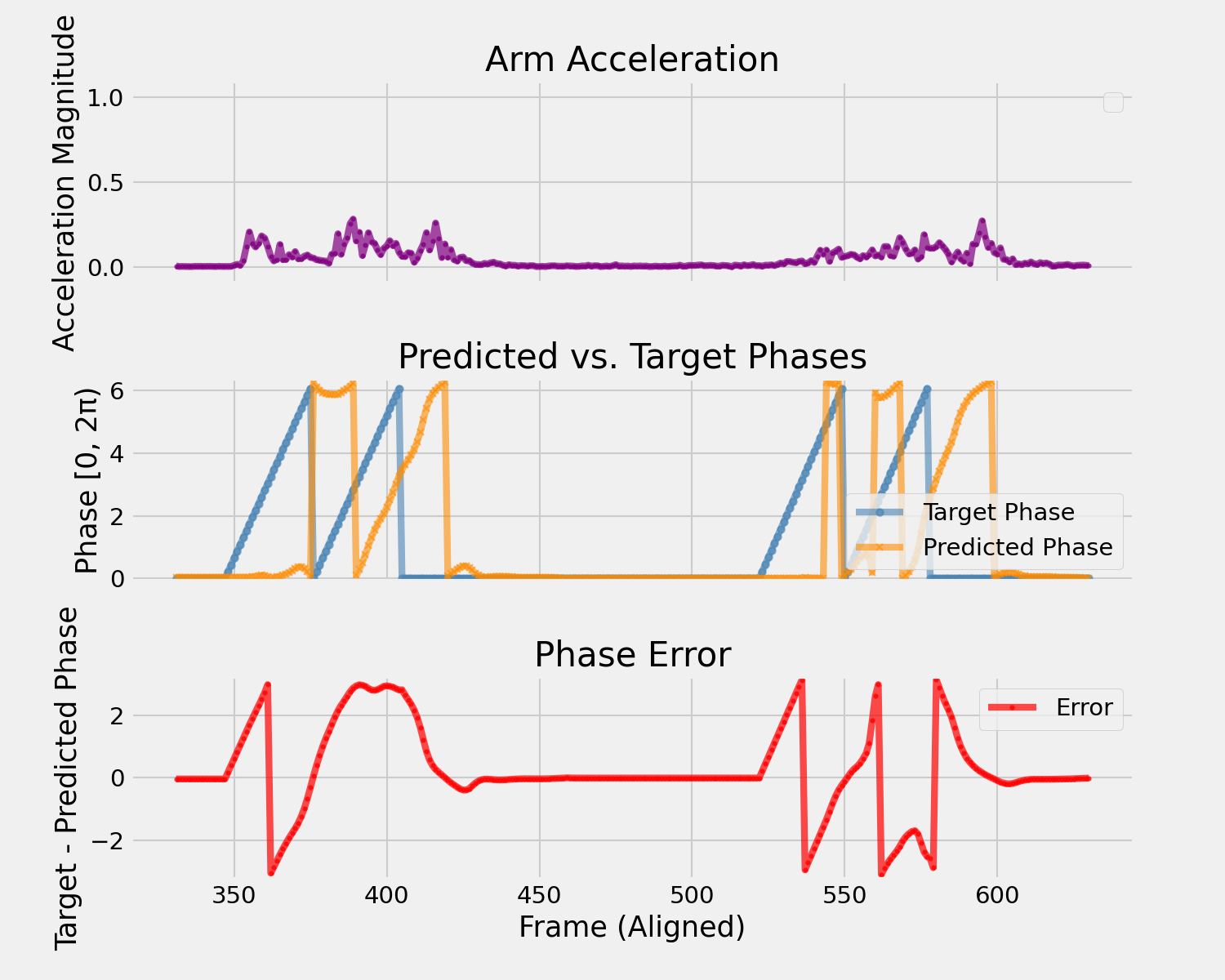}
    } 
    \subfloat[\label{fig:fermataperformanceSubject5b}\centering Regular bars, 60 Hz using $\mathbf{x}_t$]{%
        \includegraphics[width=0.48\linewidth]{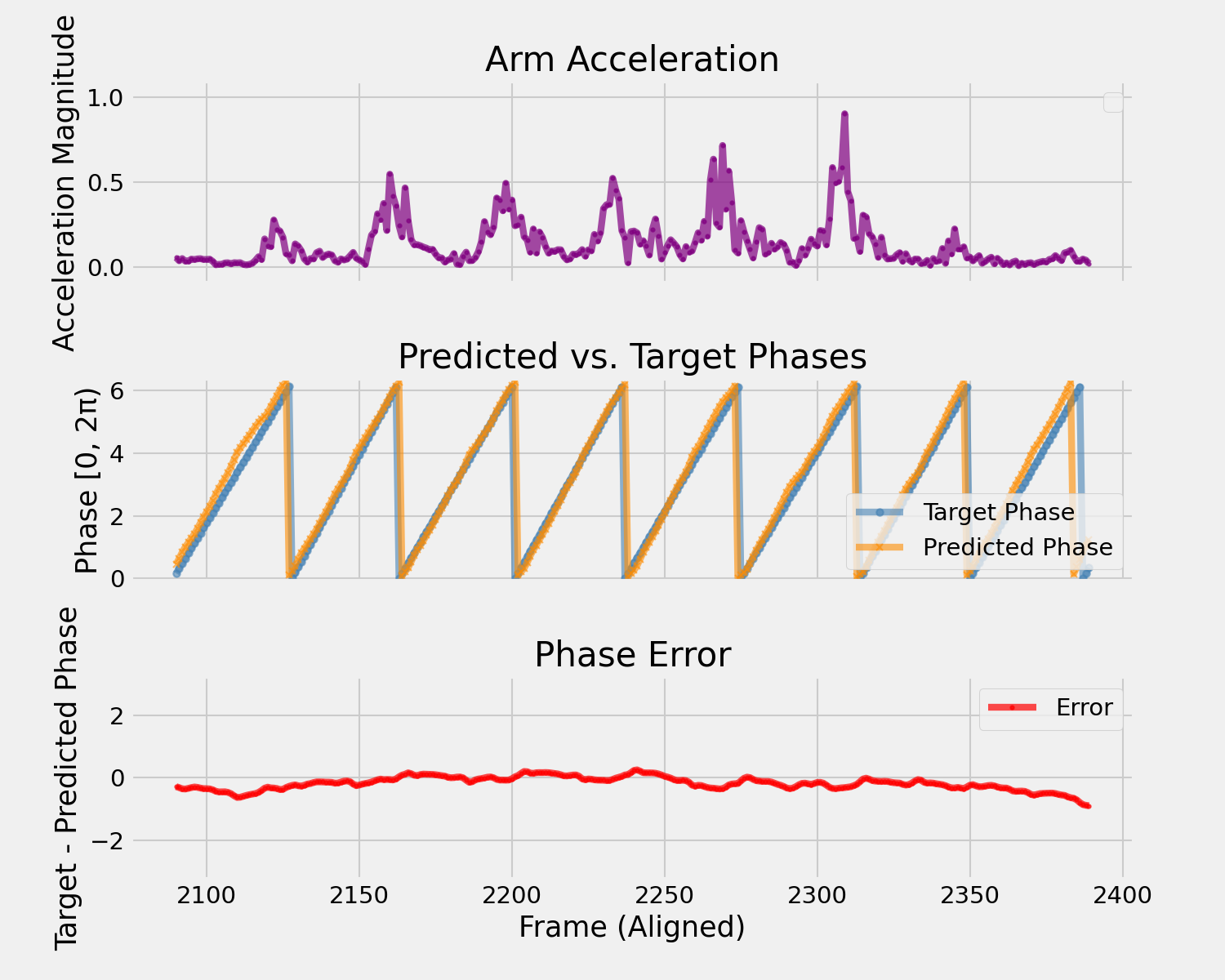}
    } \\ 

    \subfloat[\label{fig:fermataperformanceSubject5c}\centering Fermata bars, 20 Hz using $\mathbf{x}_t^\mathrm{2D}$]{%
        \includegraphics[width=0.48\linewidth]{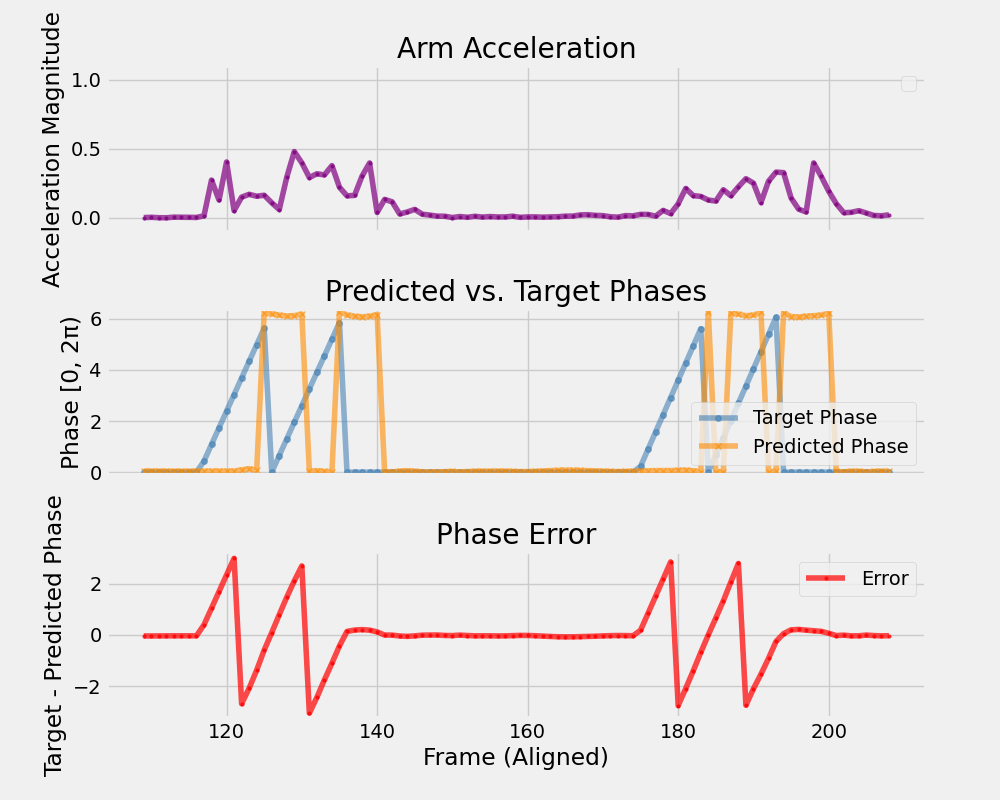}
        }
    \subfloat[\label{fig:fermataperformanceSubject5d}\centering Regular bars, 20 Hz using $\mathbf{x}_t^\mathrm{2D}$]{%
        \includegraphics[width=0.48\linewidth]{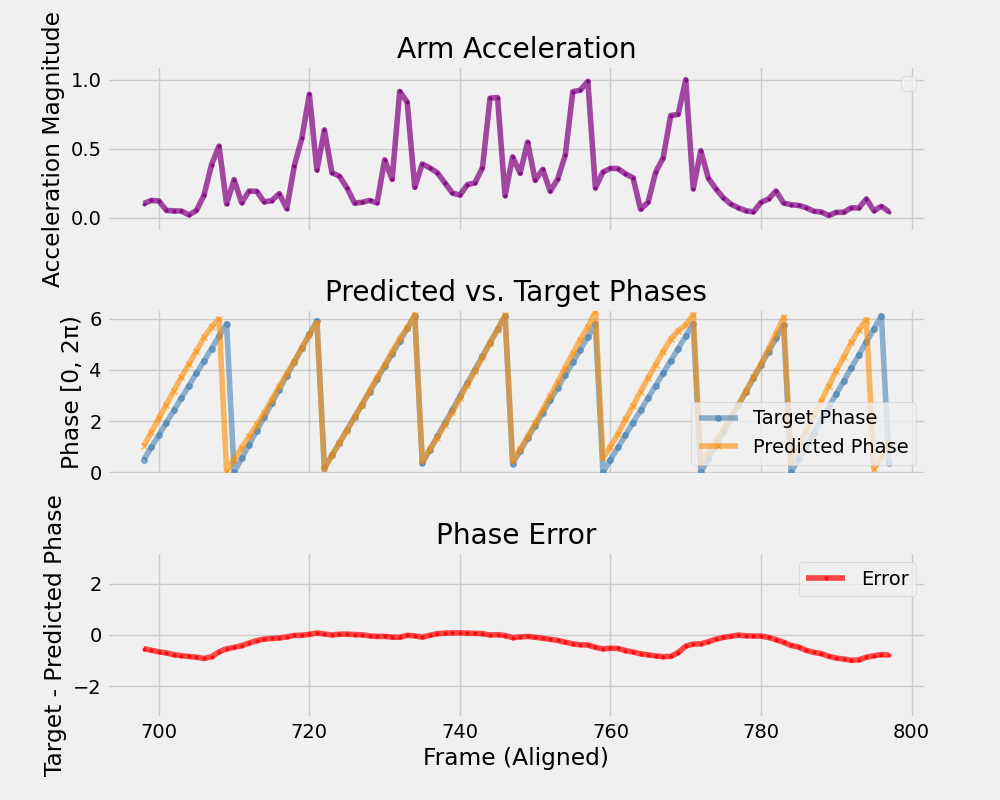}}
    
    \caption{Example of phase estimation for Subject 5. (a) and (b) correspond to column 1 in Table \ref{tab:cv_metrics_by_test_detail}, (c) and (d) correspond to column 4 in Table \ref{tab:cv_metrics_by_test_detail}. Top: Arm acceleration $\rho_t$. Middle: Ground truth phase $\varphi^\mathrm{gt}_t$ and estimated phase $\hat{\varphi}_t$. Bottom: Phase error $\varphi^\mathrm{gt}_t - \hat{\varphi}_t$. The three graphs are temporally aligned.}
    \label{fig:fermataperformanceSubject5}
\end{figure*}
\begin{figure*}[t]
    \centering
    \subfloat[\label{fig:fermataperformanceSubject6a}\centering Fermata bars, 60 Hz using $\mathbf{x}_t$]{%
        \includegraphics[width=0.48\linewidth]{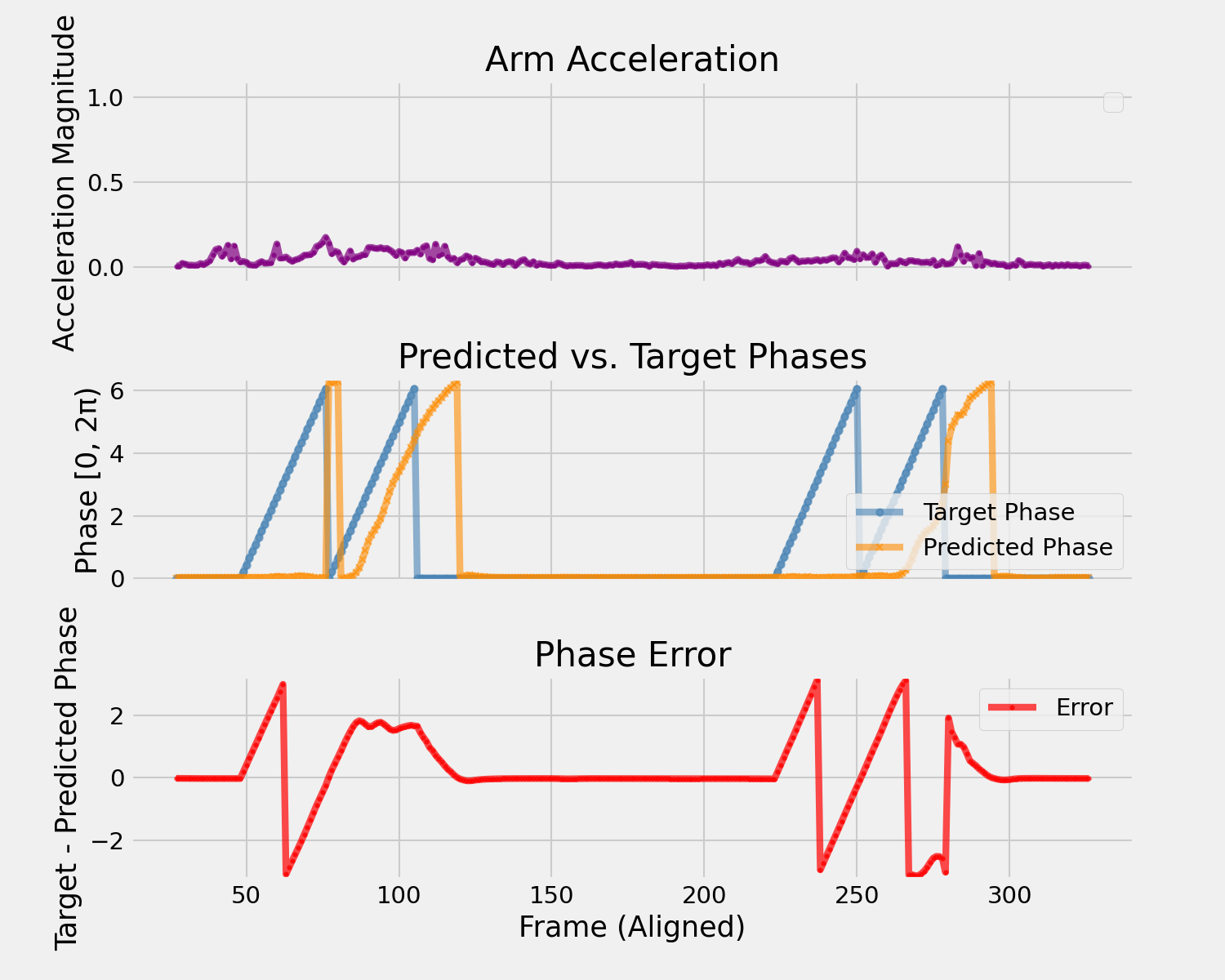}
    } 
    \subfloat[\label{fig:fermataperformanceSubject6b}\centering Regular bars, 60 Hz using $\mathbf{x}_t$]{%
        \includegraphics[width=0.48\linewidth]{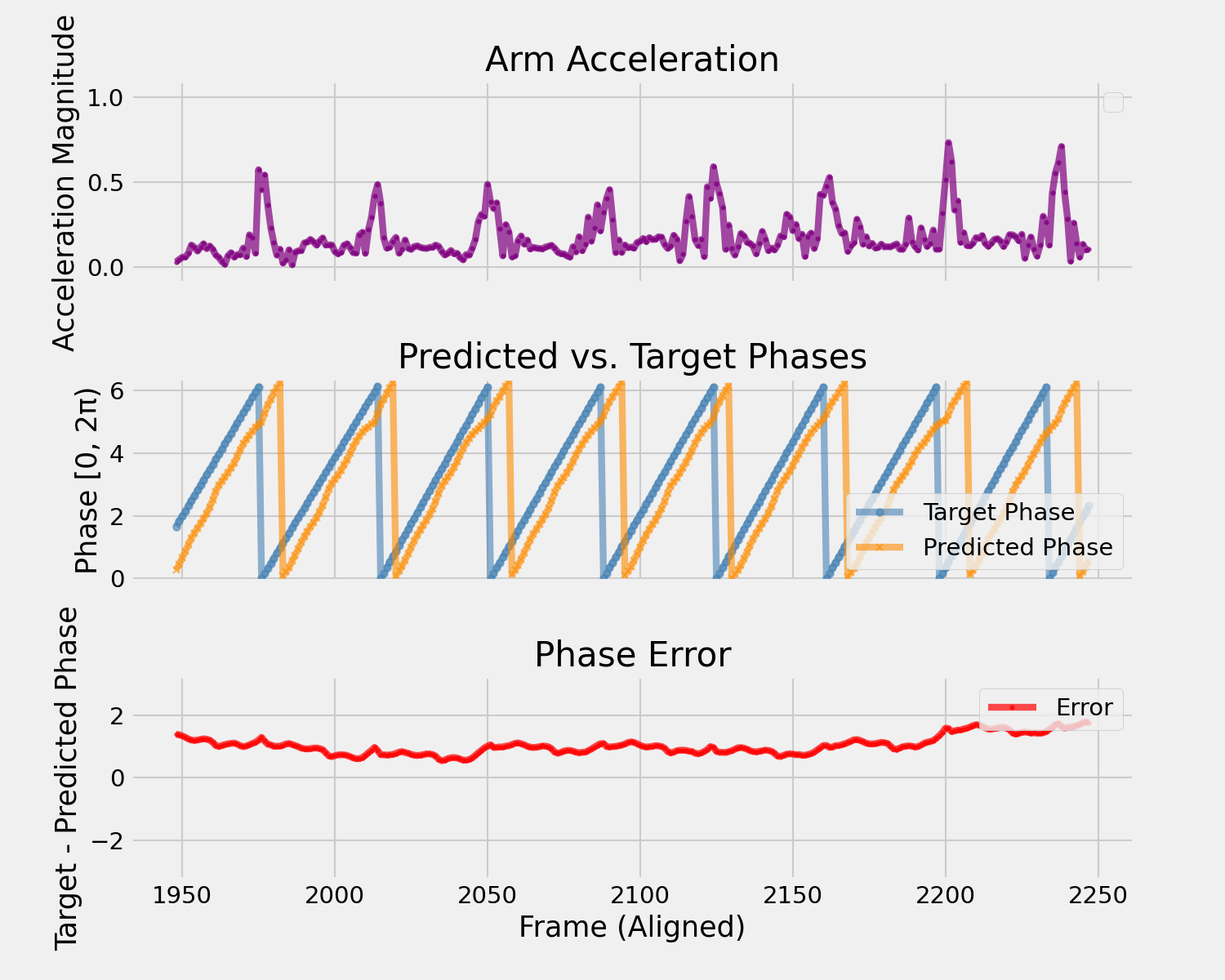}
    } \\ 

    \subfloat[\label{fig:fermataperformanceSubject6c}\centering Fermata bars, 20 Hz using $\mathbf{x}_t^\mathrm{2D}$]{%
        \includegraphics[width=0.48\linewidth]{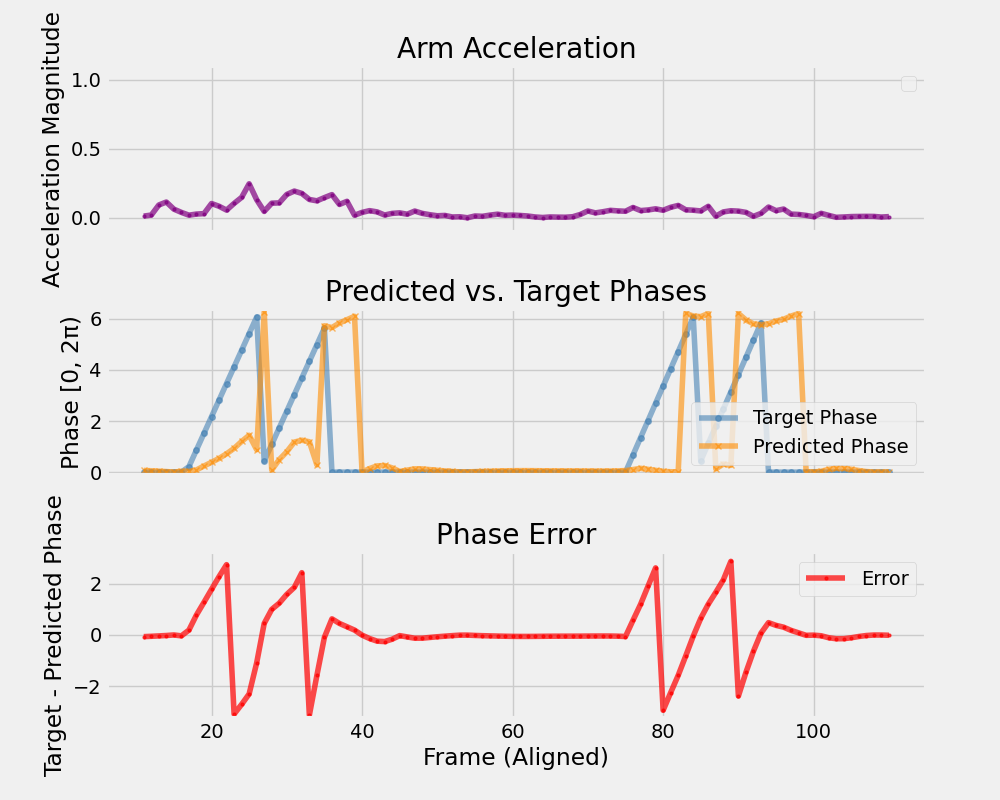}
        }
    \subfloat[\label{fig:fermataperformanceSubject6d}\centering Regular bars, 20 Hz using $\mathbf{x}_t^\mathrm{2D}$]{%
        \includegraphics[width=0.48\linewidth]{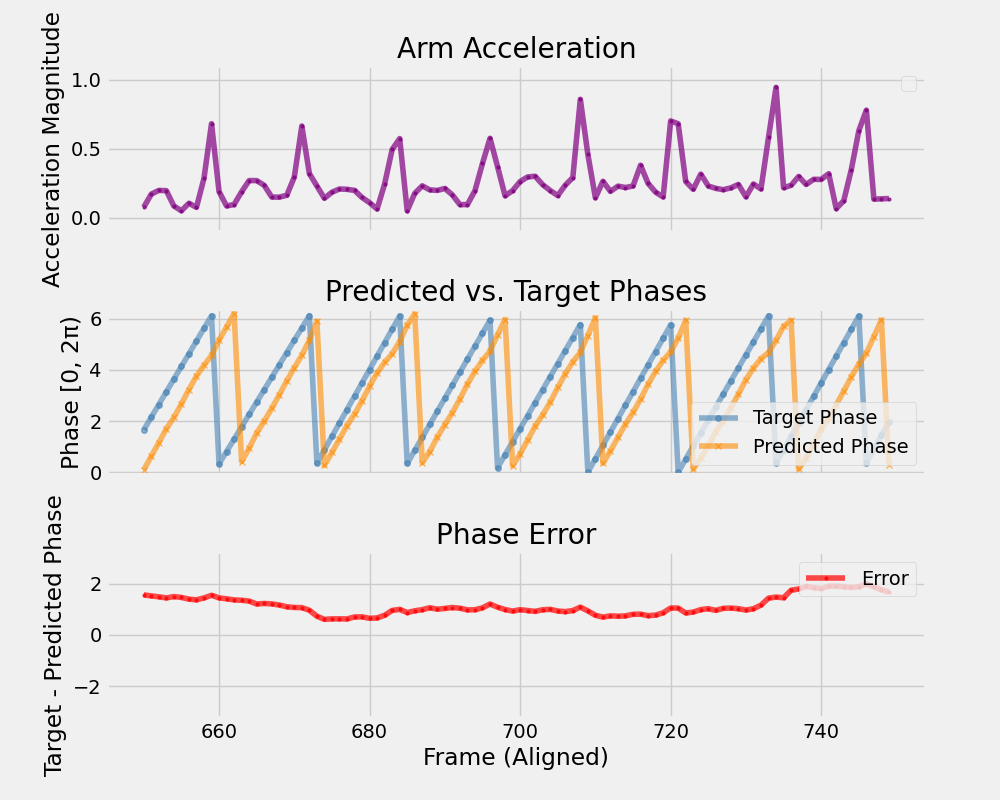}}
    
    \caption{Example of phase estimation for Subject 6. (a) and (b) correspond to column 1 in Table \ref{tab:cv_metrics_by_test_detail}, (c) and (d) correspond to column 4 in Table \ref{tab:cv_metrics_by_test_detail}. Top: Arm acceleration $\rho_t$. Middle: Ground truth phase $\varphi^\mathrm{gt}_t$ and estimated phase $\hat{\varphi}_t$. Bottom: Phase error $\varphi^\mathrm{gt}_t - \hat{\varphi}_t$. The three graphs are temporally aligned.}
    \label{fig:fermataperformanceSubject6}
\end{figure*}
\begin{figure*}[t]
    \centering
    \subfloat[\label{fig:fermataperformanceSubject7a}\centering Fermata bars, 60 Hz using $\mathbf{x}_t$]{%
        \includegraphics[width=0.48\linewidth]{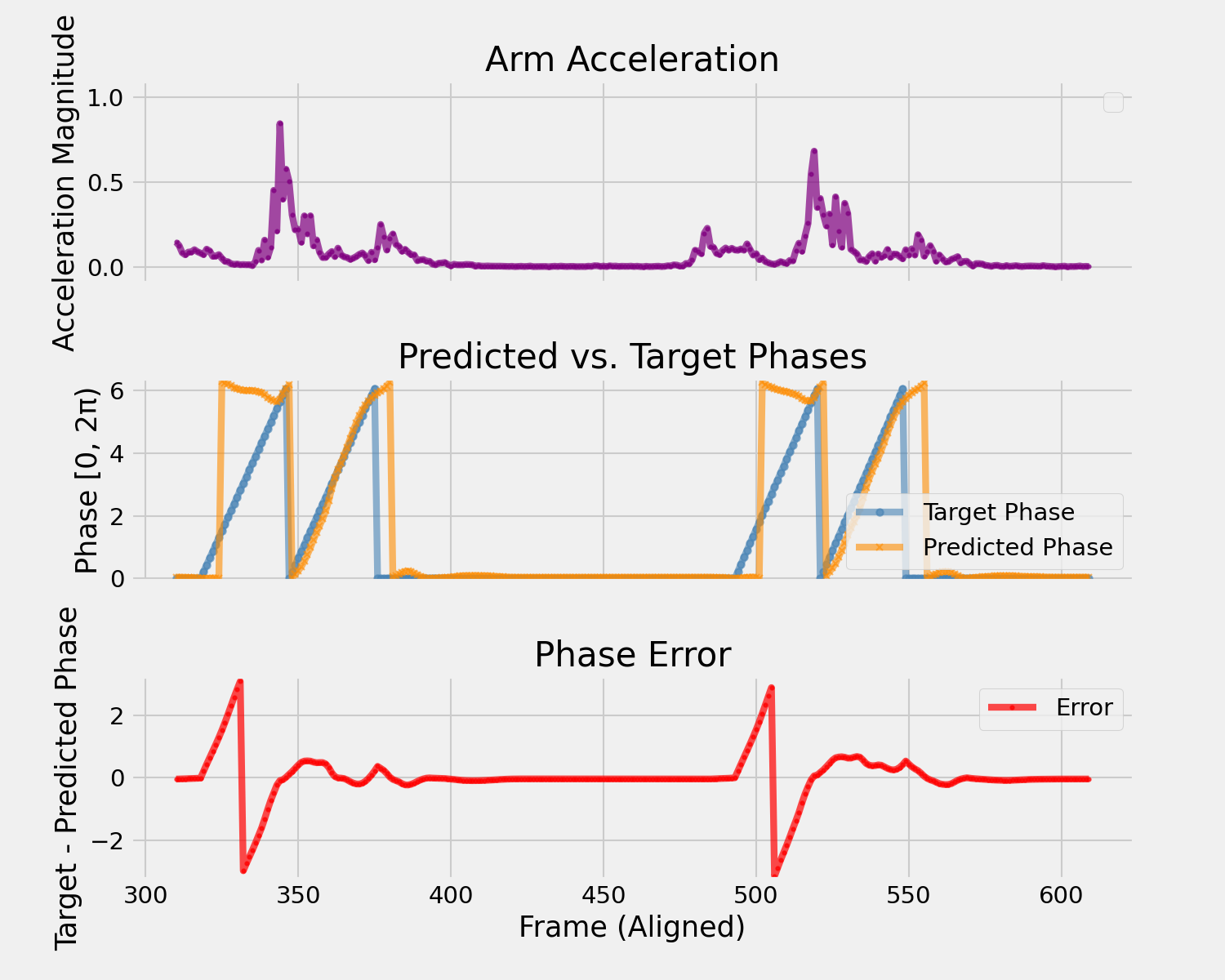}
    } 
    \subfloat[\label{fig:fermataperformanceSubject7b}\centering Regular bars, 60 Hz using $\mathbf{x}_t$]{%
        \includegraphics[width=0.48\linewidth]{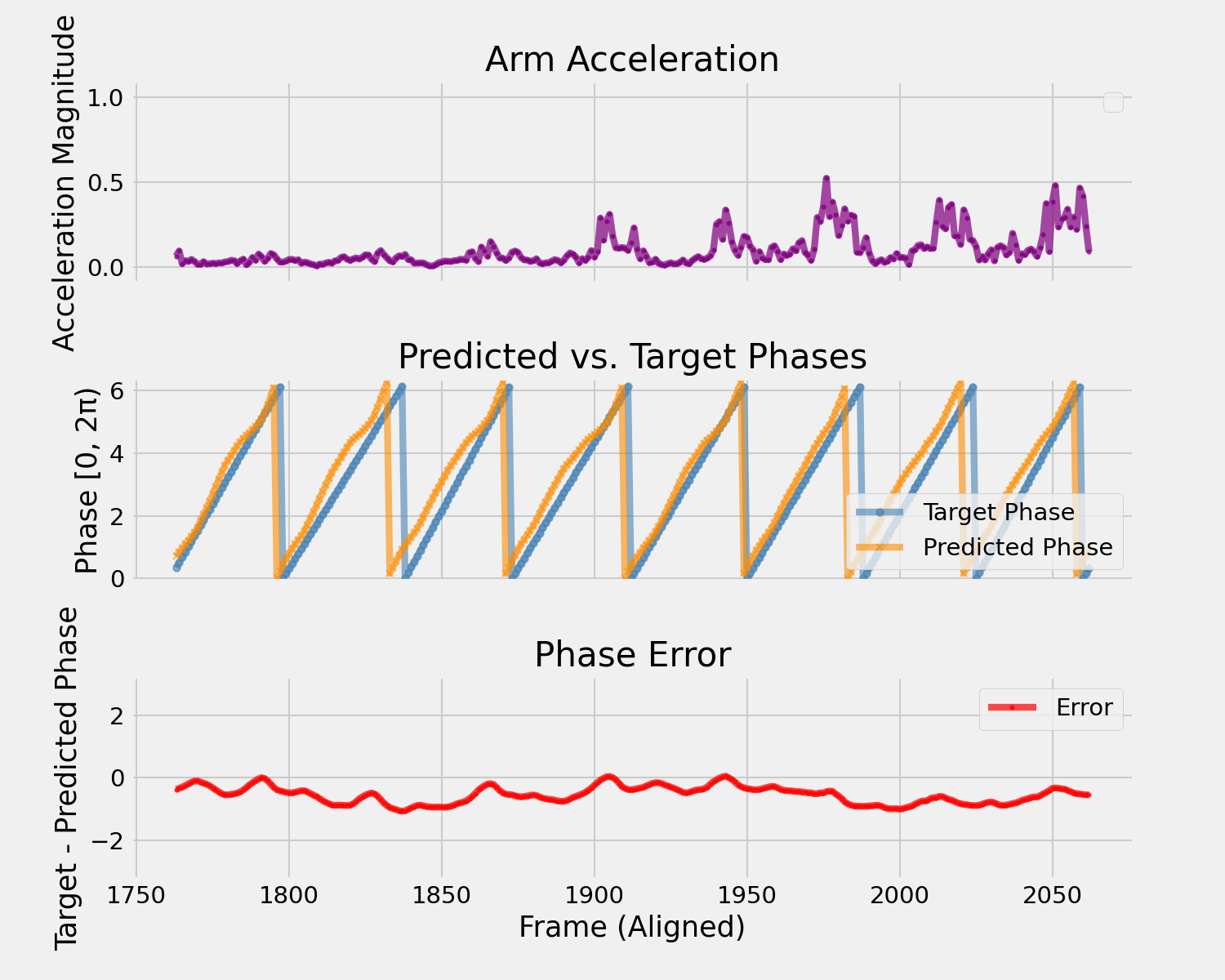}
    } \\ 

    \subfloat[\label{fig:fermataperformanceSubject7c}\centering Fermata bars, 20 Hz using $\mathbf{x}_t^\mathrm{2D}$]{%
        \includegraphics[width=0.48\linewidth]{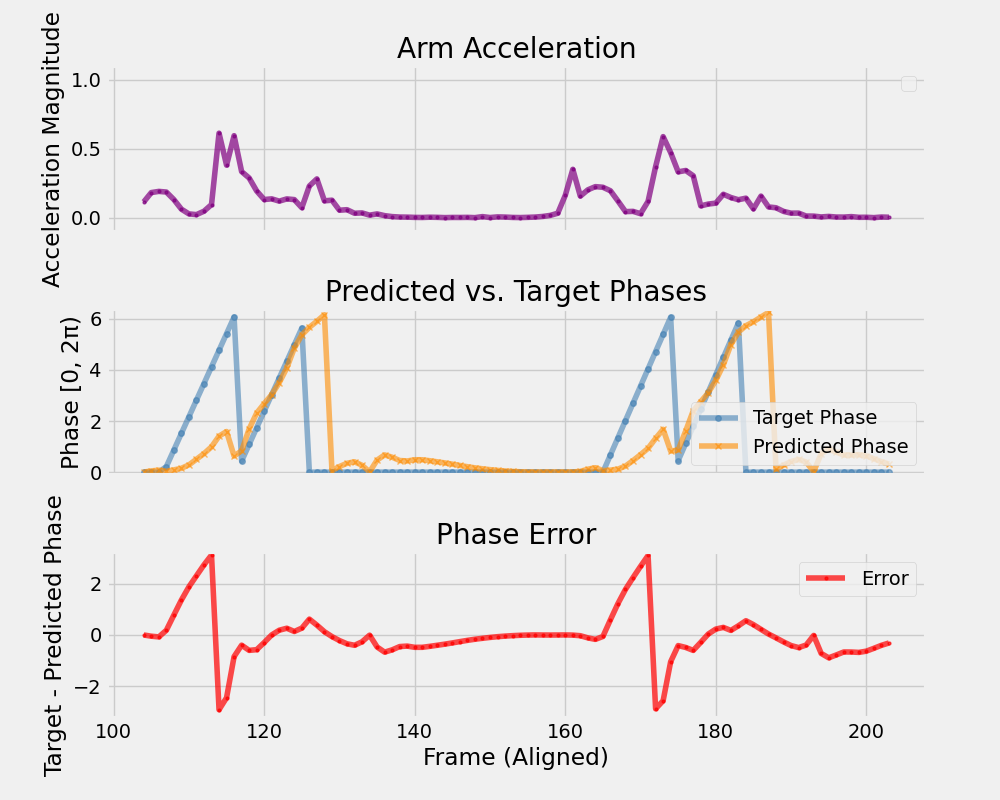}
        }
    \subfloat[\label{fig:fermataperformanceSubject7d}\centering Regular bars, 20 Hz using $\mathbf{x}_t^\mathrm{2D}$]{%
        \includegraphics[width=0.48\linewidth]{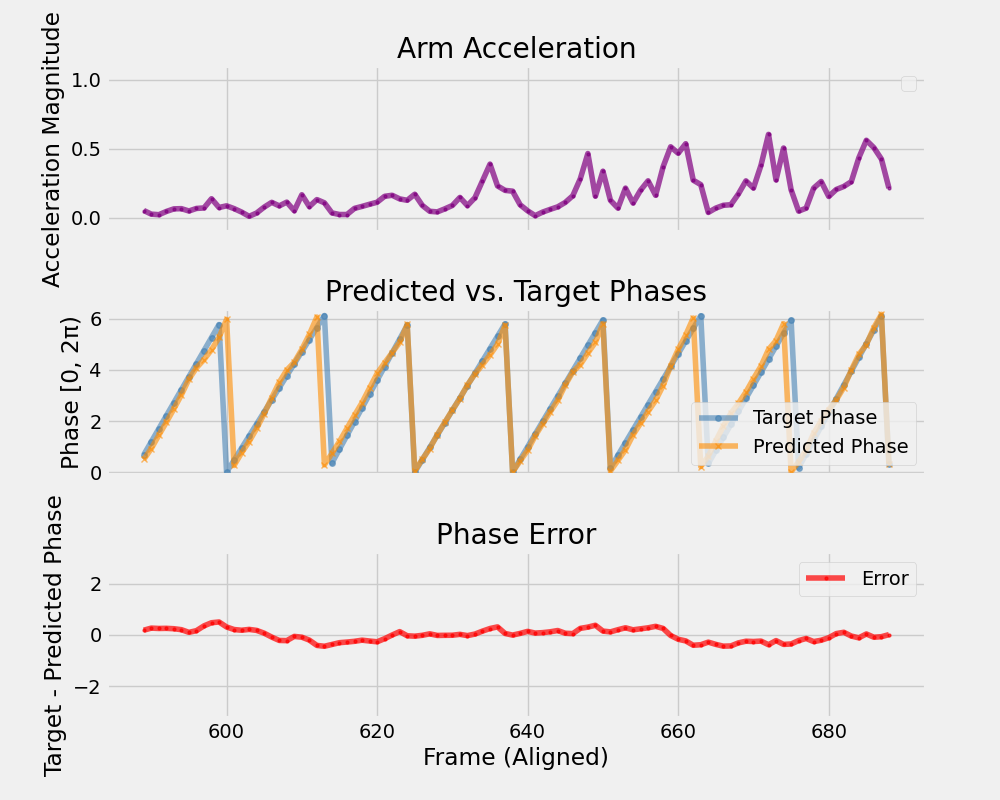}}
    
    \caption{Example of phase estimation for Subject 7. (a) and (b) correspond to column 1 in Table \ref{tab:cv_metrics_by_test_detail}, (c) and (d) correspond to column 4 in Table \ref{tab:cv_metrics_by_test_detail}. Top: Arm acceleration $\rho_t$. Middle: Ground truth phase $\varphi^\mathrm{gt}_t$ and estimated phase $\hat{\varphi}_t$. Bottom: Phase error $\varphi^\mathrm{gt}_t - \hat{\varphi}_t$. The three graphs are temporally aligned.}
    \label{fig:fermataperformanceSubject7}
\end{figure*}
\begin{figure*}[t]
    \centering
    \subfloat[\label{fig:fermataperformanceSubject8a}\centering Fermata bars, 60 Hz using $\mathbf{x}_t$]{%
        \includegraphics[width=0.48\linewidth]{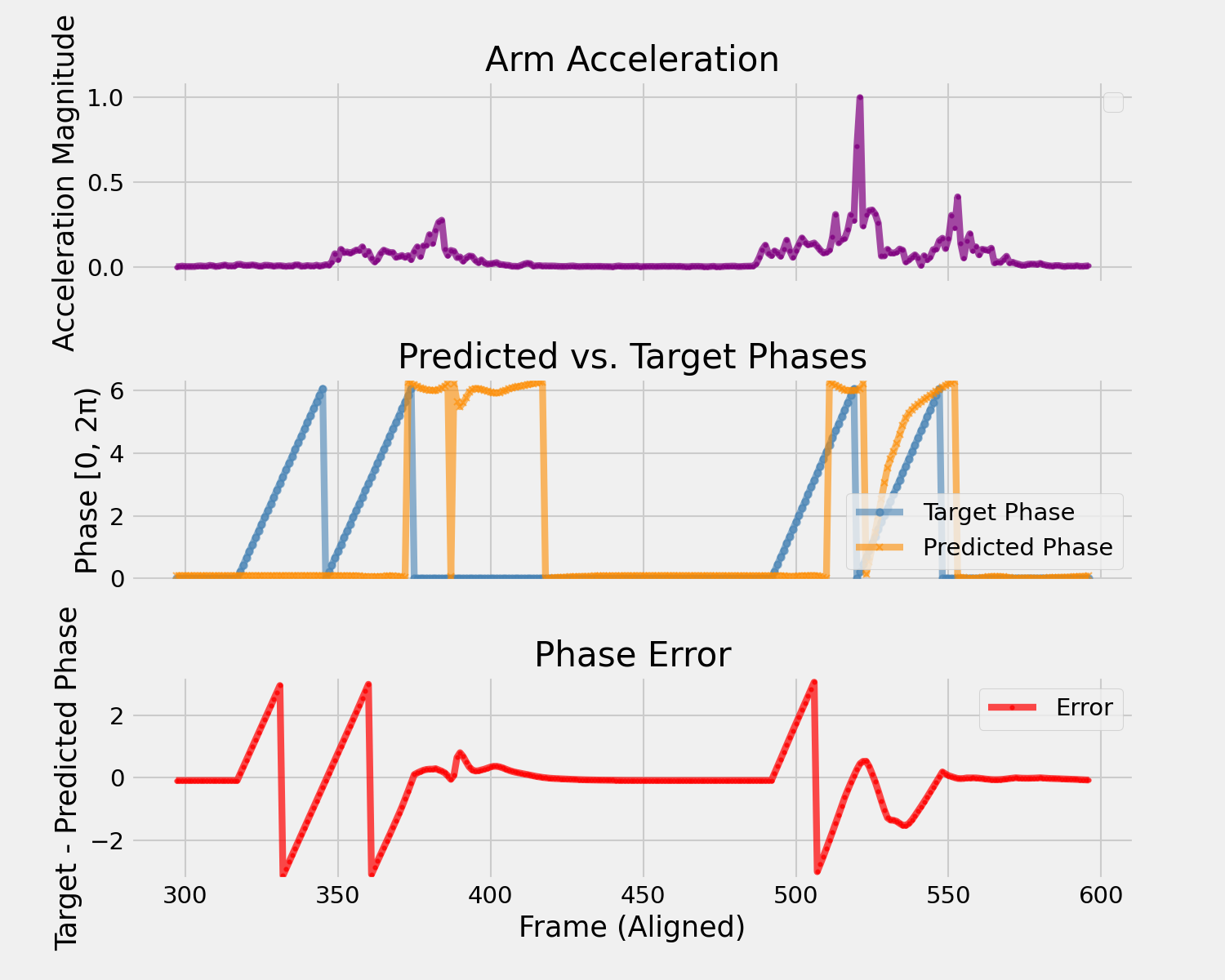}
    } 
    \subfloat[\label{fig:fermataperformanceSubject8b}\centering Regular bars, 60 Hz using $\mathbf{x}_t$]{%
        \includegraphics[width=0.48\linewidth]{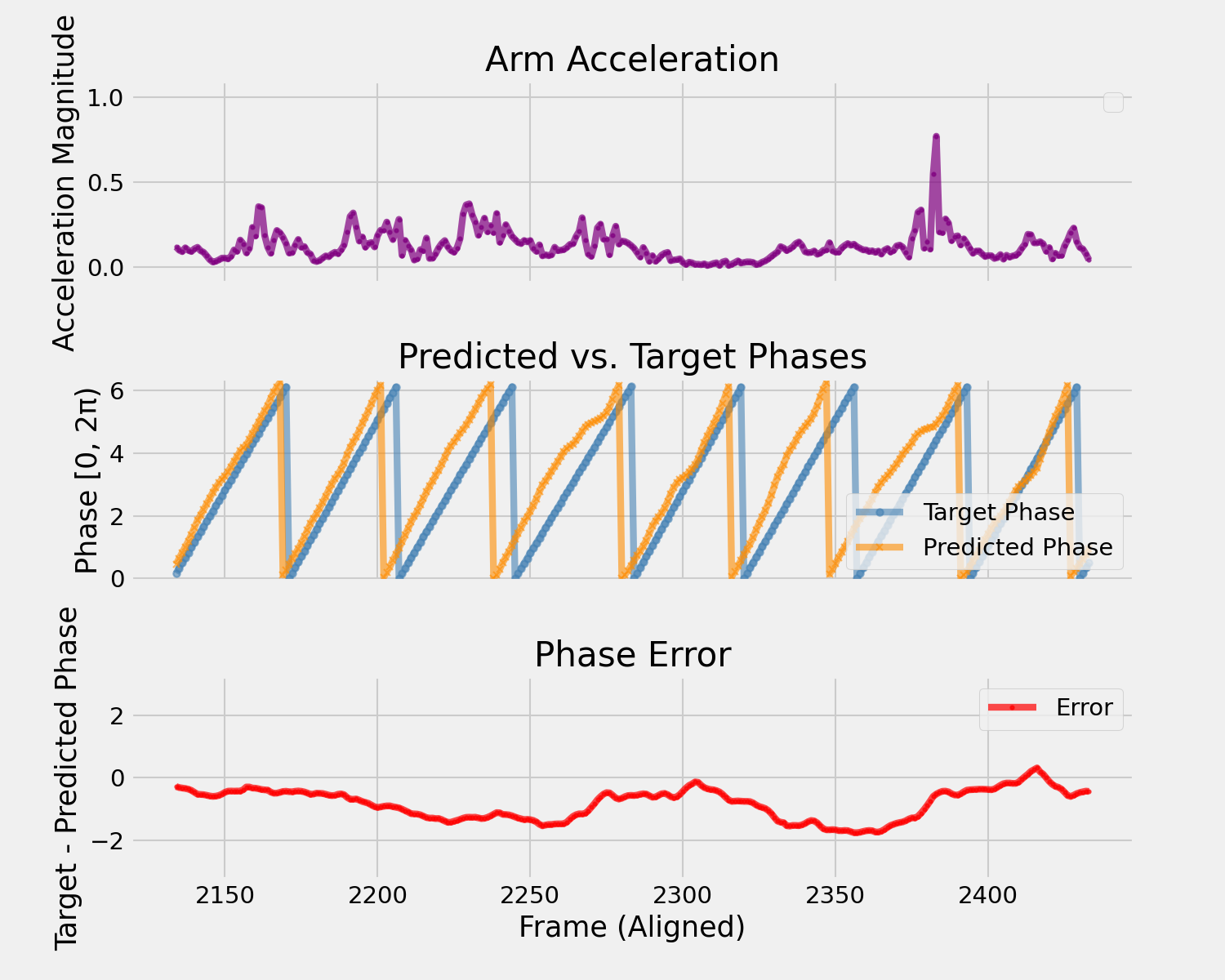}
    } \\ 

    \subfloat[\label{fig:fermataperformanceSubject8c}\centering Fermata bars, 20 Hz using $\mathbf{x}_t^\mathrm{2D}$]{%
        \includegraphics[width=0.48\linewidth]{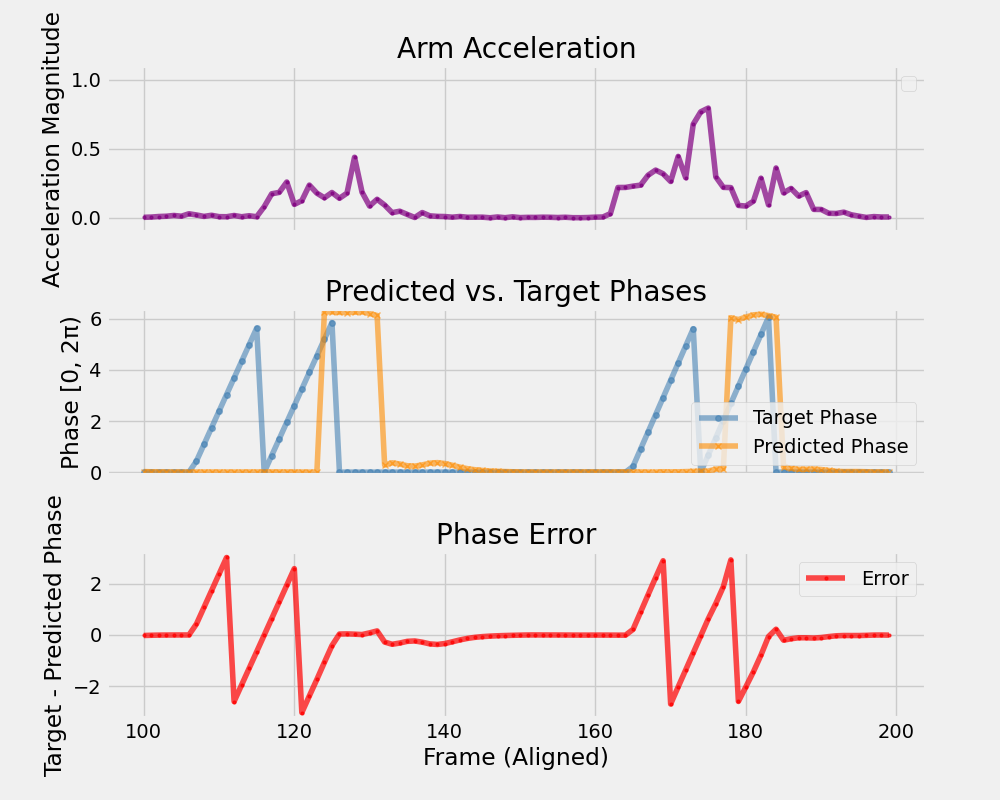}
        }
    \subfloat[\label{fig:fermataperformanceSubject8d}\centering Regular bars, 20 Hz using $\mathbf{x}_t^\mathrm{2D}$]{%
        \includegraphics[width=0.48\linewidth]{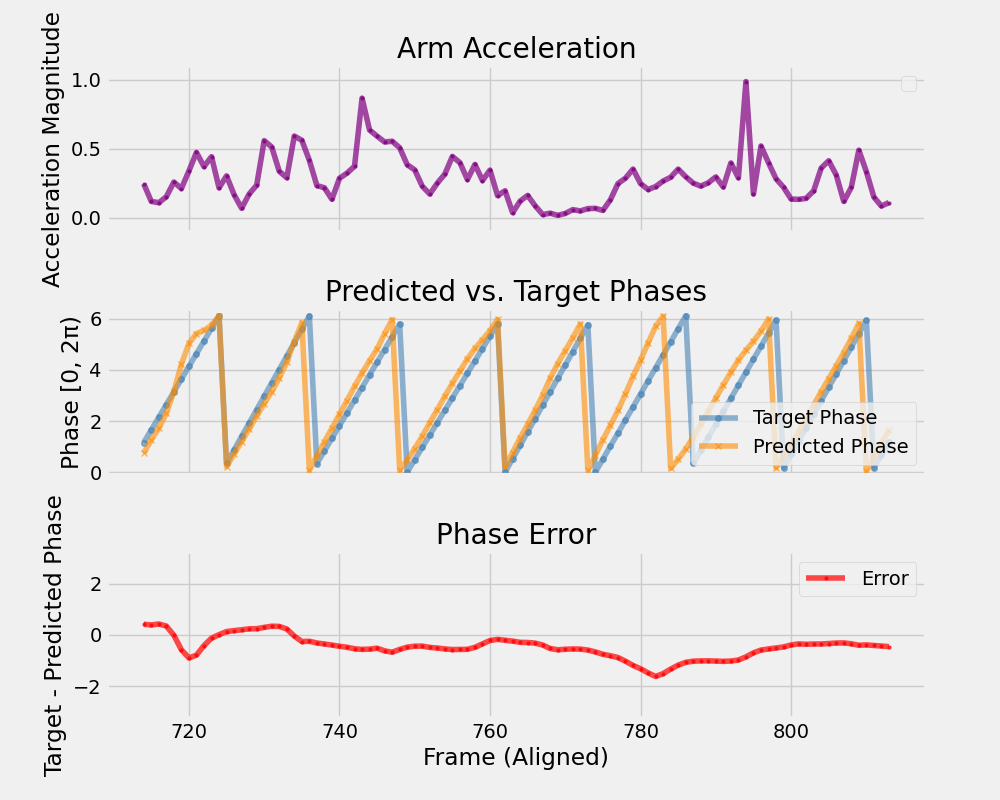}}
    
    \caption{Example of phase estimation for Subject 8. Top: Arm acceleration $\rho_t$. (a) and (b) correspond to column 1 in Table \ref{tab:cv_metrics_by_test_detail}, (c) and (d) correspond to column 4 in Table \ref{tab:cv_metrics_by_test_detail}. Middle: Ground truth phase $\varphi^\mathrm{gt}_t$ and estimated phase $\hat{\varphi}_t$. Bottom: Phase error $\varphi^\mathrm{gt}_t - \hat{\varphi}_t$. The three graphs are temporally aligned.}
    \label{fig:fermataperformanceSubject8}
\end{figure*}
\begin{figure*}[t]
    \centering
    \subfloat[\label{fig:fermataperformanceSubject9a}\centering Fermata bars, 60 Hz using $\mathbf{x}_t$]{%
        \includegraphics[width=0.48\linewidth]{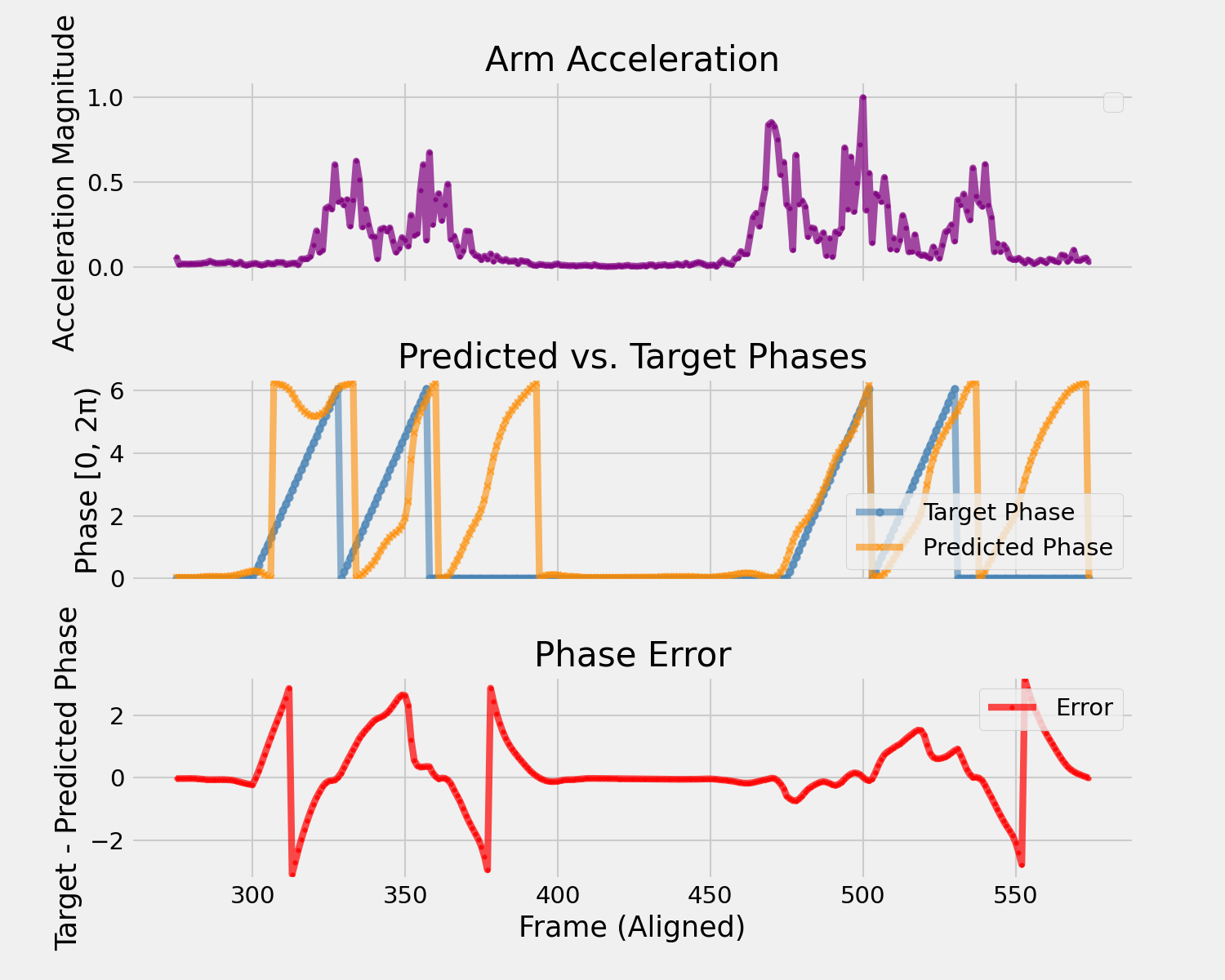}
    } 
    \subfloat[\label{fig:fermataperformanceSubject9b}\centering Regular bars, 60 Hz using $\mathbf{x}_t$]{%
        \includegraphics[width=0.48\linewidth]{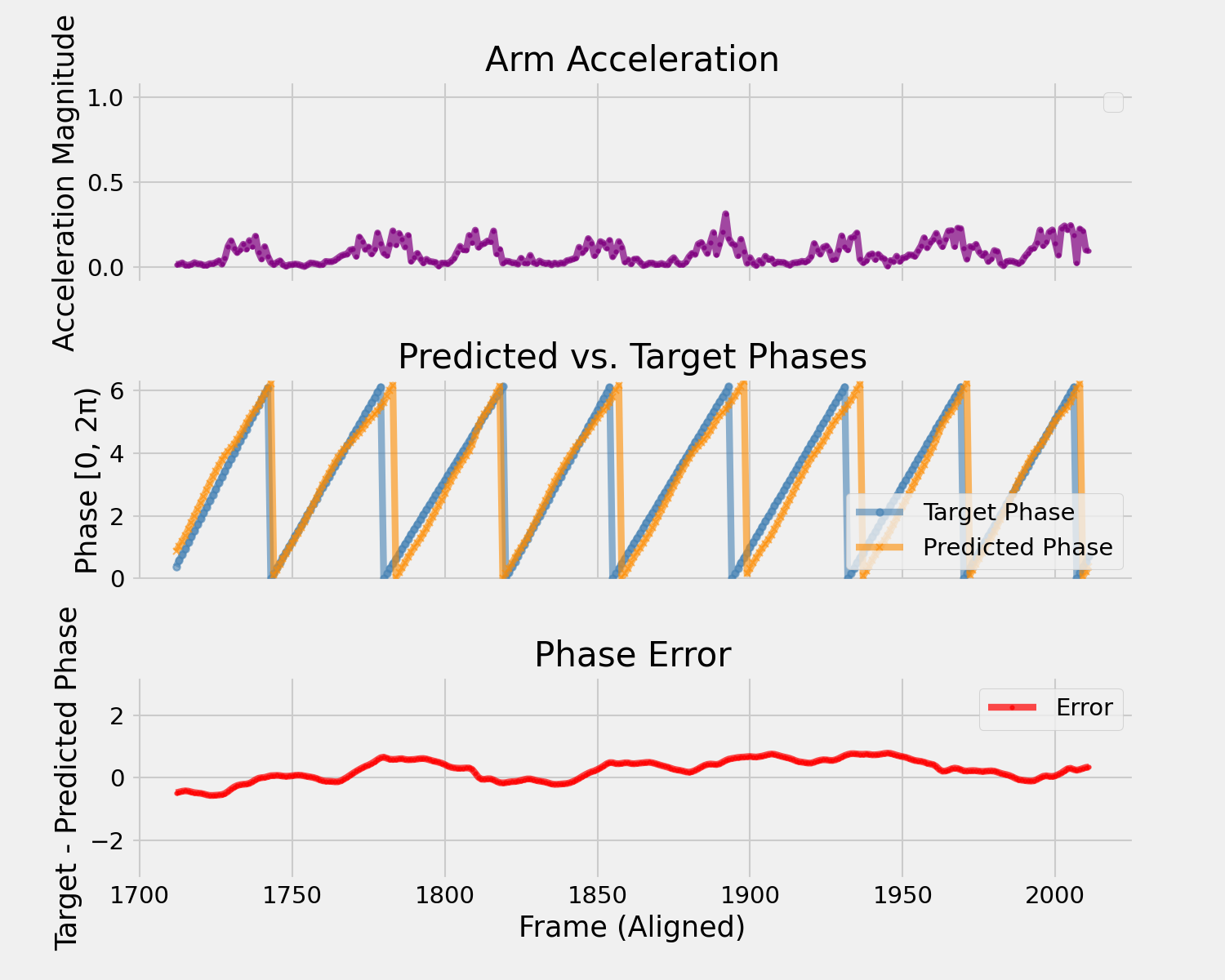}
    } \\ 

    \subfloat[\label{fig:fermataperformanceSubject9c}\centering Fermata bars, 20 Hz using $\mathbf{x}_t^\mathrm{2D}$]{%
        \includegraphics[width=0.48\linewidth]{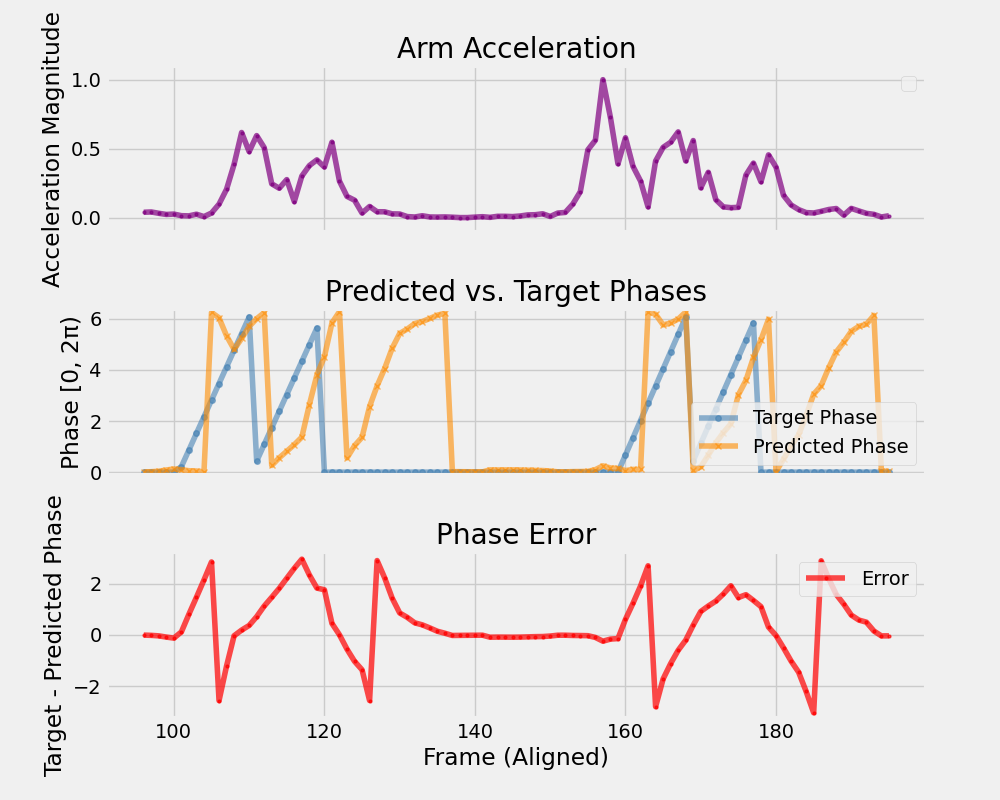}
    }
    \subfloat[\label{fig:fermataperformanceSubject9d}\centering Regular bars, 20 Hz using $\mathbf{x}_t^\mathrm{2D}$]{%
        \includegraphics[width=0.48\linewidth]{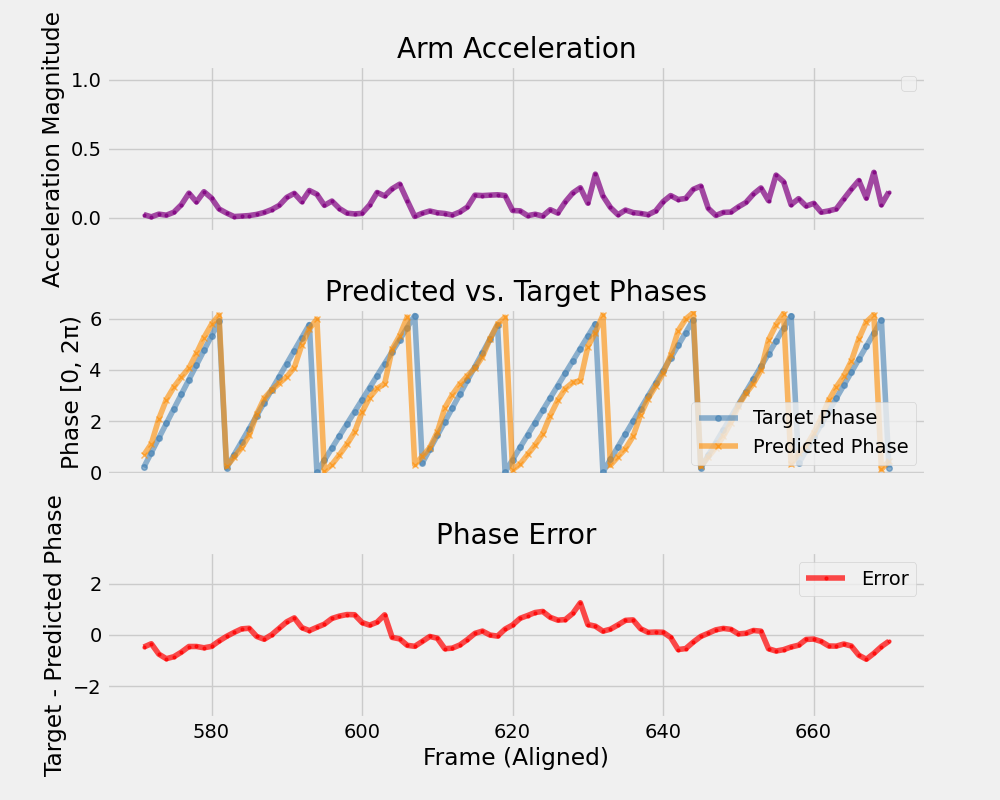}
    }
    
    \caption{Example of phase estimation for Subject 9. (a) and (b) correspond to column 1 in Table \ref{tab:cv_metrics_by_test_detail}, (c) and (d) correspond to column 4 in Table \ref{tab:cv_metrics_by_test_detail}. Top: Arm acceleration $\rho_t$. Middle: Ground truth phase $\varphi^\mathrm{gt}_t$ and estimated phase $\hat{\varphi}_t$. Bottom: Phase error $\varphi^\mathrm{gt}_t - \hat{\varphi}_t$. The three graphs are temporally aligned.}
    \label{fig:fermataperformanceSubject9}
\end{figure*}
\begin{figure*}[t]
    \centering
    \subfloat[\label{fig:fermataperformanceSubject10a}\centering Fermata bars, 60 Hz using $\mathbf{x}_t$]{%
        \includegraphics[width=0.48\linewidth]{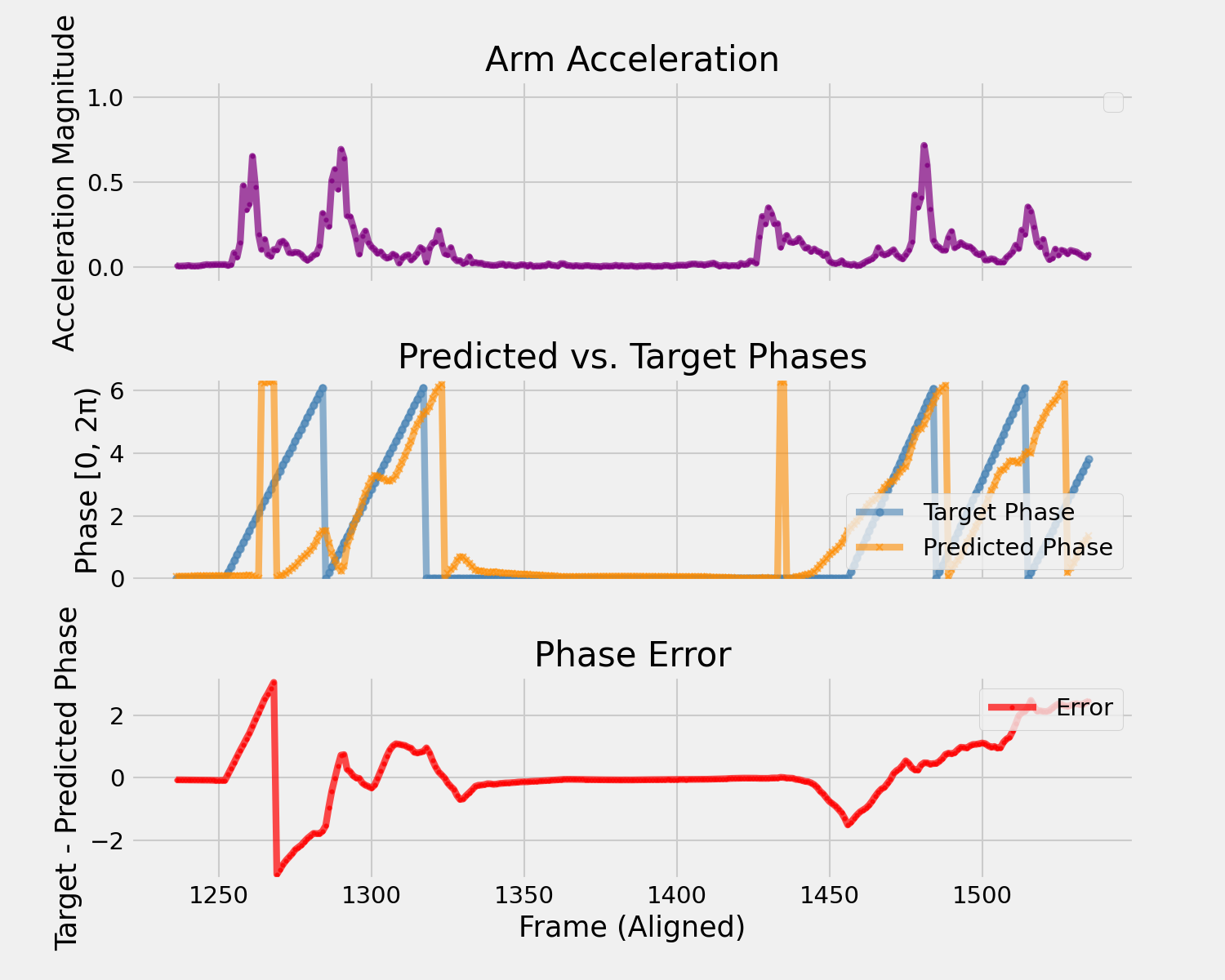}
    } 
    \subfloat[\label{fig:fermataperformanceSubject10b}\centering Regular bars, 60 Hz using $\mathbf{x}_t$]{%
        \includegraphics[width=0.48\linewidth]{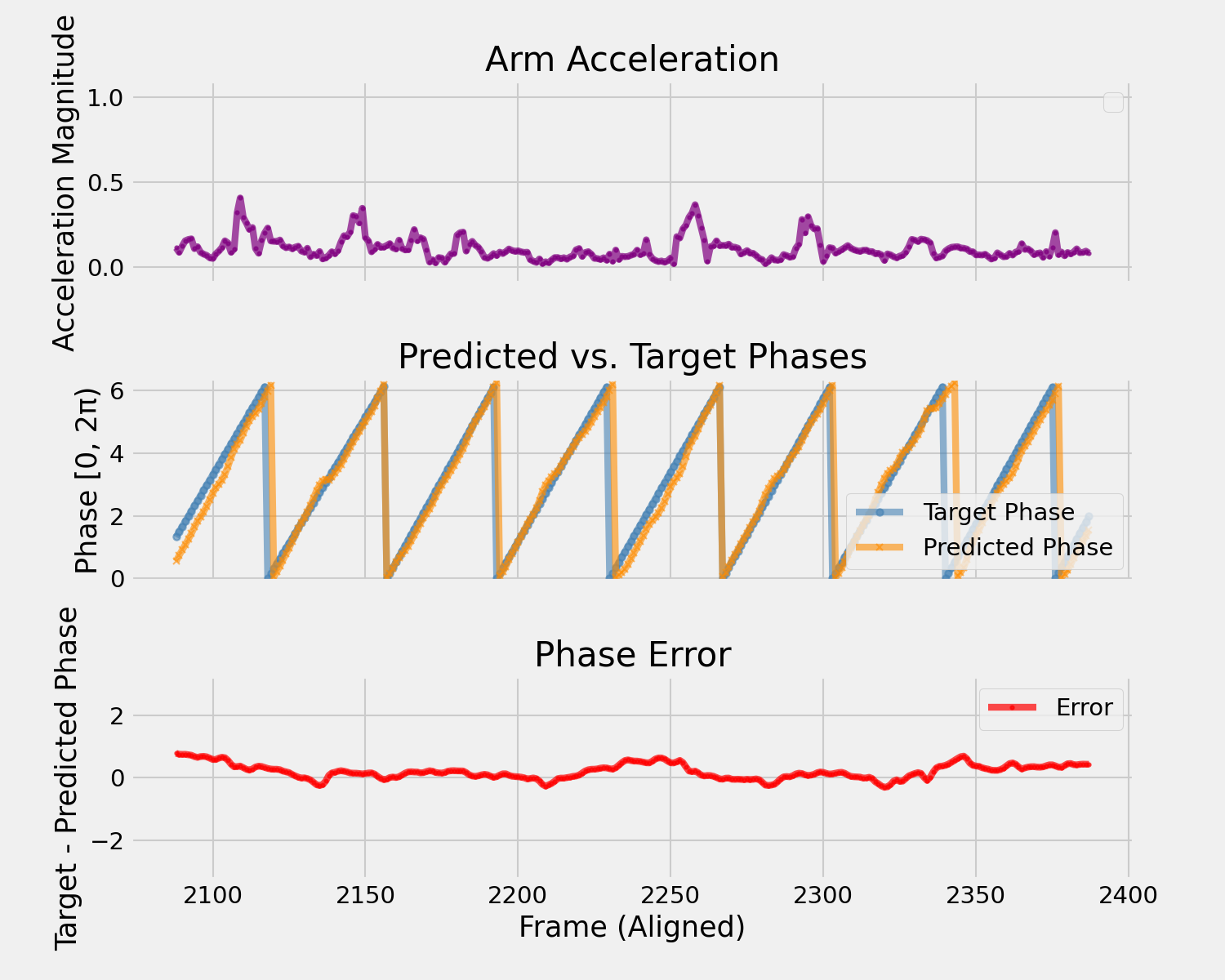}
    } \\ 

    \subfloat[\label{fig:fermataperformanceSubject10c}\centering Fermata bars, 20 Hz using $\mathbf{x}_t^\mathrm{2D}$]{%
        \includegraphics[width=0.48\linewidth]{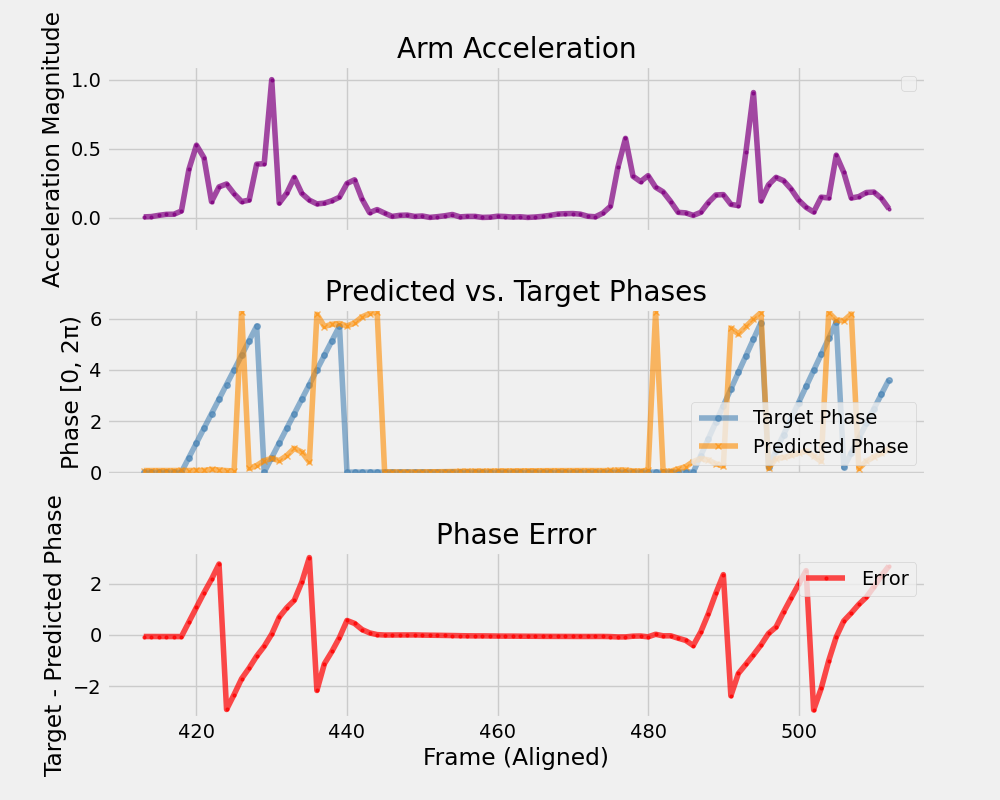}
    }
    \subfloat[\label{fig:fermataperformanceSubject10d}\centering Regular bars, 20 Hz using $\mathbf{x}_t^\mathrm{2D}$]{%
        \includegraphics[width=0.48\linewidth]{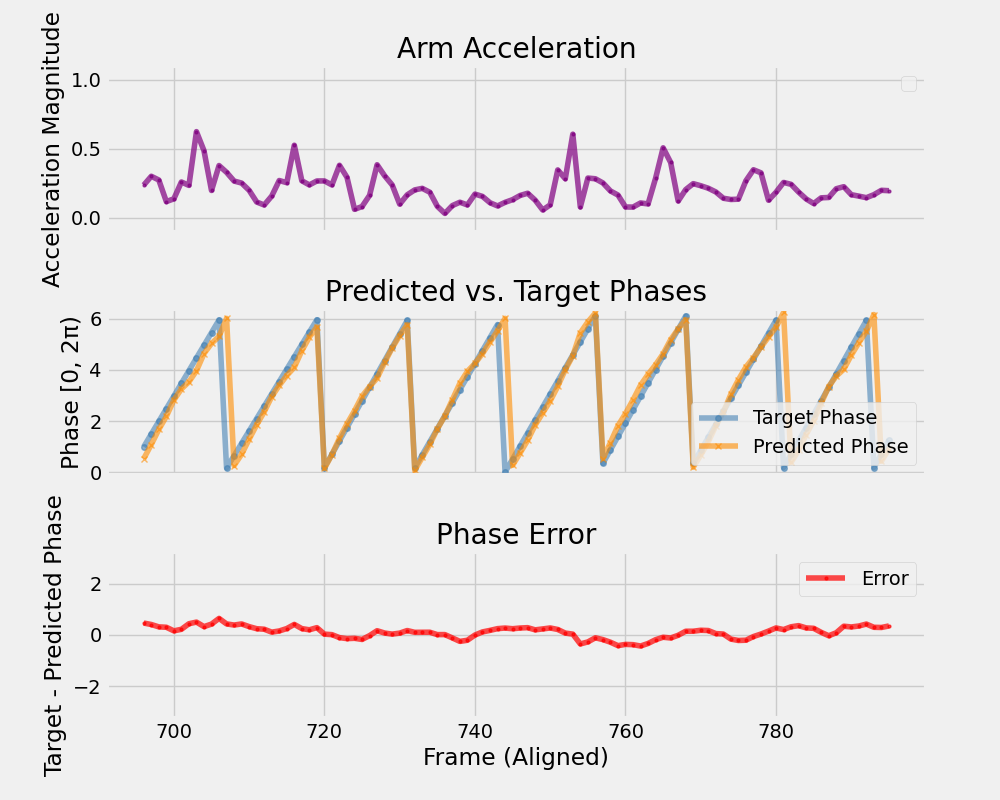}
    }
    
    \caption{Example of phase estimation for Subject 10. (a) and (b) correspond to column 1 in Table \ref{tab:cv_metrics_by_test_detail}, (c) and (d) correspond to column 4 in Table \ref{tab:cv_metrics_by_test_detail}. Top: Arm acceleration $\rho_t$. Middle: Ground truth phase $\varphi^\mathrm{gt}_t$ and estimated phase $\hat{\varphi}_t$. Bottom: Phase error $\varphi^\mathrm{gt}_t - \hat{\varphi}_t$. The three graphs are temporally aligned.}
    \label{fig:fermataperformanceSubject10}
\end{figure*}
\begin{figure*}[t]
    \centering
    \subfloat[\label{fig:fermataperformanceSubject11a}\centering Fermata bars, 60 Hz using $\mathbf{x}_t$]{%
        \includegraphics[width=0.48\linewidth]{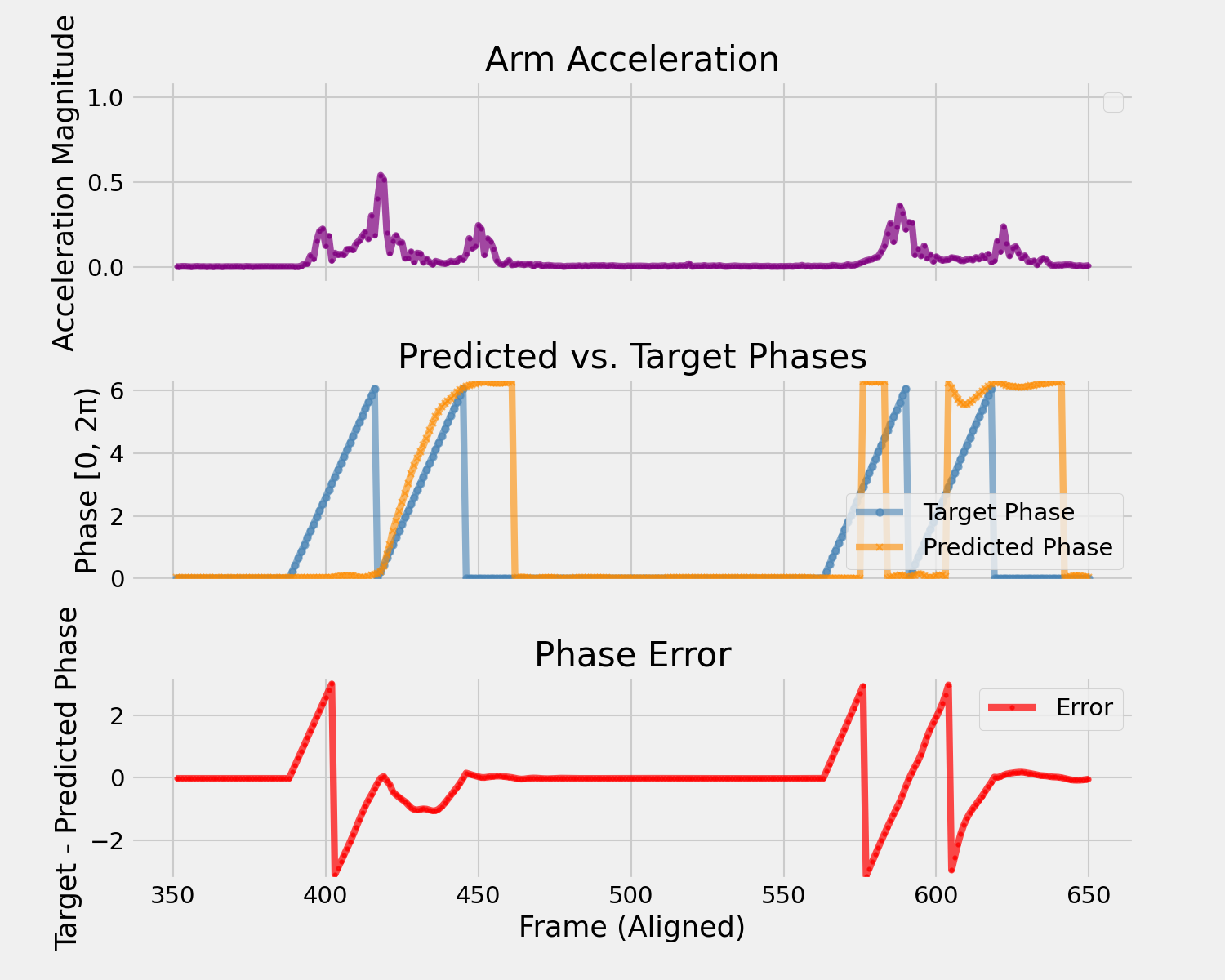}
    } 
    \subfloat[\label{fig:fermataperformanceSubject11b}\centering Regular bars, 60 Hz using $\mathbf{x}_t$]{%
        \includegraphics[width=0.48\linewidth]{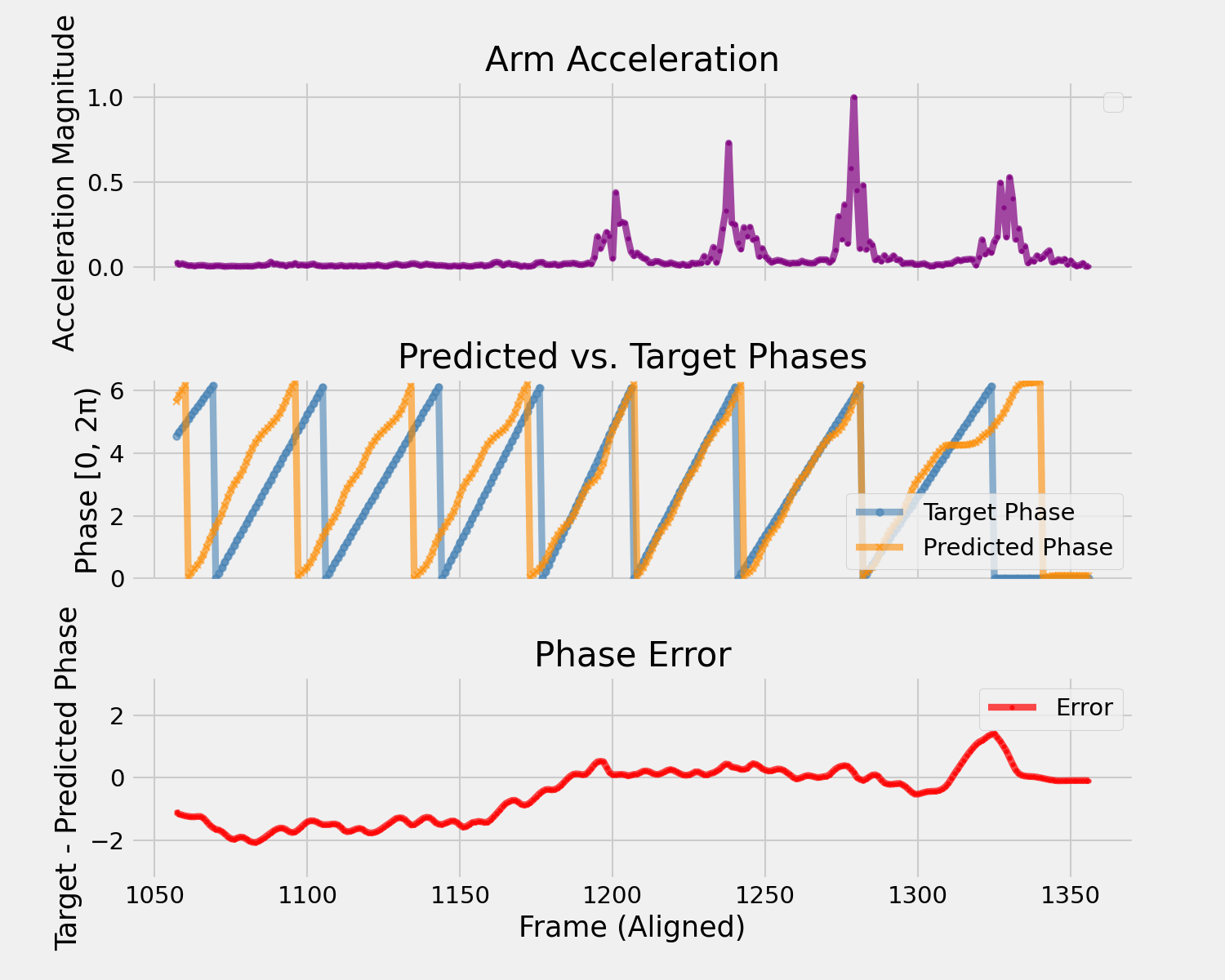}
    } \\ 

    \subfloat[\label{fig:fermataperformanceSubject11c}\centering Fermata bars, 20 Hz using $\mathbf{x}_t^\mathrm{2D}$]{%
        \includegraphics[width=0.48\linewidth]{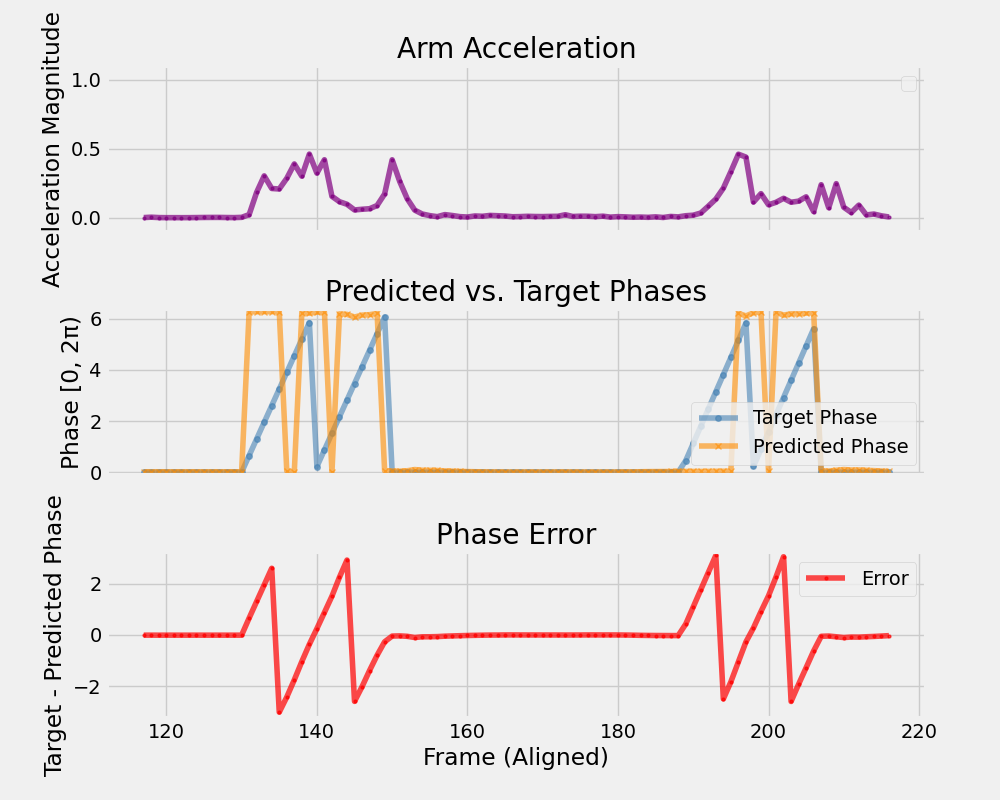}
    }
    \subfloat[\label{fig:fermataperformanceSubject11d}\centering Regular bars, 20 Hz using $\mathbf{x}_t^\mathrm{2D}$]{%
        \includegraphics[width=0.48\linewidth]{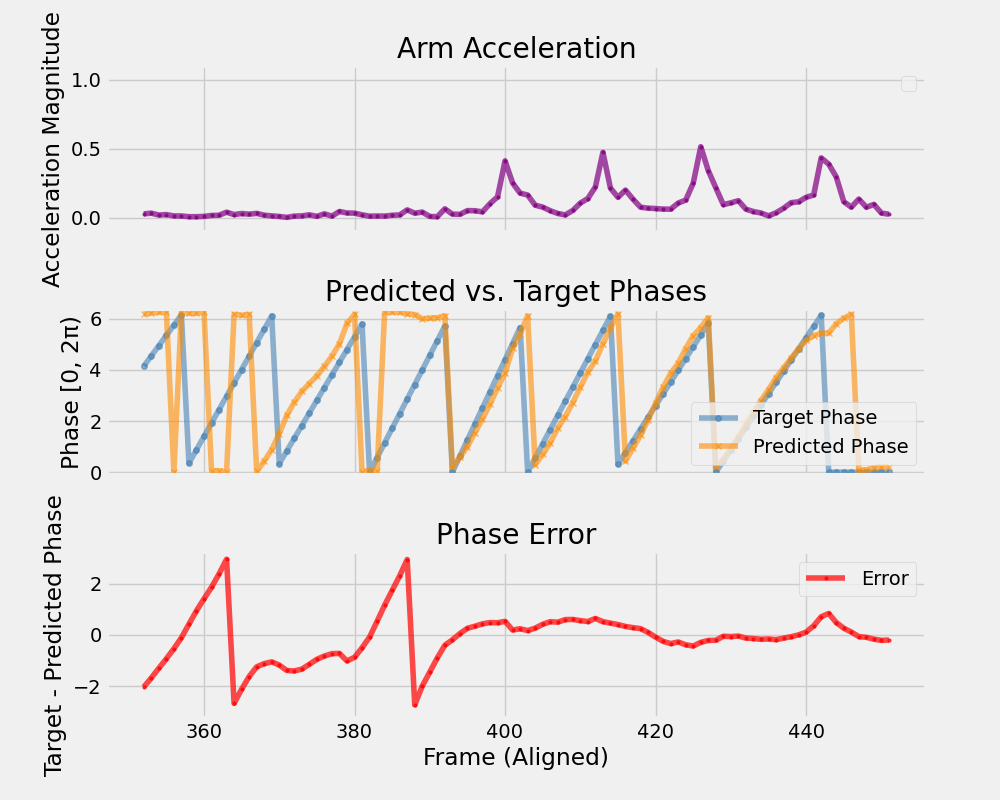}
    }
    
    \caption{Example of phase estimation for Subject 11. (a) and (b) correspond to column 1 in Table \ref{tab:cv_metrics_by_test_detail}, (c) and (d) correspond to column 4 in Table \ref{tab:cv_metrics_by_test_detail}. Top: Arm acceleration $\rho_t$. Middle: Ground truth phase $\varphi^\mathrm{gt}_t$ and estimated phase $\hat{\varphi}_t$. Bottom: Phase error $\varphi^\mathrm{gt}_t - \hat{\varphi}_t$. The three graphs are temporally aligned.}
    \label{fig:fermataperformanceSubject11}
\end{figure*}
\begin{figure*}[t]
    \centering
    \subfloat[\label{fig:fermataperformanceSubject12a}\centering Fermata bars, 60 Hz using $\mathbf{x}_t$]{%
        \includegraphics[width=0.48\linewidth]{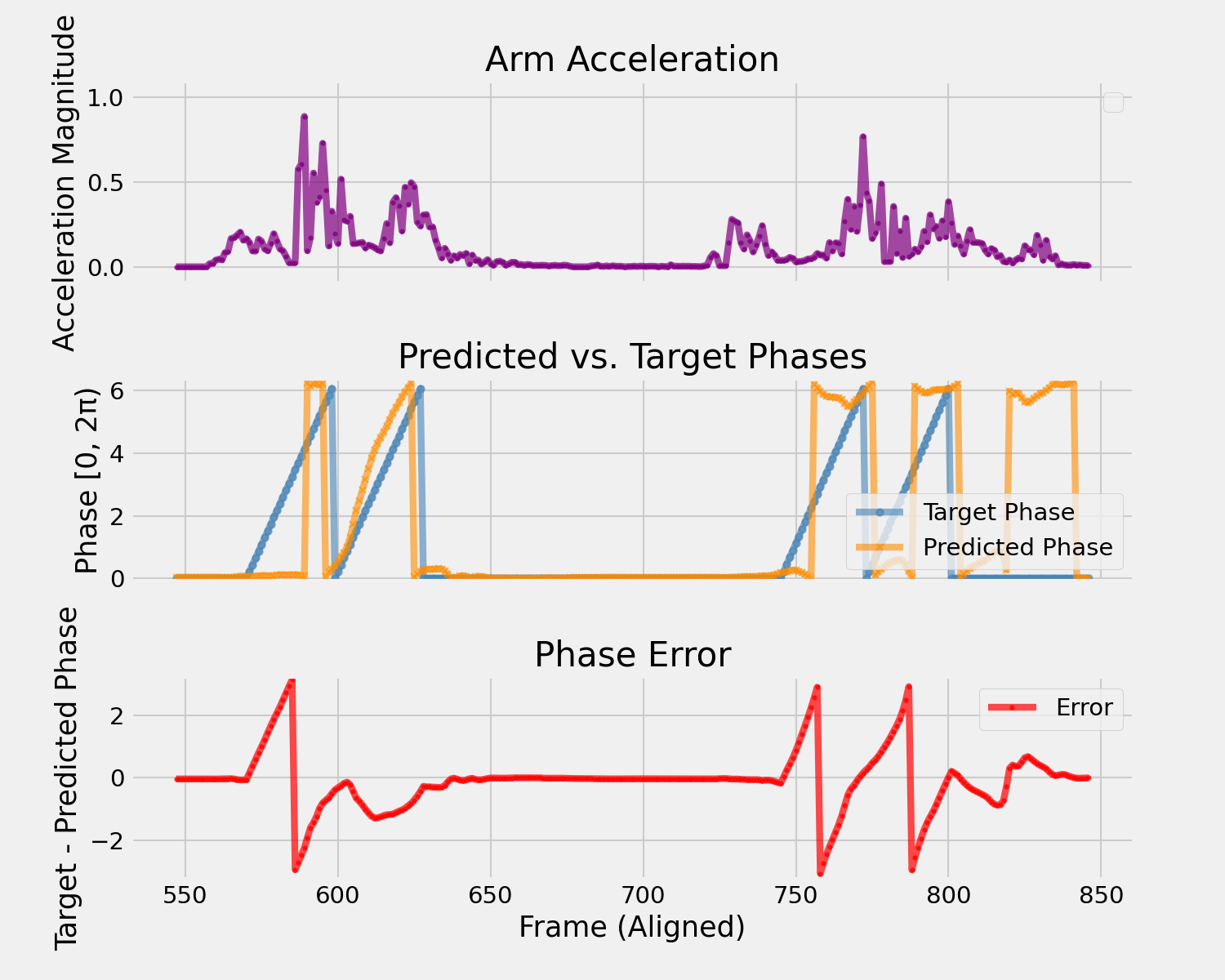}
    } 
    \subfloat[\label{fig:fermataperformanceSubject12b}\centering Regular bars, 60 Hz using $\mathbf{x}_t$]{%
        \includegraphics[width=0.48\linewidth]{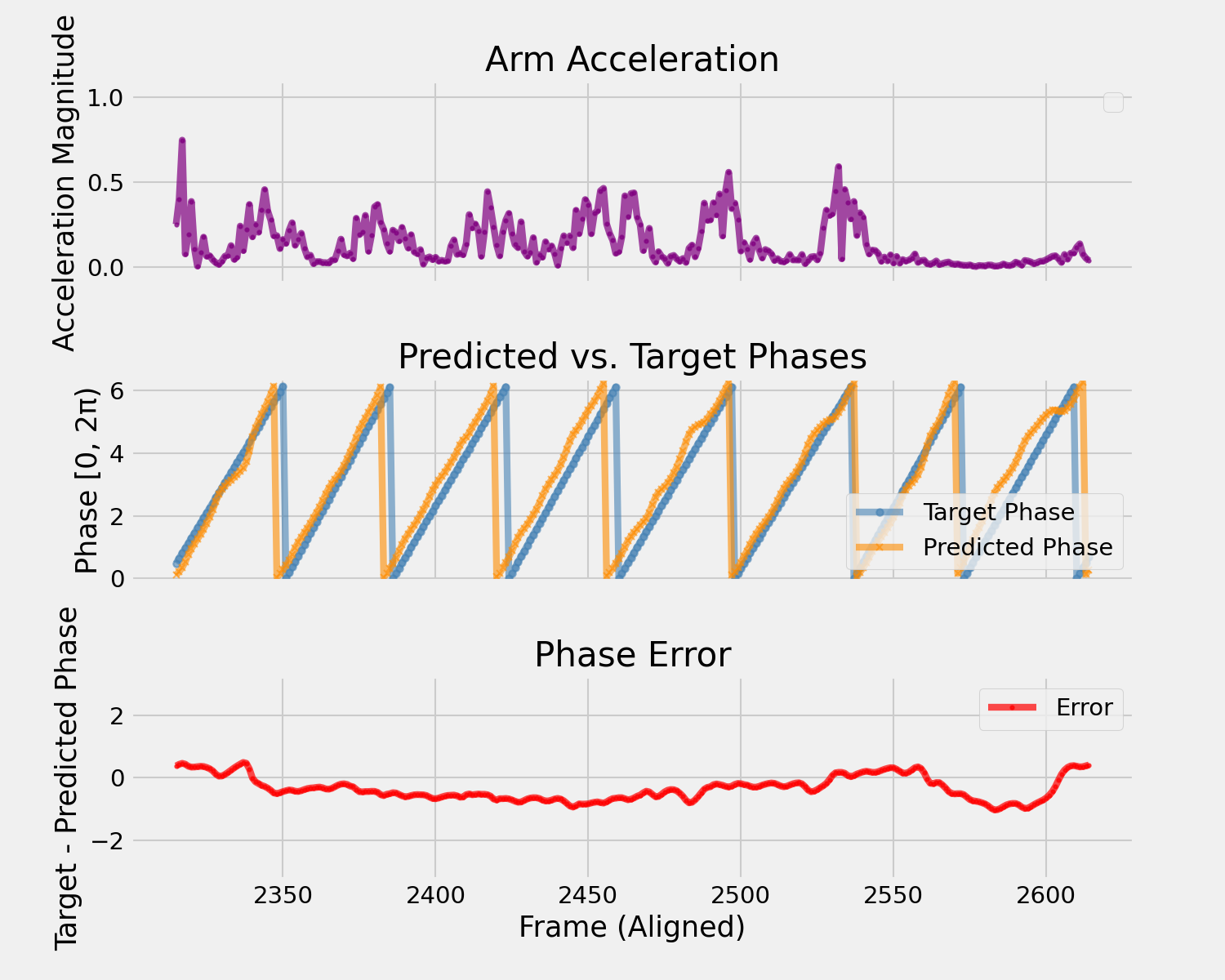}
    } \\ 

    \subfloat[\label{fig:fermataperformanceSubject12c}\centering Fermata bars, 20 Hz using $\mathbf{x}_t^\mathrm{2D}$]{%
        \includegraphics[width=0.48\linewidth]{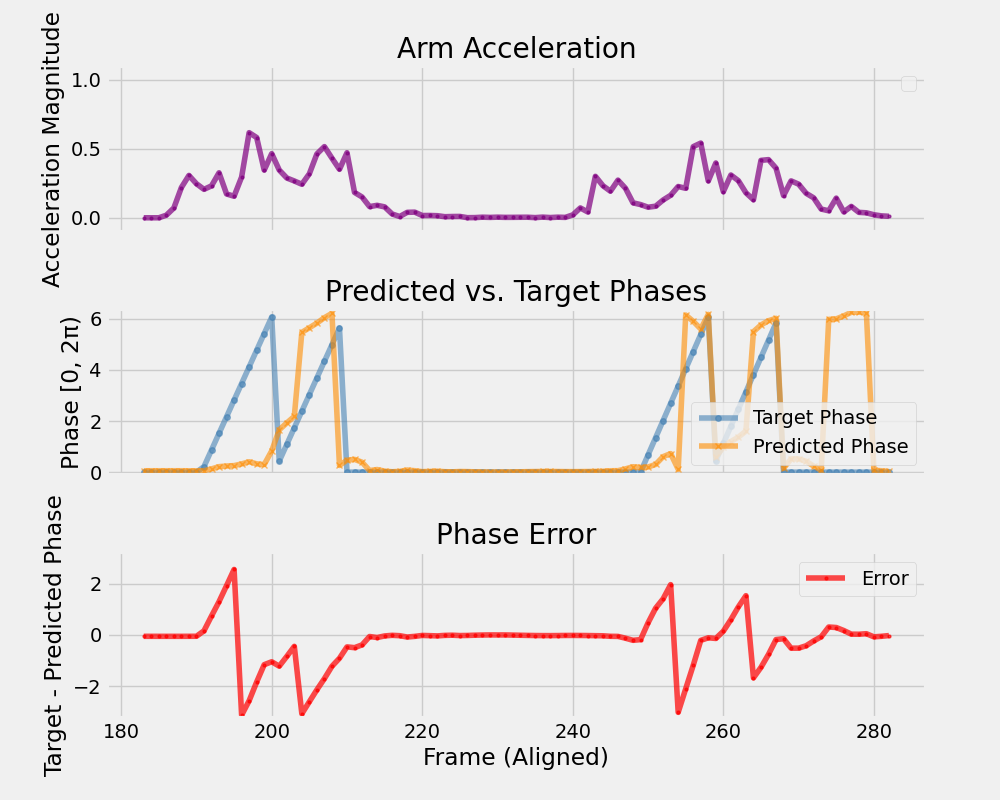}
    }
    \subfloat[\label{fig:fermataperformanceSubject12d}\centering Regular bars, 20 Hz using $\mathbf{x}_t^\mathrm{2D}$]{%
        \includegraphics[width=0.48\linewidth]{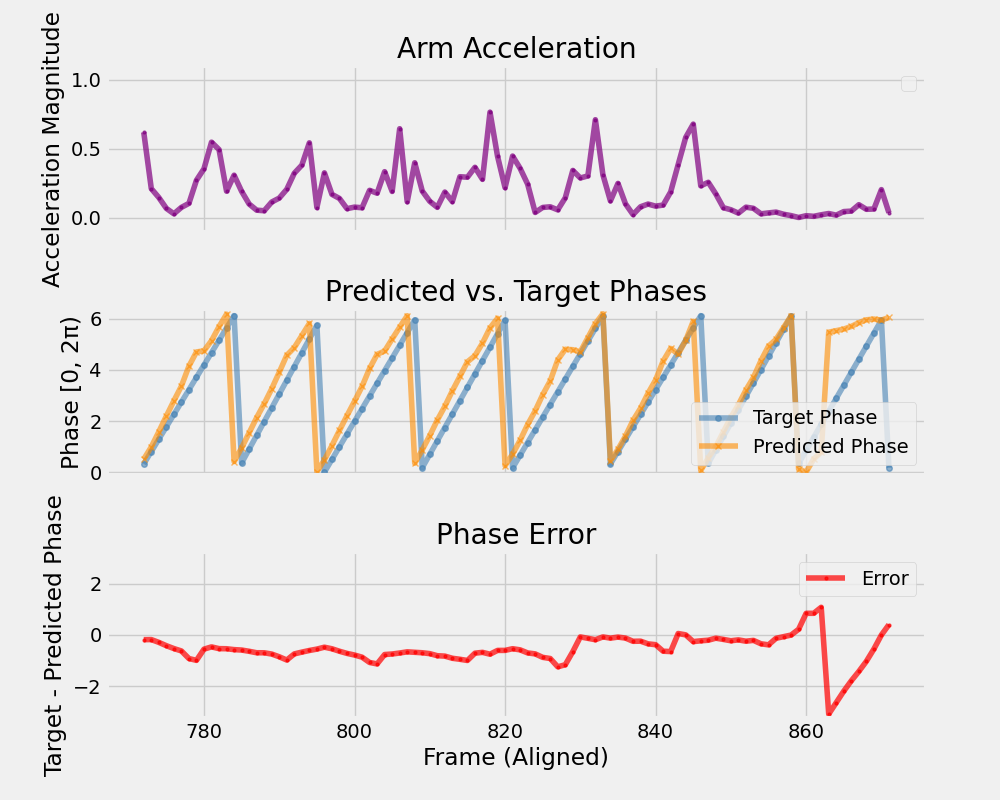}
    }
    
    \caption{Example of phase estimation for Subject 12. (a) and (b) correspond to column 1 in Table \ref{tab:cv_metrics_by_test_detail}, (c) and (d) correspond to column 4 in Table \ref{tab:cv_metrics_by_test_detail}. Top: Arm acceleration $\rho_t$. Middle: Ground truth phase $\varphi^\mathrm{gt}_t$ and estimated phase $\hat{\varphi}_t$. Bottom: Phase error $\varphi^\mathrm{gt}_t - \hat{\varphi}_t$. The three graphs are temporally aligned.}
    \label{fig:fermataperformanceSubject12}
    \vspace{-12mm}
\end{figure*}

Fig.~\ref{fig:fermataperformanceSubject2}-\ref{fig:fermataperformanceSubject12} are equivalent to Fig.~\ref{fig:fermataperformance} in the main paper, and present further examples of the performance of the gesture recognition module. 
These plots confirm the findings in Sec.~\ref{sec:evaluation_grm} of the main paper, based on the example from Subject 1 in Fig.~\ref{fig:fermataperformance}. 

While Fig.~\ref{fig:fermataperformanceSubject2b}, \ref{fig:fermataperformanceSubject3b}, \ref{fig:fermataperformanceSubject5b}, \ref{fig:fermataperformanceSubject7b}, \ref{fig:fermataperformanceSubject9b}, and \ref{fig:fermataperformanceSubject12b} represent near perfect alignment between estimated and actual within-bar phase, some misaligned sections can be detected in \ref{fig:fermataperformanceSubject4b}, \ref{fig:fermataperformanceSubject6b}, \ref{fig:fermataperformanceSubject8b}, \ref{fig:fermataperformanceSubject10b}, and \ref{fig:fermataperformanceSubject12b}. The reason is that the subjects are off-beat during the take. 

Note also that a higher prevalence of phase estimation error is observed with lower resolution data, as illustrated in Fig.~\ref{fig:fermataperformanceSubject11d} and \ref{fig:fermataperformanceSubject12d}. This phenomenon is also discussed in Sec.~\ref{sec:evaluation_grm}.

In Fig.~\ref{fig:fermataperformancea}, \ref{fig:fermataperformanceSubject2a}, \ref{fig:fermataperformanceSubject3a},
\ref{fig:fermataperformanceSubject5a},
\ref{fig:fermataperformanceSubject7a},
\ref{fig:fermataperformanceSubject9a}, and
\ref{fig:fermataperformanceSubject12a}, we see a delay in the detection of the silent upbeats.
In Fig.~\ref{fig:fermataperformanceSubject4a}, \ref{fig:fermataperformanceSubject6a}, \ref{fig:fermataperformanceSubject8a}, \ref{fig:fermataperformanceSubject10a}, \ref{fig:fermataperformanceSubject11a}, we see that the the network missed the silent upbeats before fermata bars. This is coherent with the discussion in Sec.~\ref{sec:evaluation_grm}. The way to address this problem, as discussed in Sec.~\ref{sec:future}, is more training data and a more precise human motion tracking.

This difficulty with silent upbeat detection in this piece is however not solely attributable to the neural network itself: The timing coordination in the start of Beethoven's Fifth Sympony is well known for its difficulty for both conductor and orchestra. This is partly the reason to choose this piece for a gamified experience.

Some recorded subjects achieved much higher precision in silent upbeat detection (e.g., Fig.~\ref{fig:fermataperformanceSubject3}), though no clear correlation was observed between conducting experience and upbeat accuracy in this constrained recording scenario, where the task was essentially reduced to following overlaid clicks on the audio track. 
The delays visible in Fig.~\ref{fig:fermataperformanceSubject4}, \ref{fig:fermataperformanceSubject6}, and \ref{fig:fermataperformanceSubject8} further illustrate how the artificial data collection setup, where participants followed a playback rather than leading a live ensemble, particularly affected the timing of upbeats. Nonetheless, the network successfully recognizes not only the downbeat that starts a bar, but also the motion pattern within the bar, which indicates whether the bar progresses at a constant tempo or contains a fermata.

\subsection{Additional Examples from the User Study}
\label{sec:appendixRealtime}

\begin{figure*}
    \centering
    \subfloat[\label{fig:userstudy1a_raw}\centering Fermata bars]{%
        \includegraphics[width=5.5cm]{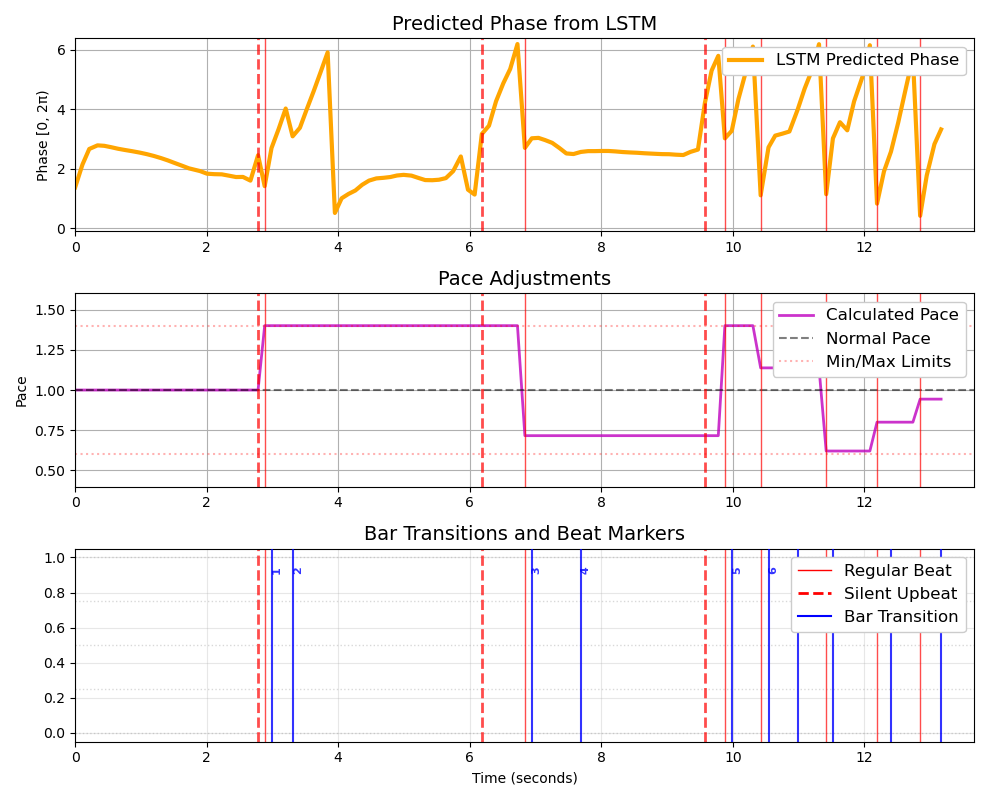}
    } 
    \subfloat[\label{fig:userstudy1b_raw}\centering Regular bars, steady pace]{%
        \includegraphics[width=5.5cm]{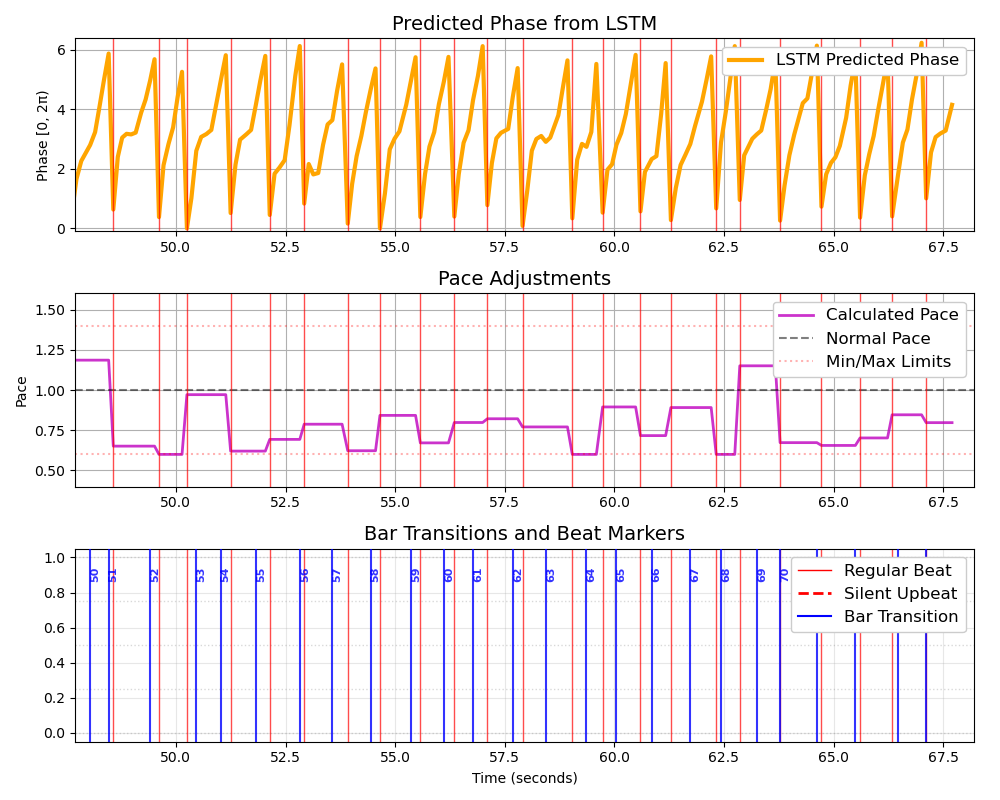}
    } 
    \subfloat[\label{fig:userstudy1c_raw}\centering Regular bars, varying pace]{%
        \includegraphics[width=5.5cm]{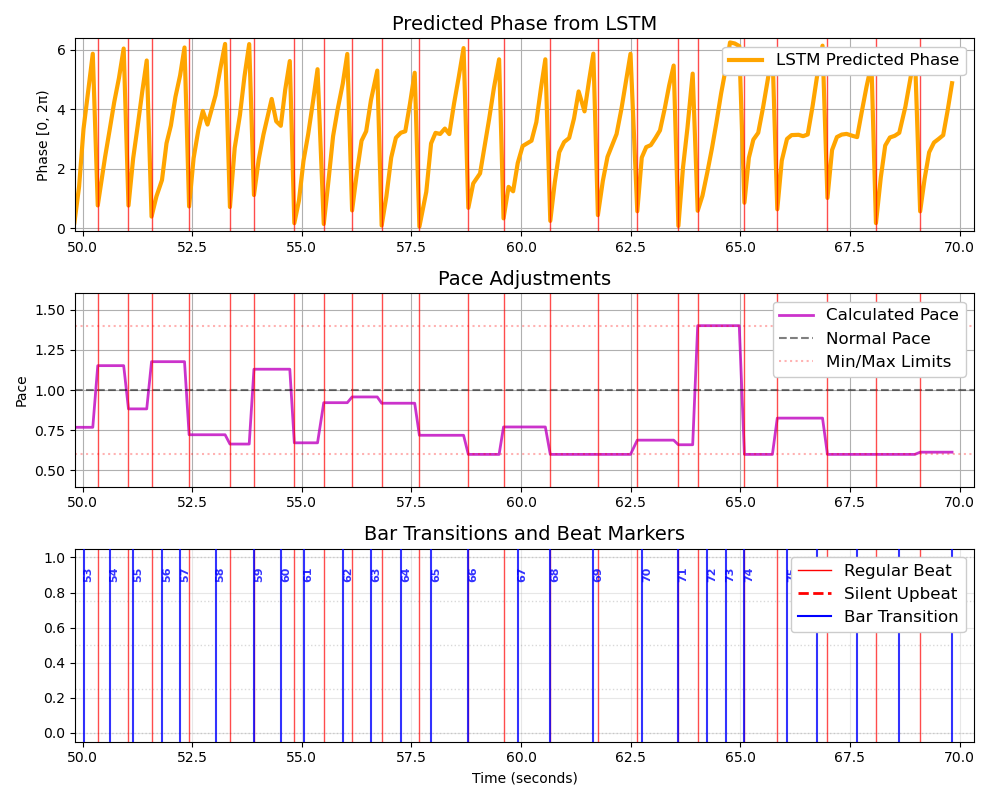}
    }
    \caption{Real-time interaction of User 1 in Raw setting. Top: Predicted phase $\hat{\varphi}_t$. Middle: Pace adjustments according to the detected beats. Bottom: Beats and bar beginnings.}
    \label{fig:userstudy1_raw_app}
\end{figure*}

\begin{figure*}
\centering
    \subfloat[\label{fig:userstudy1a_wa}\centering Fermata bars]{%
        \includegraphics[width=5.5cm]{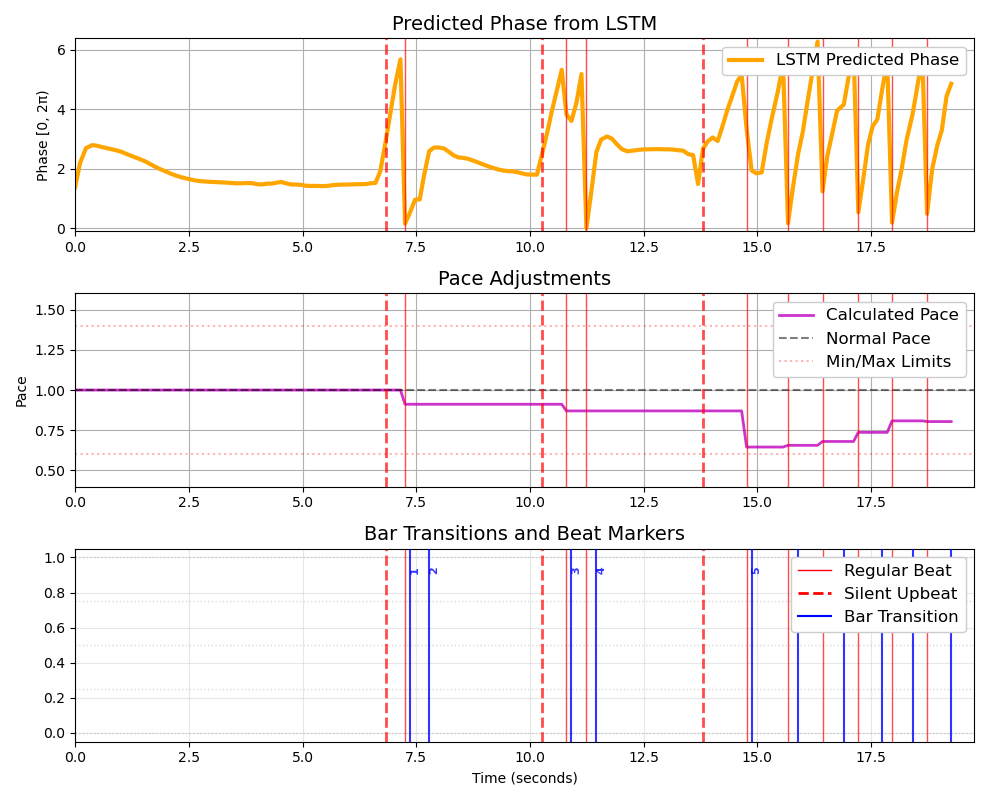}
    } 
    \subfloat[\label{fig:userstudy1b_wa}\centering Regular bars, steady pace]{%
        \includegraphics[width=5.5cm]{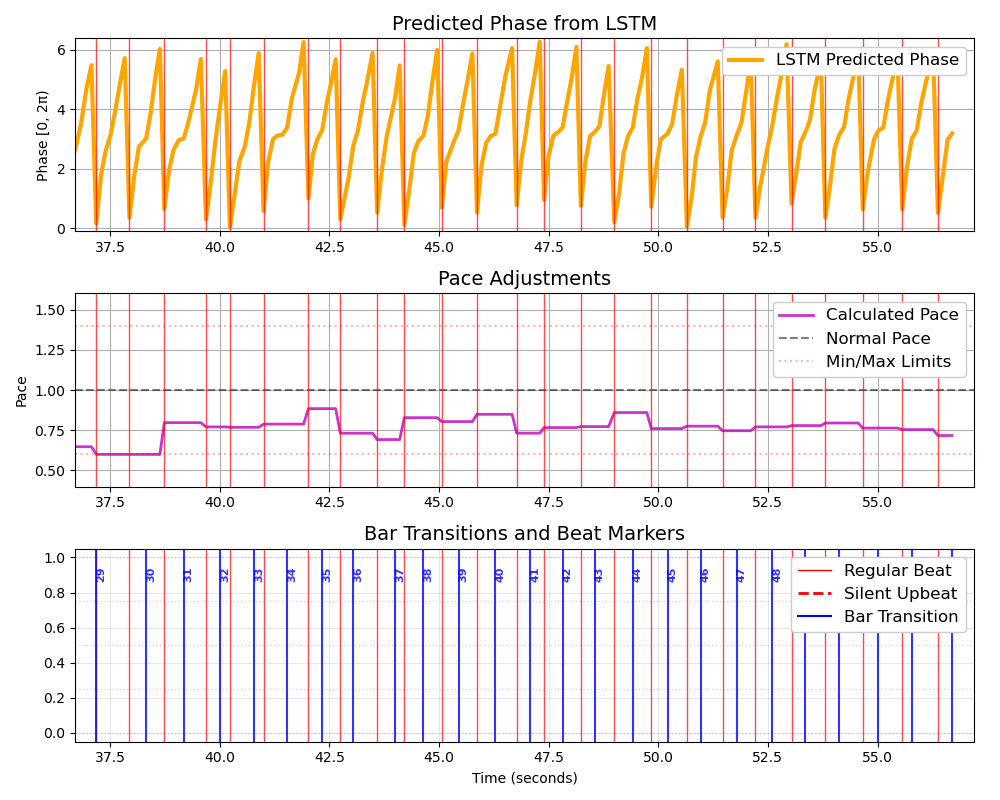}
    } 
    \subfloat[\label{fig:userstudy1c_wa}\centering Regular bars, varying pace]{%
        \includegraphics[width=5.5cm]{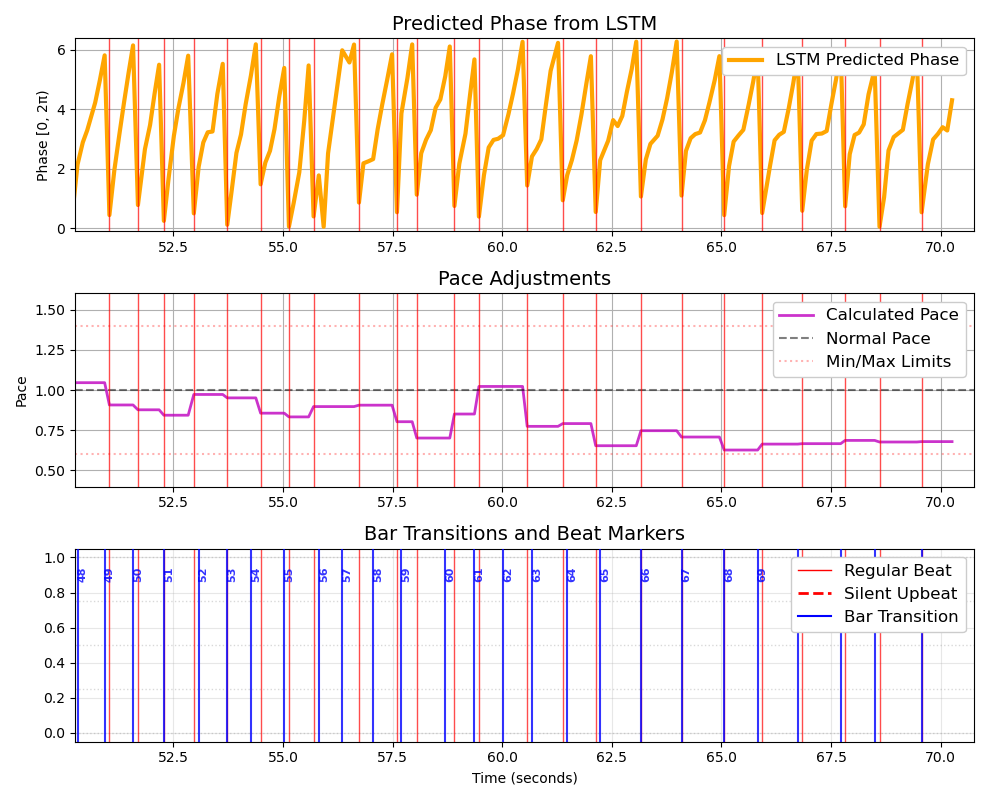}
    }
    \caption{Real-time interaction of User 1 in Average setting. Top: Predicted phase $\hat{\varphi}_t$. Middle: Pace adjustments according to the detected beats. Bottom: Beats and bar beginnings.}
    \label{fig:userstudy1_wa_app}
\end{figure*}

\begin{figure*}
    \centering
    \subfloat[\label{fig:userstudy2a_raw}\centering Fermata bars]{%
        \includegraphics[width=5.5cm]{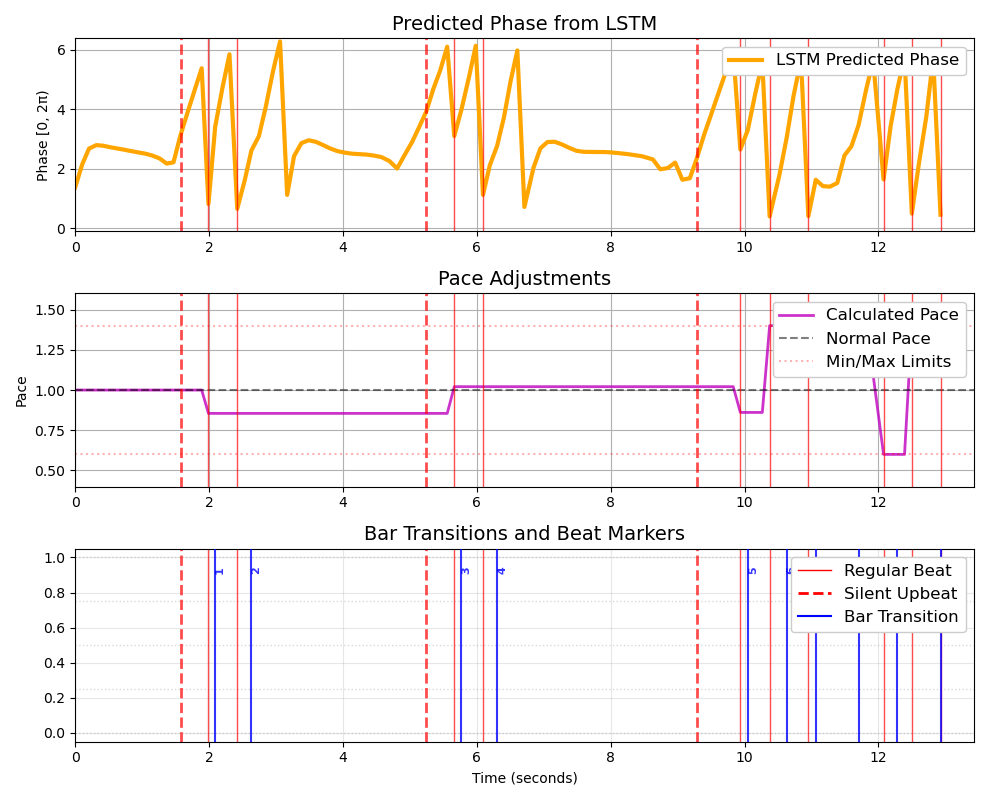}
    } 
    \subfloat[\label{fig:userstudy2_braw}\centering Regular bars, steady pace]{%
        \includegraphics[width=5.5cm]{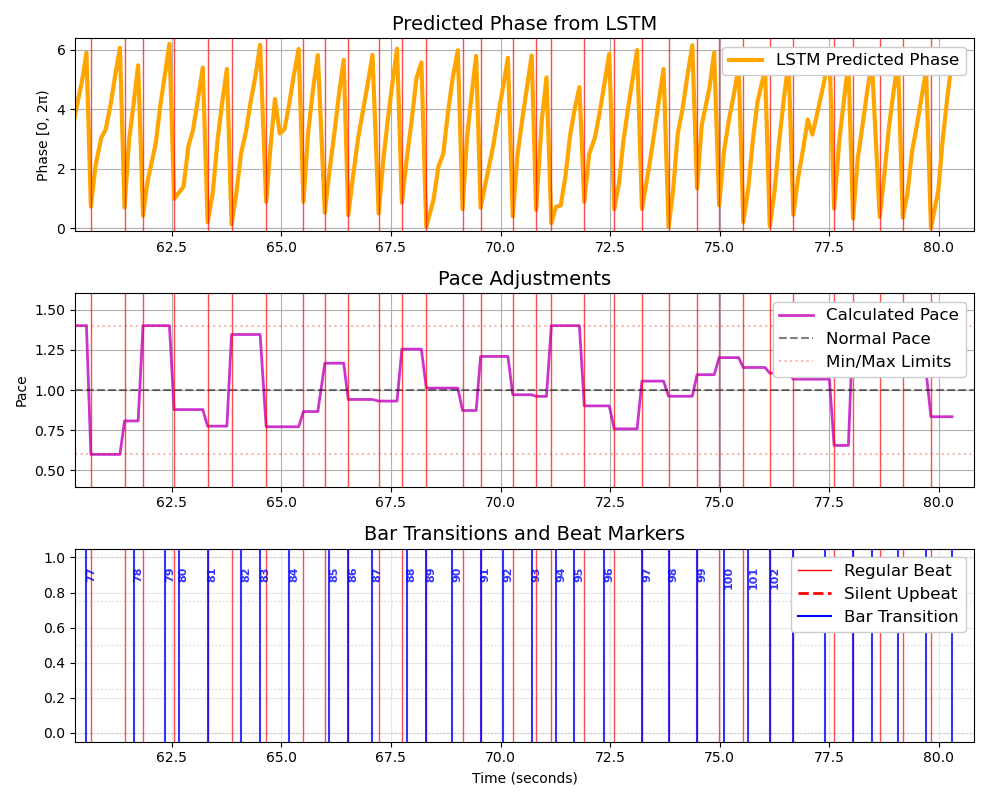}
    } 
    \subfloat[\label{fig:userstudy2c_raw}\centering Regular bars, varied pace]{%
        \includegraphics[width=5.5cm]{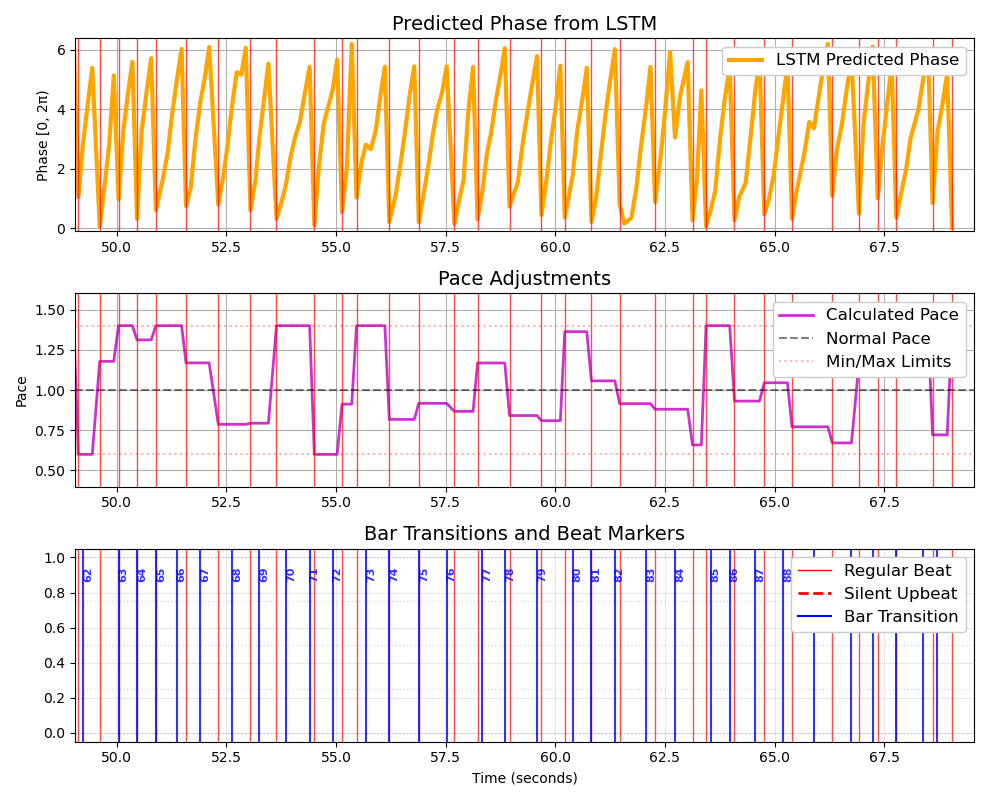}
    }
\caption{Real-time interaction of User 2 in Raw setting. Top: Predicted phase $\hat{\varphi}_t$. Middle: Pace adjustments according to the detected beats. Bottom: Beats and bar beginnings.}
\label{fig:userstudy2_raw_app}
\end{figure*}

\begin{figure*}
    \centering
    \subfloat[\label{fig:userstudy2a_median}\centering Fermata bars]{%
        \includegraphics[width=5.5cm]{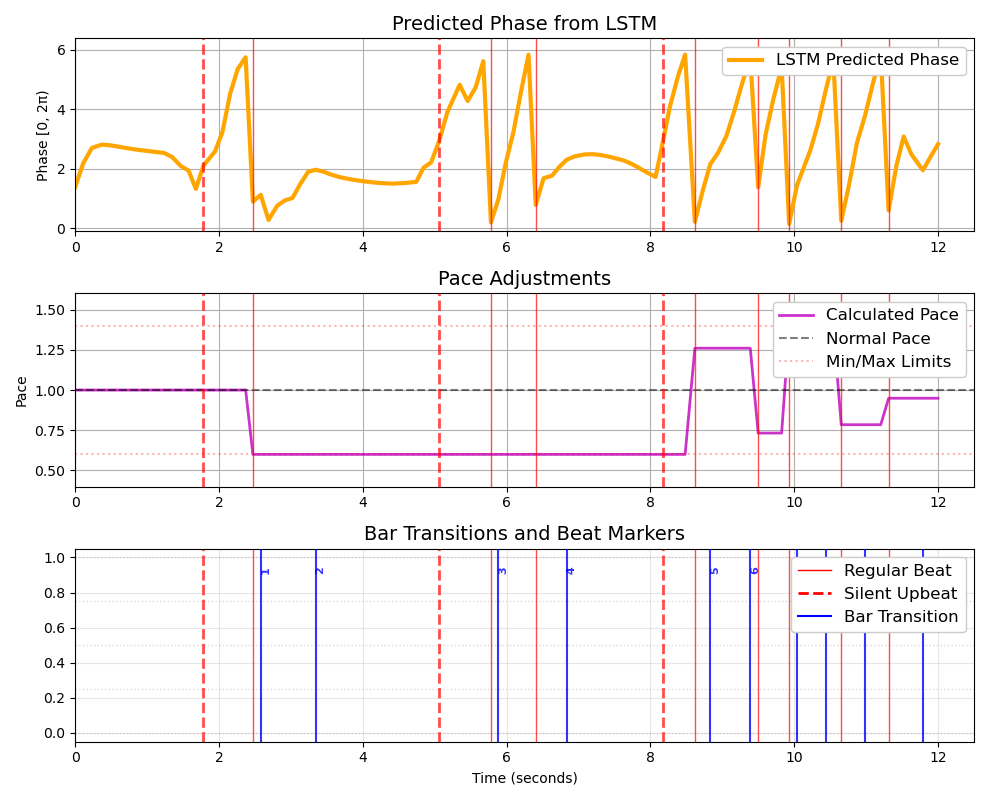}
    } 
    \subfloat[\label{fig:userstudy2b_median}\centering Regular bars, steady pace]{%
        \includegraphics[width=5.5cm]{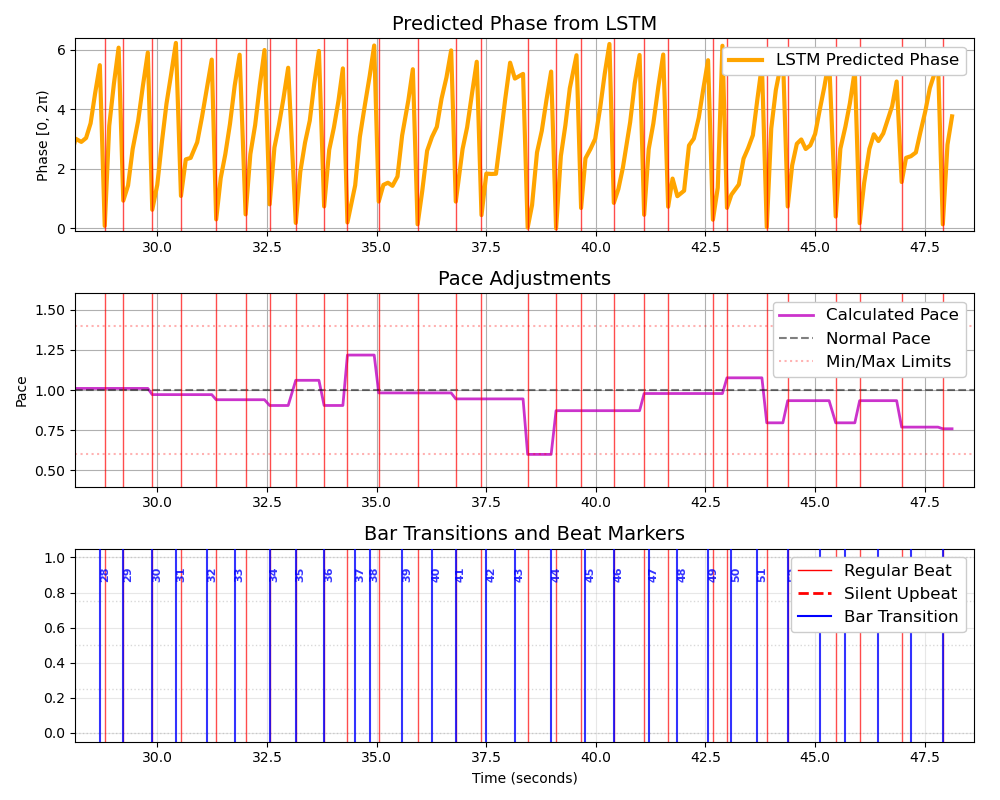}
    } 
    \subfloat[\label{fig:userstudy2c_median}\centering Regular bars, varying pace]{%
        \includegraphics[width=5.5cm]{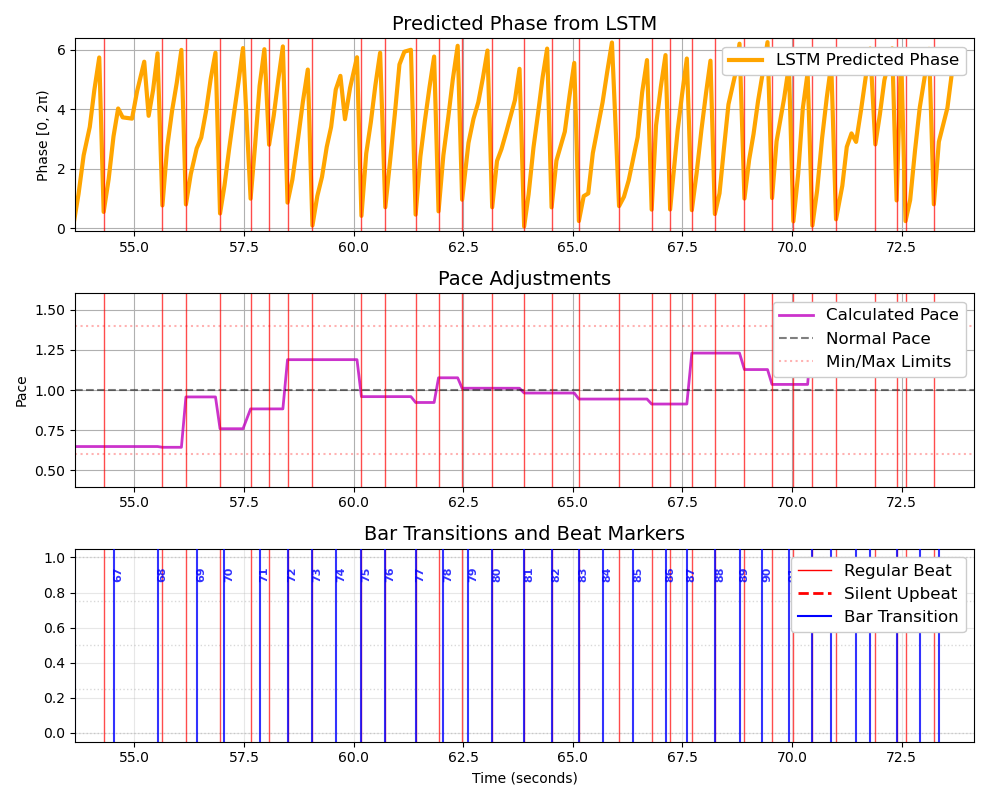}
    }
\caption{Real-time interaction of User 2 in Median setting. Top: Predicted phase $\hat{\varphi}_t$. Middle: Pace adjustments according to the detected beats. Bottom: Beats and bar beginnings.}
\label{fig:userstudy2_median_app}
\end{figure*}

\begin{figure*}
    \centering
    \subfloat[\label{fig:userstudy2a_wa}\centering Fermata bars]{%
        \includegraphics[width=5.5cm]{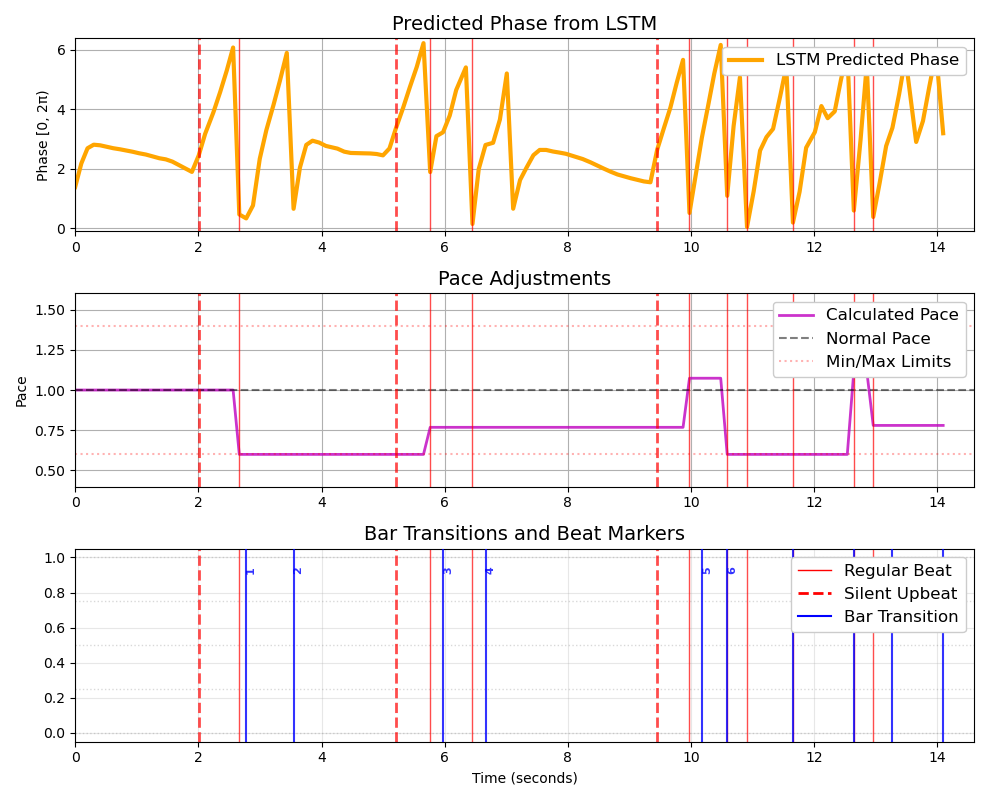}
    } 
    \subfloat[\label{fig:userstudy2b_wa}\centering Regular bars, steady pace]{%
        \includegraphics[width=5.5cm]{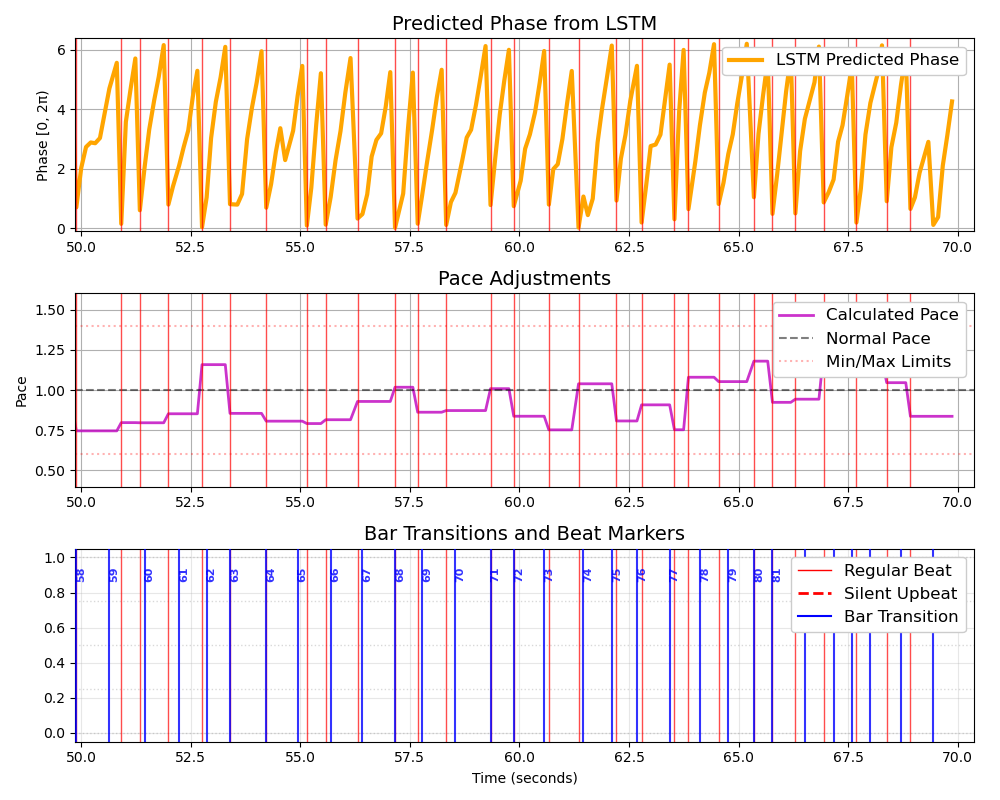}
    } 
    \subfloat[\label{fig:userstudy2c_wa}\centering Regular bars, varying pace]{%
        \includegraphics[width=5.5cm]{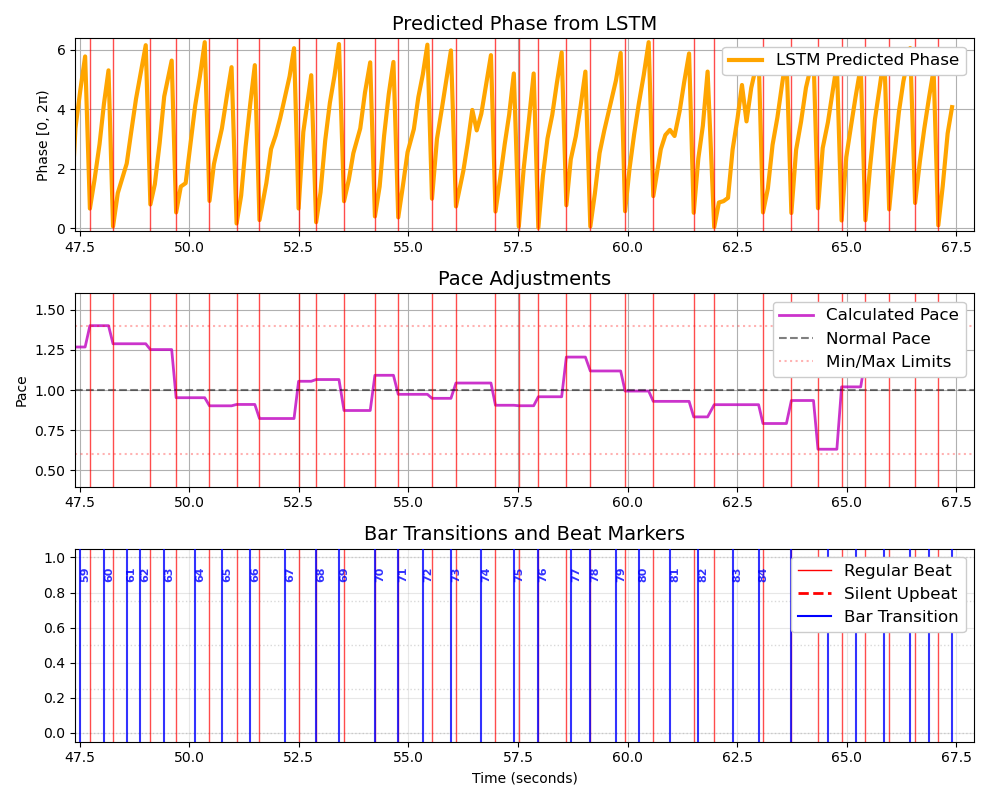}
    }
\caption{Real-time interaction of User 2 in  Average setting. Top: Predicted phase $\hat{\varphi}_t$. Middle: Pace adjustments according to the detected beats. Bottom: Beats and bar beginnings.}
\label{fig:userstudy2_wa_app}
\end{figure*}

\begin{figure*}
    \centering
    \subfloat[\label{fig:userstudy3a_raw}\centering Fermata bars]{%
        \includegraphics[width=5.5cm]{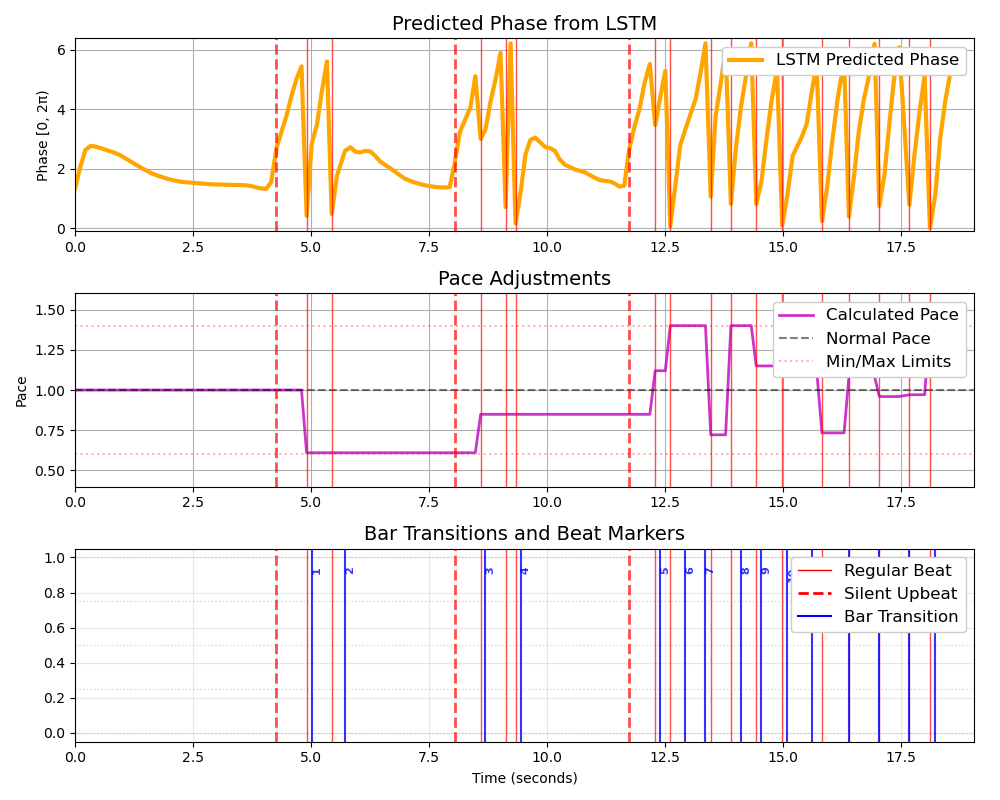}
    } 
    \subfloat[\label{fig:userstudy3b_raw}\centering Regular bars, steady pace]{%
        \includegraphics[width=5.5cm]{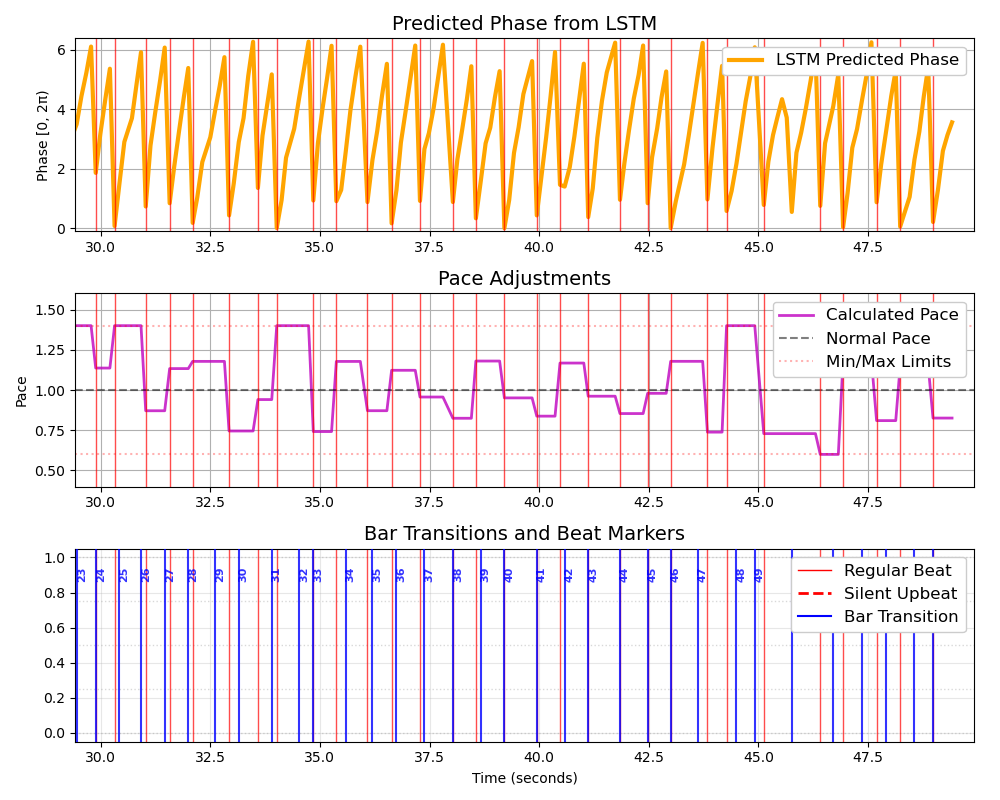}
    } 
    \subfloat[\label{fig:userstudy3c_raw}\centering Regular bars, varying pace]{%
        \includegraphics[width=5.5cm]{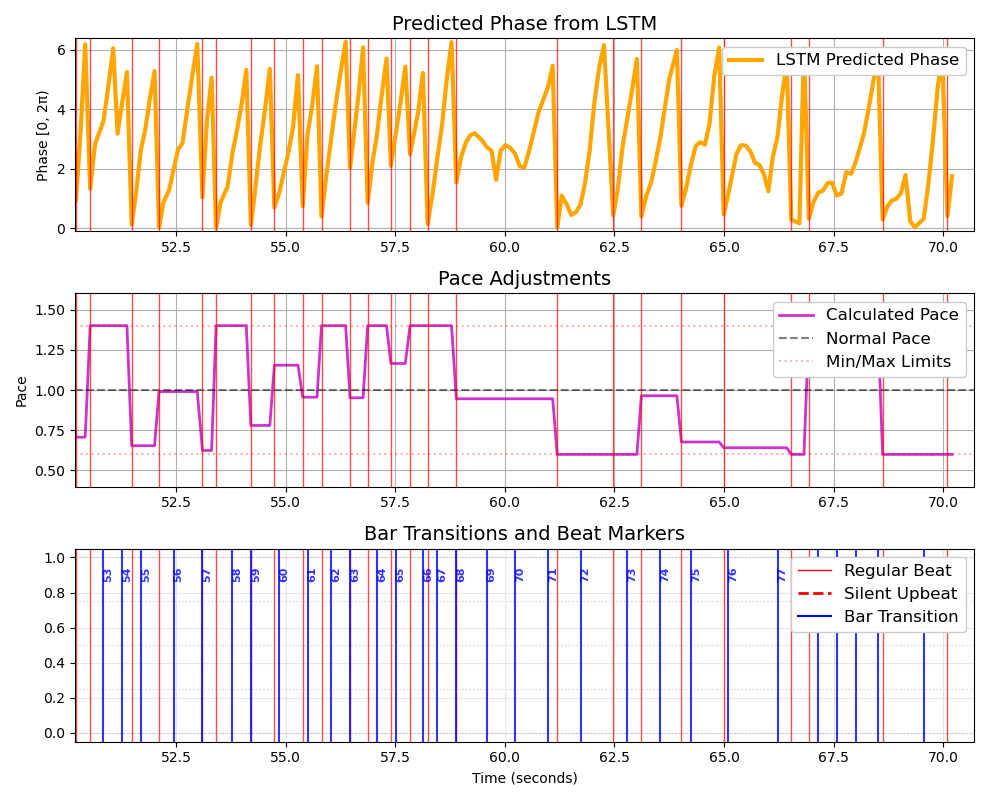}
    }
\caption{Real-time interaction of User 3 in Raw setting. Top: Predicted phase $\hat{\varphi}_t$. Middle: Pace adjustments according to the detected beats. Bottom: Beats and bar beginnings.}
\label{fig:userstudy3_raw_app}
\end{figure*}

\begin{figure*}
    \centering
    \subfloat[\label{fig:userstudy3a_median}\centering Fermata bars]{%
        \includegraphics[width=5.5cm]{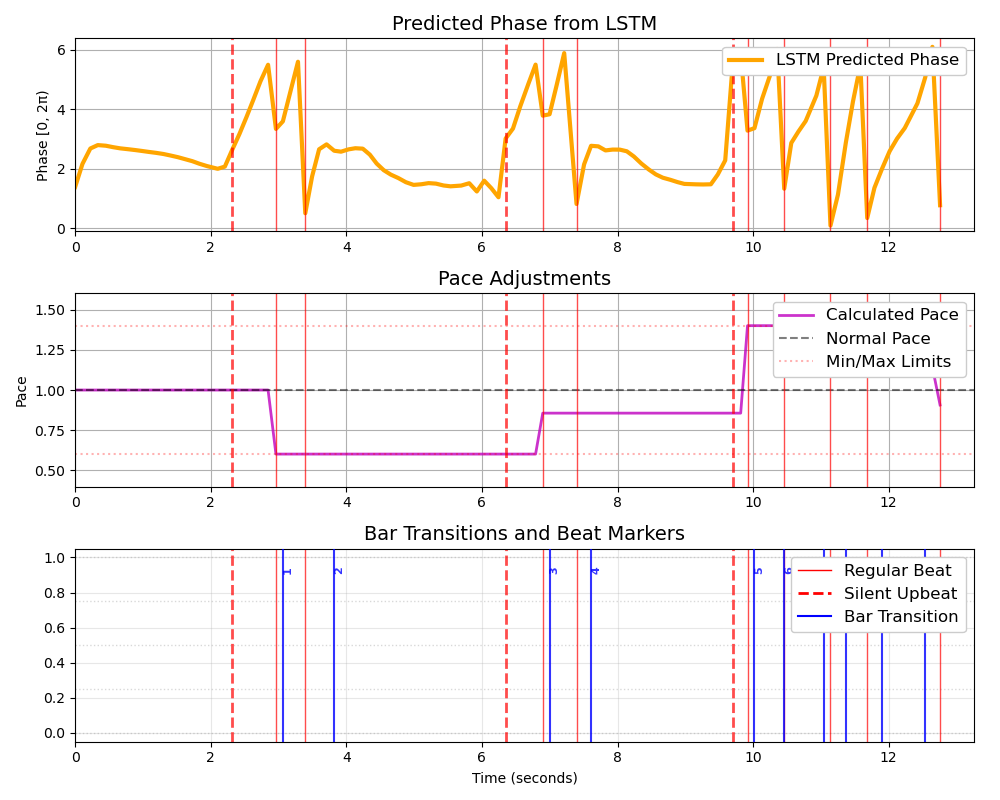}
    } 
    \subfloat[\label{fig:userstudy3b_median}\centering Regular bars, steady pace]{%
        \includegraphics[width=5.5cm]{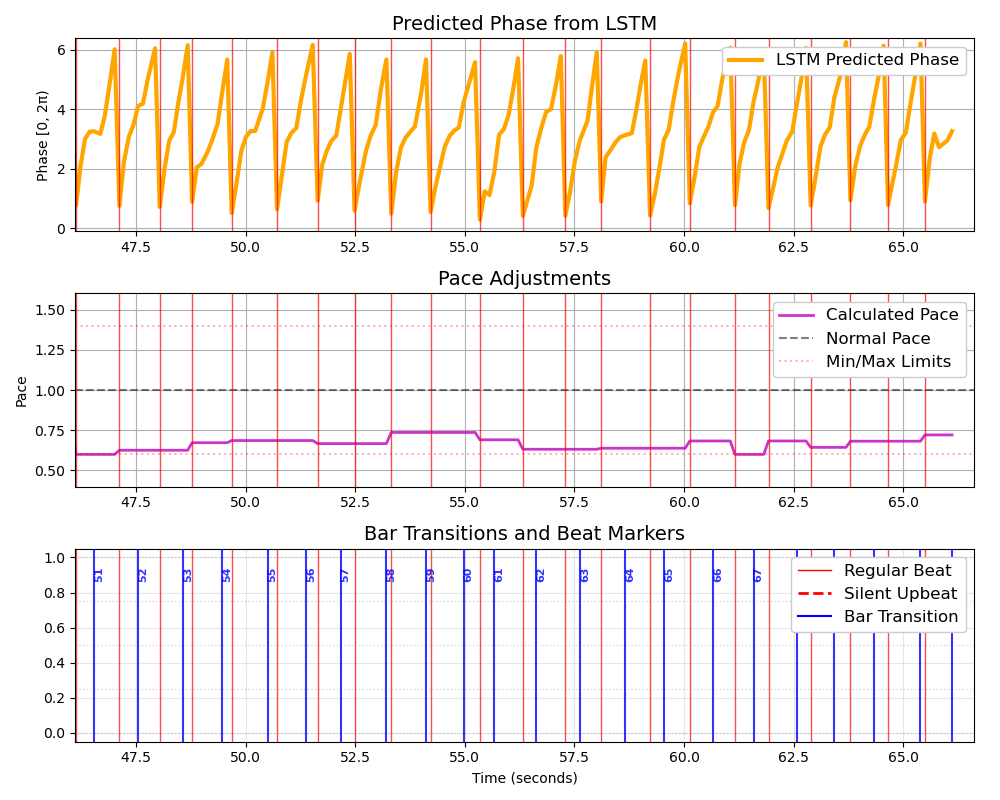}
    } 
    \subfloat[\label{fig:userstudy3c_median}\centering Regular bars, varying pace]{%
        \includegraphics[width=5.5cm]{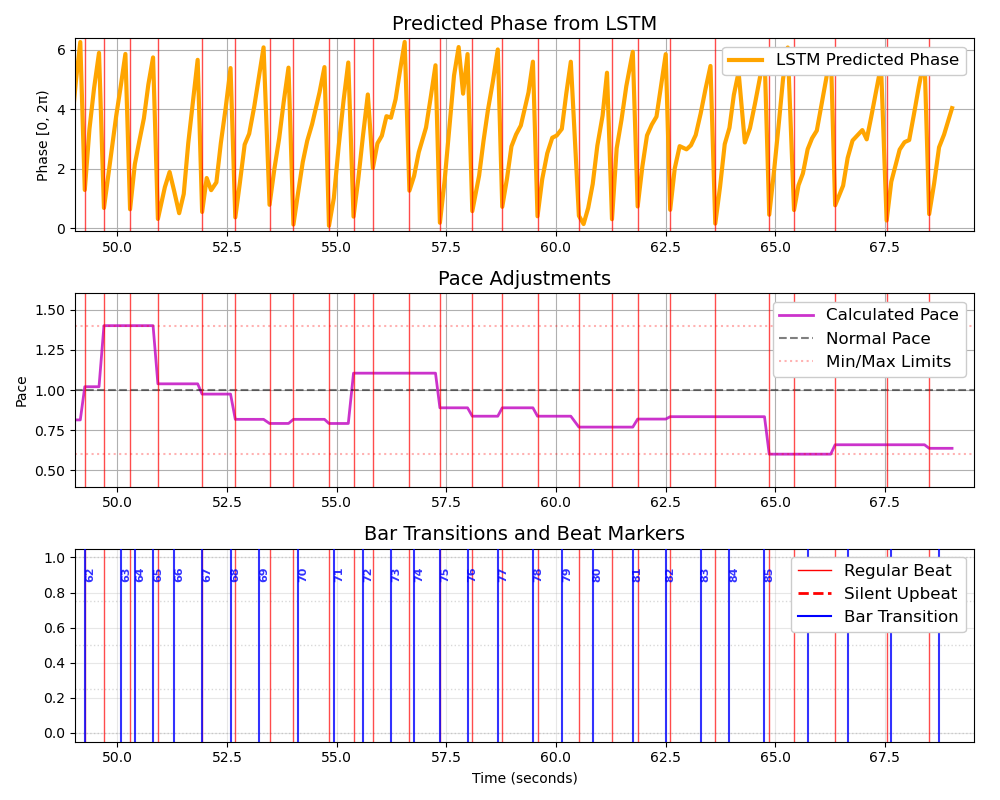}
    }
\caption{Real-time interaction of User 3 in Median setting. Top: Predicted phase $\hat{\varphi}_t$. Middle: Pace adjustments according to the detected beats. Bottom: Beats and bar beginnings.}
\label{fig:userstudy3_median_app}
\end{figure*}

\begin{figure*}
    \centering
    \subfloat[\label{fig:userstudy3a_wa}\centering Fermata bars]{%
        \includegraphics[width=5.5cm]{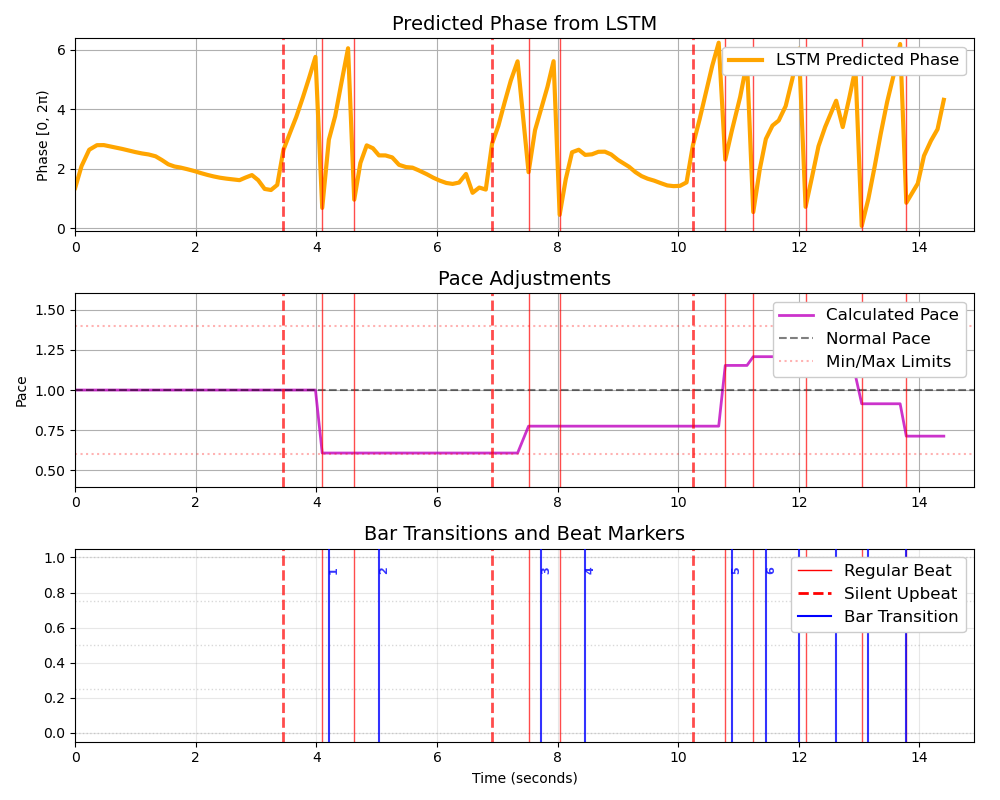}
    } 
    \subfloat[\label{fig:userstudy3b_wa}\centering Regular bars, steady pace]{%
        \includegraphics[width=5.5cm]{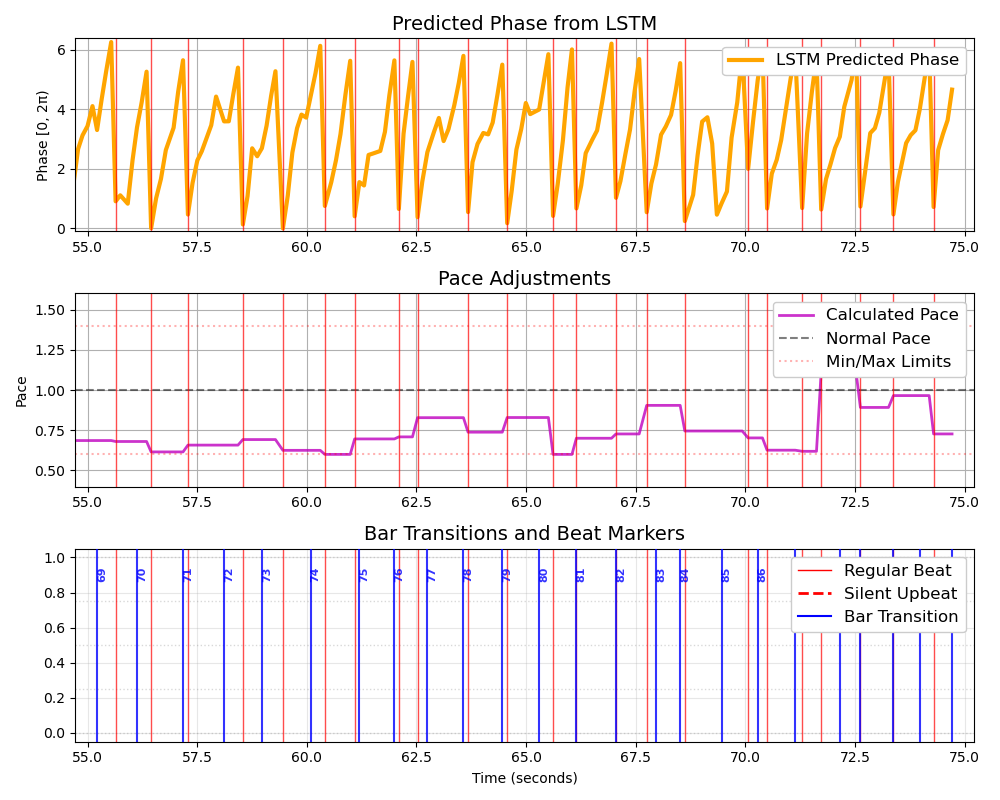}
    } 
    \subfloat[\label{fig:userstudy3c_wa}\centering Regular bars, varying pace]{%
        \includegraphics[width=5.5cm]{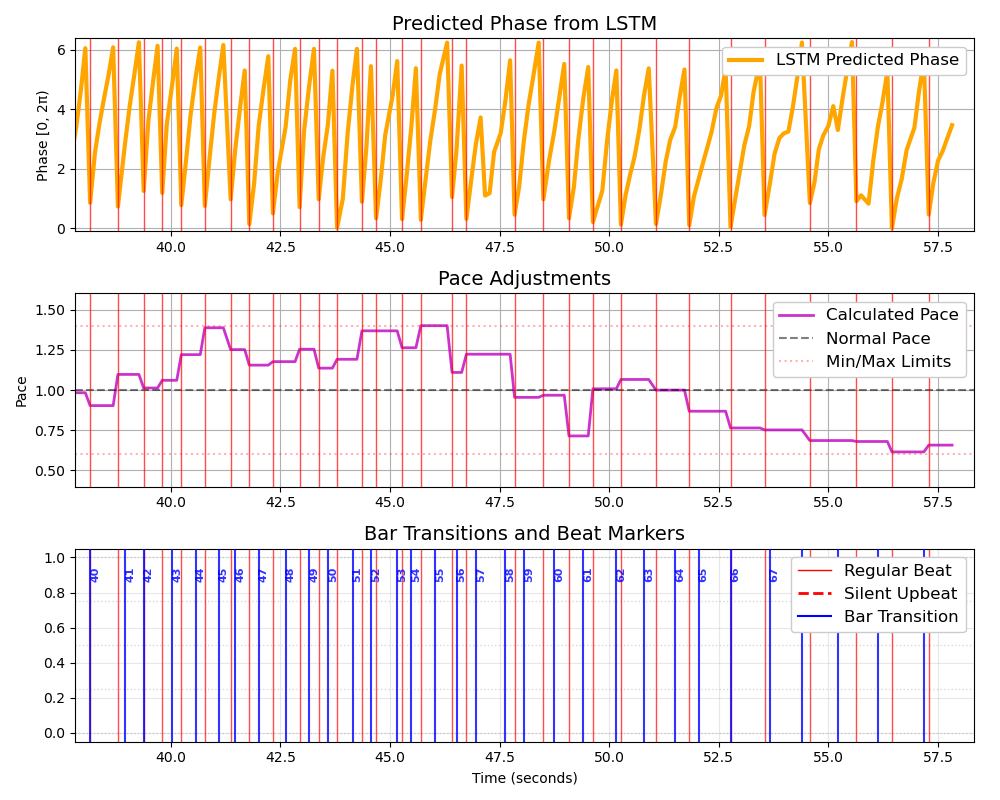}
    }
\caption{Real-time interaction of subject User 3 in Average setting. Top: Predicted phase $\hat{\varphi}_t$. Middle: Pace adjustments according to the detected beats. Bottom: Beats and bar beginnings.}
\label{fig:userstudy3_wa_app}
\end{figure*}

We here provide more examples from the user study described in Sec.~6.2.
Fig.~\ref{fig:userstudy1_raw_app}–\ref{fig:userstudy3_wa_app} show examples from all user recordings in Sec.~\ref{sec:RealTimeController}. 

Upbeats are generally well detected (Fig.~\ref{fig:userstudy1a_wa}, \ref{fig:userstudy2a_raw}, \ref{fig:userstudy2a_median}, \ref{fig:userstudy2a_wa}, \ref{fig:userstudy3a_raw}, \ref{fig:userstudy3a_median}, \ref{fig:userstudy3a_wa}), with the exception of Fig.~\ref{fig:userstudy1a_raw}, where the first upbeat is misdetected, leading to an unintended higher tempo.

User 1 and User 3 tended to select a slower, steadier pace, which is reflected in the stable playback of the Median setting (Fig.~\ref{fig:userstudy1b_median}, \ref{fig:userstudy3b_median}). In contrast, User 2 (the orchestra conductor) conducted with faster gestures, causing deviations in Fig.~\ref{fig:userstudy2b_median}. The Average setting performs similarly to Median (Fig.~\ref{fig:userstudy1c_wa}, \ref{fig:userstudy3c_wa}) but responds more quickly to tempo changes (Fig.~\ref{fig:userstudy1b_wa}, \ref{fig:userstudy2b_wa}, \ref{fig:userstudy3b_wa}). Raw shows the largest deviations overall (Fig.~\ref{fig:userstudy1a_raw}–\ref{fig:userstudy3a_raw}). All is coherent with the discussion in Sec.~\ref{sec:handsonperformance}.


\subsection{Additional Results from the Field Study}
\label{sec:AppendixFieldStudy}
We here present some additional measurements from the field study (Sec.~6.3). 
The distances between the beats and the nearest bar starts, and the average bar length for each subject in the field study is presented in Table \ref{tab:field_study}.

The results indicate that the system consistently follows the 
user's conducting gestures across all takes. The mean distance as a percentage of the average 
bar length has an overall mean of $24.2 \pm 2.2\%$, which is considerably lower than the $50\%$ 
expected from a system advancing playback arbitrarily, suggesting that the system is genuinely 
tracking the user's gestures rather than progressing independently. The low standard deviation 
of $2.2\%$ further confirms that performance is highly consistent across takes, regardless of 
individual differences in conducting style or tempo.


\clearpage

\clearpage
\end{document}